\documentclass[a4paper,12pt]{article}
\usepackage{mathrsfs}
\usepackage{soul}
\usepackage[usenames,dvipsnames]{color}
\usepackage{jheppub}
\usepackage{amsmath, amssymb, slashed, epsf, color, graphicx, latexsym}
\usepackage{epsfig}
\usepackage{graphics}
    \usepackage[utf8]{inputenc}

\begin{document}
	
	\preprint{TIFR/TH/17-53}
	\title{An Action for and Hydrodynamics from \\ the improved Large $D$ membrane}
	\author[a]{Yogesh Dandekar} 
	\author[a]{, Suman Kundu}
	\author[a,b]{, Subhajit Mazumdar}
	\author[a]{, Shiraz Minwalla}
	\author[a] {, Amiya Mishra}
	\author[a] {and Arunabha Saha}
	\affiliation[a]{Tata Institute of Fundamental Research, Mumbai, India-400005}
	\affiliation[b]{The Racah Institute of Physics, The Hebrew University of Jerusalem, Jerusalem 91904, Israel}
	\emailAdd{yogesh@theory.tifr.res.in}
	\emailAdd{suman.kundu\_290@tifr.res.in}
	\emailAdd{mazumdar.subhajit@mail.huji.ac.il}
	\emailAdd{minwalla@theory.tifr.res.in}
	\emailAdd{amiya.mishra@theory.tifr.res.in}
	\emailAdd{arunabha@theory.tifr.res.in}
	
	\abstract{It has recently been demonstrated that black hole dynamics at large $D$ is dual to the motion 
	of a probe membrane propagating in the background of a spacetime that solves Einstein's equations. The equation 
	of motion of this membrane is determined by the membrane stress tensor. In this paper we `improve' the membrane 
	stress tensor derived in earlier work to ensure that it defines consistent probe membrane dynamics even at finite $D$ while 
	reducing to previous results at large $D$. Our improved stress tensor is the 
	sum of a Brown York term and a fluid energy momentum tensor. The fluid has an unusual equation of 
	state; its pressure is nontrivial but its energy density vanishes. We demonstrate that
	all stationary solutions of our membrane equations are produced by the extremization of an action functional 
	of the membrane shape. Our action is an offshell generalization of the membrane's	
	thermodynamical partition function. We demonstrate that the thermodynamics of static spherical 
	membranes in flat space and global AdS space exactly reproduces the thermodynamics of the 
	dual Schwarzschild black holes even at finite $D$. We study the long wavelength dynamics of membranes in AdS 
	space, and demonstrate that the boundary `shadow' of this membrane dynamics is boundary hydrodynamics with 
	with a definite constitutive relation. We determine the explicit form of shadow dual boundary stress 
        tensor upto second order in derivatives of the boundary temperature and velocity,  and verify that this 
        stress tensor agrees exactly with the fluid gravity stress tensor to first order in derivatives, 
        but deviates from the later at second order and finite $D$.}

\maketitle
\section{Introduction}

It has recently been demonstrated that the dynamics of black holes in a large 
number of dimensions is `dual' to the motion of a probe membrane \footnote{
The development of this  `membrane-gravity' correspondence was motivated by early observations and computations 
\cite{Emparan:2013moa,Emparan:2013xia, Emparan:2014cia,Emparan:2014aba} by Emparan, Suzuki and Tanabe (EST) (see also \cite{Giribet:2013wia,Prester:2013gxa,Emparan:2013oza}). A precise formulation of the duality between black hole motion 
and the solutions of an initial value problem for membrane motion was presented in 
\cite{Bhattacharyya:2015dva,Bhattacharyya:2015fdk,Dandekar:2016fvw,Dandekar:2016jrp,Bhattacharyya:2016nhn,Bhattacharyya:2017hpj}. 
Parallel work developing the effective description of black hole dynamics at large $D$ in various special limits 
and using it to address physical questions of interest 
can be found in
\cite{Emparan:2014jca,Emparan:2015rva,Emparan:2015hwa,Suzuki:2015iha,Suzuki:2015axa,Emparan:2015gva,Tanabe:2015hda,Tanabe:2015isb,Chen:2015fuf,Emparan:2016sjk,Sadhu:2016ynd,Herzog:2016hob,Tanabe:2016pjr,Tanabe:2016opw,Rozali:2016yhw,Chen:2016fuy,Chen:2017wpf,Chen:2017hwm,Rozali:2017bll,Chen:2017rxa}.}
propagating without back reaction on any background that solves Einstein's equations. 
\footnote{The reason that the membrane does not correct the spacetime in which it moves is essentially kinematical . It follows from 
Newton's law that the `Coulombic' fields of the membrane die off with distance away from the membrane like $1/r^{D-3} \sim e^{-(D-3)\ln r} $ 
and so are exponentially small at fixed distances away from the membrane. It turns out that radiation fields from the membrane 
die off even more rapidly  - like $\frac{1}{D^D}$ \cite{Bhattacharyya:2016nhn}. Consequently the effect of the membrane on the background 
geometry is extremely small at distances larger than those of order $\frac{1}{D}$ away from the membrane; this is the case even though the 
membrane stress tensor is not small at large $D$.} The degrees of freedom of this probe membrane are its shape
(one degree of freedom)  and a velocity  field ($D-2$ degrees of freedom)  that 
lives on its world volume. The membrane hosts a stress tensor which is given in terms of the shape and velocity field. 
The equations of motion for the membrane variables are generated by the requirement that the membrane stress tensor 
is conserved. This requirement yields as many equations as variables - and so presumably defines well posed probe dynamics 
- as we now explain in more detail. 

The membrane stress tensor $T_{MN}$ - viewed as a tensor field in the background space time on which the membrane propagates 
- is delta function localized on the membrane world volume. The tensor indices $M$ and $N$ lie purely 
`within' the membrane world volume (i.e. $T_{MN} n^M=0$ where $n^M$ is the normal to the membrane world volume), 
so this stress tensor is equally well characterized by its restriction, $T_{\mu\nu}$, to the membrane world volume 
of the membrane. The membrane is a  consistent source for gravitational fluctuations about the background 
spacetime in which it propagates 
if and only if its stress tensor field is conserved in spacetime  i.e. if 
\begin{equation}\label{emcons}
E_{M} = \nabla^NT_{NM}=0
\end{equation}

The projection of \eqref{emcons} tangent to the membrane world volume imposes the world volume
stress tensor  conservation equations \footnote{In the equation below 
$\nabla_\mu$ is the covariant derivative on the world volume of the membrane.}
\begin{equation}\label{sttr}
\nabla^\mu T_{\mu\nu}=0
\end{equation}

On the other hand the normal component of the equation of motion yields
\begin{equation}\label{introkt}
n_M E^M \propto T_{\mu\nu} K^{\mu\nu}=0
\end{equation} 
where $K_{\mu\nu}$ is the extrinsic curvature of the membrane. 

\eqref{sttr} and \eqref{introkt} are $D$ equations for the 
$D-1$ independent membrane variables. These equations nonetheless 
define consistent membrane dynamics at large $D$ because it turns out 
that the form of the large $D$ membrane stress tensor is such that 
\eqref{introkt} is obeyed as an identity order by order in the  $\frac{1}{D}$ expansion.
If, for instance we insert the leading order membrane stress tensor \cite{Bhattacharyya:2016nhn} 
into the LHS of \eqref{introkt} we find that the RHS is of a low enough order in $\frac{1}{D}$ 
that it can - and presumably will - cancel against the contribution
of subleading terms in $T_{\mu\nu}$.  In other words the conservation equations \eqref{emcons} applied to 
the leading order stress tensor of \cite{Bhattacharyya:2016nhn} yields consistent probe membrane dynamics 
in a power series expansion in $\frac{1}{D}$. However if the equations of motion are taken literally at any finite 
$D$, no matter how large, they are inconsistent and generically have no solutions. 

This paper is devoted to a study of the near equilibrium properties of our membrane. 
 We will find it instructive to perform our analysis at finite $D$, even though 
 our results are guaranteed to reproduce black hole physics only at large $D$. 
This is only possible once we have a formulation of probe membrane dynamics that 
is self consistent at finite $D$.  It turns out to be not too difficult 
to find such a formulation. In this paper we present an `improved' version of the 
leading order membrane stress tensor of  \cite{Bhattacharyya:2016nhn}.
Our improved stress tensor reduces to the results of  \cite{Bhattacharyya:2016nhn} at large $D$, 
but differs from it at subleading orders in $\frac{1}{D}$. The improvement is 
chosen to ensure that the new stress tensor obeys the equation \eqref{introkt} 
as an identity even at finite $D$. It follows that the equations of motion 
that follow from the conservation of this stress tensor constitute $D-1$ equations 
for the $D-1$ membrane variables even at finite $D$ and so presumably define 
consistent membrane dynamics even at finite $D$. Moreover the improved stress tensor 
turns out also to  exactly obey a local form of the second law of thermodynamics  under certain assumptions. 
More precisely our improved stress tensor quantitatively reproduces 
the entropy production equation reported in \cite{Dandekar:2016fvw} 
at leading order in large $D$. 

In the rest of this paper we first present our improved version of the leading order large $D$ membrane 
stress tensor of \cite{Bhattacharyya:2016nhn}. We then use this stress tensor to study of the properties of the 
membrane in equilibrium. In particular we demonstrate 
that all stationary solutions of the resultant membrane equations can be obtained from the 
extremisation of an action functional of the shape of the membrane. We apply this formalism to 
simple stationary solutions. Finally, in the case of a background $AdS$ spacetime, we  proceed to study 
the dynamics of our membrane in near equilibrium situations and investigate relationship between 
our improved large $D$ membrane equations and the equation of fluid gravity. 

In the rest of this introduction we present a more detailed outline of the contents of this paper. 
To end this subsection, we re emphasize that - as in previous work - the membranes of this paper 
reproduce  black hole motion only at large $D$ limit even though their dynamics is well defined 
even at finite $D$. The membrane equations presented in this paper are just the first term in a  
systematically improvable approximation to black hole dynamics. Given this fact it is somewhat 
surprising that the membrane equations presented in this paper turn out - in simple situations - to reproduce
 black hole physics better than we had the right to expect, getting some results exactly right 
 even at finite values of $D$ - as we explain below.

\subsection{The improved membrane stress tensor and resultant equations of motion}

Consider a $D$ dimensional bulk spacetime with metric $G_{MN}$ that obeys Einstein's equations with a 
cosmological constant 
\begin{equation}\label{eeq}
{\bar R}_{MN}+(D-1) \lambda G_{MN} =0
\end{equation}
\footnote{The constant $\lambda$ in \eqref{eeq} is proportional to (minus of) the 
usual cosmological constant. We have chosen the normalization of 
$\lambda$ to ensure that $AdS_{D}$ with radius 
$\frac{1}{\sqrt{\lambda}}$ is a solution the equations 
\eqref{eeq} when $\lambda$ is positive, while de Sitter space with radius 
$\frac{1}{\sqrt{-\lambda}}$ solves \eqref{eeq} when $\lambda$ is negative. 
Upon setting $\lambda=0$ \eqref{eeq} reduces to the usual (flat space) 
vacuum Einstein equations. } Consider a codimension one membrane propagating in this spacetime.  
The membrane stress tensor obtained from the analysis of Einstein's equations at large $D$ was reported in 
equation 1.10 of \cite{Bhattacharyya:2016nhn} as
\begin{equation} \label{std}
16 \pi T_{\mu\nu}= \mathcal{{K}} u_\mu u_\nu -2 \sigma_{\mu\nu} +  K_{\mu\nu} 
\end{equation} 
upto corrections that are subleading in $1/D$. Here  $u_\mu$ is a velocity field on the membrane, 
$\sigma_{\mu\nu}$ is the shear tensor of this velocity field (see \eqref{sigdef} for a definition), 
$K_{\mu\nu}$  is the extrinsic curvature of the membrane world volume (see \eqref{kdef} for a definition), $\mathcal{K}$ is the trace of the 
extrinsic curvature. 

\eqref{std} may be rewritten in the form 
\begin{equation}\label{stdr} 
16 \pi T_{\mu\nu}= \mathcal{{K}} {\cal P}_{\mu\nu} -2 \sigma_{\mu\nu} + \left( K_{\mu\nu} - {\mathcal K} g_{\mu\nu}
\right)
\end{equation}
where $g_{\mu\nu}$ is the induced metric on the membrane world volume and 
${\cal P}_{\mu\nu}= g_{\mu\nu} + u_\mu u_\nu$ is the projector orthogonal to the membrane velocity.

In this paper we study the dynamics of membranes governed by the improved stress tensor 
\begin{equation}\label{stf} 
16 \pi T_{\mu\nu}= \mathcal{{\tilde K}} {\cal P}_{\mu\nu} -2 \sigma_{\mu\nu} + \left( K_{\mu\nu} - {\mathcal K} g_{\mu\nu}
\right)
\end{equation}
where
\begin{equation}\label{km}
\begin{split}
		\mathcal{{\tilde K}} &= \frac{\mathcal{K}^2-K^{\mu\nu}K_{\mu\nu}+2K^{\mu\nu}\sigma_{\mu\nu}}{\mathcal{K}+u.K.u}\\
	\end{split}
\end{equation}

It is easily verified that $\mathcal{{\tilde K}}$ reduces to  ${\mathcal K}$ 
in the large $D$ limit defined in \cite{Bhattacharyya:2015dva,Bhattacharyya:2015fdk,Bhattacharyya:2017hpj}, and so it follows  that \eqref{stf} reduces to \eqref{stdr} 
at leading order in the large $D$ limit. Moreover it is easily verified that the stress 
tensor \eqref{stf} obeys the equation
\begin{equation}\label{KT}
K_{\mu\nu} T^{\mu\nu}=0
\end{equation}
as an exact algebraic identity (the same is not true for the 
stress tensor \eqref{std}).

We emphasize that \eqref{stf} is the stress tensor that lives on a probe brane that does not back react on the background spacetime. \footnote{In other words, in working with \eqref{stf} we multiply the full stress tensor by 
$\epsilon$, work only to first order in the $\epsilon$ expansion and then set $\epsilon$ to unity at the end of the 
computation. The order $\epsilon$  back reaction of the membrane on the background spacetime produces an order 
$\epsilon^2$ correction to the membrane equations, which we ignore.}

Note that the stress tensor \eqref{stf} consists of the sum of the  identically conserved Brown York stress tensor
\begin{equation}\label{bysty} 
16 \pi T_{\mu\nu}^{BY}=K_{\mu\nu} - {\mathcal K} g_{\mu\nu}
\end{equation}
and the `fluid' stress tensor
\begin{equation}\label{flst}
16 \pi T_{\mu\nu}^{fluid} = \mathcal{{\tilde K}} {\cal P}_{\mu\nu} -2 \sigma_{\mu\nu}
\end{equation}
Comparing \eqref{flst} to the standard fluid form of the stress tensor 
\begin{equation}\label{flstg}
 T_{\mu\nu}^{fluid} = \rho u_\mu u_\nu + p {\cal P}_{\mu\nu} -2 \eta \sigma_{\mu\nu}
\end{equation}
(here $\rho$ is the fluid energy density, $p$ is its pressure and $\eta$ its shear viscosity)
 we see that our membrane fluid has 
\begin{equation}\label{depv}
\rho=0,~~~ p=\frac{\mathcal{{\tilde K}}}{16 \pi}, ~~~ \eta=\frac{1}{16 \pi}
\end{equation}
It is striking that fluid energy density vanishes 
identically; it follows immediately that the notion of an intrinsic fluid temperature $T$ 
is ambiguous and that the fluid entropy density $s$ is a pure number. 
\footnote{Usually, the entropy density is a function of the energy density. However our  fluid has vanishing 
energy density. It follows that in this special case the entropy density 
has nothing to be a function of and so is a pure number.}
However the dynamics of the membrane 
is defined by an interaction between the membrane `fluid' and its shape - this interaction 
apparently endows any bit of the membrane with a definite temperature. Indeed the formula 
 for the  membrane pressure  \eqref{depv} - together with the vanishing of the fluid energy density 
 plus standard thermodynamics - allows us to conclude that $Ts = \frac{\mathcal{{\tilde K}}}{16 \pi}$.
 As we have explained above we expect the entropy density $s$ to be a constant.
 Below we will see that $s = \frac{1}{4}$ so that $T= \frac{\mathcal{{\tilde K}}}{4 \pi}$, 
 where $T$ is the local temperature of the membrane. 
 Note that the temperature - which was left undetermined by the fluid equation of state - is 
determined by the membrane's local extrinsic geometry. 
\footnote{It is easy to cook up systems with the unusual thermodynamics of our fluid. Consider a 
substance consisting of $\frac{1}{4 \ln 2}$ qubits per unit volume. Let the 
Hamiltonian of this system simply vanish.  A volume $V$ of such a system 
is associated with a finite dimensional Hilbert space of zero energy states 
whose number is given by $e^{\frac{V}{4}}$. } 
\footnote{Had our membrane fluid been less exceptional, the energy density of the fluid as a function 
of position would have been an additional variable of our problem. Membrane motion would then have 
had $D-1$ fluid variables plus one shape variable - the additional equation of motion could then have 
come from the equation \eqref{introkt} which would no longer have been identically obeyed. 
Black hole membranes are special precisely because they are described by a fluid of vanishing 
energy density - and so a total of $D-1$ rather than $D$ variables, and so (for consistency) by 
a stress tensor that obeys \eqref{introkt} as an identity.} Note also that the viscosity 
of our membrane obeys the KSS relation \cite{Kovtun:2004de}
\begin{equation}\label{kss}
\frac{\eta}{s}=\frac{1}{4 \pi}
\end{equation}

Simple algebraic manipulations (see the next section) reveal that 
\begin{equation}\begin{split} \label{entcur}
\nabla\cdot J_S &=\frac{1}{2\tilde{\mathcal{K}}}\sigma_{\alpha\beta}\sigma^{\alpha\beta}\\
J_S^\mu&= \frac{u^\mu}{4}\\
\end{split}
\end{equation}
We identify $J_S^\mu$ as the entropy current of our membrane. This definition reduces to the 
entropy current of \cite{Bhattacharyya:2016nhn} at large $D$. In the same limit \eqref{entcur} reduces 
to the entropy production equation Eq. (1.5)  of 
\cite{Dandekar:2016fvw} at large $D$. It follows that the membrane equations of this paper obey a local form of the 
second law of thermodynamics provided $\mathcal{{\tilde K}}$ is everywhere (pointwise) positive.
In this paper we simply restrict attention to those solutions - large classes of which 
certainly exist - that obey this condition \footnote{This condition is always met in the strict large $D$ limit. Even at finite 
$D$ it is possible that this condition is stable under time evolution (configurations that obey this condition 
never evolve to those that do not). The investigation whether - and when - this is true is an interesting 
problem for the future. } leaving the analysis of the dynamical closure of this condition to later work. 

The derivation \eqref{entcur} used the conservation of the Brown York part of the membrane stress tensor. As this conservation applies
only in spacetimes that  obeys Einstein's equation, it follows that, in general, the local form of the second law 
\eqref{entcur} is valid only when the membrane probes solutions of Einstein's equations rather than general smooth manifolds. 

\subsection{Stationary Solutions and Thermodynamics} \label{ssol}

In papers written over two years ago, Emparan, Suzuki and Tanabe \cite{Emparan:2015hwa,Suzuki:2015iha} demonstrated that 
stationary black holes are governed by simple effective equations in a power series 
expansion in $\frac{1}{D}$. The formulation of \cite{Emparan:2015hwa,Suzuki:2015iha}, while very convenient for 
the study of stationary solutions, has not previously been shown to generalize in a simple way 
to allow for the study of dynamical phenomena. In this  paper we rederive (suitably 
generalized versions of) the equations of \cite{Emparan:2015hwa,Suzuki:2015iha} starting 
with the membrane equations that follow from the conservation of our improved membrane stress tensor. 
It follows that (suitable generalizations of) the beautiful results of 
\cite{Emparan:2015hwa,Suzuki:2015iha} follow from the restriction of our general 
dynamical membrane equations to stationary situations.

Having obtained the equations of motion that govern stationary solutions we proceed to elucidate their structure. In particular 
we demonstrate that these equations follow from the extremization of an intriguing 
action, and uncover their thermodynamical significance.

In order to focus on stationary solutions, in this subsection we restrict attention to 
background spacetimes $G_{MN}$ that have a timelike killing vector $k^M$.  \footnote{A large class of interesting examples of such backgrounds  
are the `vacuum' solutions of Einstein's equations with a negative cosmological constant that are asymptotically locally $AdS$, and 
that tend, at small $z$, to the metric 
$$ ds^2= \frac{dz^2 + g_{\alpha \beta}dx^\alpha dx^\beta}{z^2}$$
where $g_{\alpha \beta}$ is an arbitrary field theory metric that admits a timelike killing vector.} 

Let $J^E_M$ denote the conserved `energy current'
\begin{equation} \label{cce} J_M^E= k^N T_{MN}
\end{equation} 
and let $J^E_\mu$ denote the restriction of this current to the membrane world volume. The conserved 
energy of the membrane is given by 
\begin{equation} \label{ende}
E= \int \sqrt{h} ~q\cdot J^E
\end{equation} 
where the integral in \eqref{ende} is taken over any spatial slice of the 
membrane world-volume, $h$ is the determinant of the metric on this slice, 
and $q$ is the unit normal to this slice within the membrane world volume.

Consider a membrane configuration in which $k^M$ is everywhere tangent to the membrane and so 
defines a vector field $k^\mu$ on the membrane. If, in addition 
${\cal L}_{k} u^\nu$ vanishes  (${\cal L}_{k}$ denotes Lie derivative, on the membrane world volume along $k^\mu$) 
then we say that the membrane is in a stationary configuration w.r.t the killing field $k^M$. 

 As entropy production  vanishes on any stationary solution 
 $\nabla.u=0$ and so $\sigma_{\mu\nu}=0$ (see \eqref{entcur}). 
The first of \eqref{entcur} then implies that $\nabla.u=0$. However 
a velocity field can be both shear and divergence free only if it is 
proportional to a killing vector \cite{Caldarelli:2008mv}. It follows that 
\begin{equation}\label{velf}
u^\mu= \frac{k^\mu}{\sqrt{-k.k}}
\end{equation}
Using \eqref{velf} it is not difficult to demonstrate that the stress tensor 
conservation equation projected orthogonal to the velocity $u^\mu$ reduces to 
 \begin{equation} \label{som} 
{\cal P}_\mu^\alpha \nabla_\alpha \left( {\tilde {\mathcal K}} \sqrt{-k.k} \right) =0
\end{equation} 
implying that 
\begin{equation}\label{memeo} \begin{split}
&{\tilde {\mathcal K}} = \frac{4 \pi T_0}{\sqrt{-k.k}}\\
\end{split}
\end{equation}
where $T_0$ is a constant. At large  $D$, \eqref{memeo} reduces to 
\begin{equation}\label{emst}
{\mathcal K} =\frac{4\pi T_0}{\sqrt{-k.k}}
\end{equation}
in agreement with the large $D$ results of  \cite{Emparan:2015hwa,Suzuki:2015iha} cited above.

We demonstrate in the main text below that 
 the equations of  motion \eqref{memeo} follow as the condition that the action  
\begin{equation}\label{memact}
S = \frac{1}{16 \pi} \left[  -(D-1)\lambda \int_V \sqrt{-G} + \int_{M} \sqrt{-g}~\left({\mathcal K}- \frac{4\pi T_0}{\sqrt{-k.k}}\right) \right]
\end{equation}
is extremized. Here ${\mathcal K}$ is the trace of the extrinsic curvature of the membrane,
$g_{\mu\nu}$ is the metric on the membrane world volume $M$  and $V$ denotes the region of spacetime enclosed by the membrane.  
The variation of \eqref{memact} w.r.t the induced metric on the world volume 
defines a stress tensor given by 
\begin{equation}\label{mst}
T_{\mu\nu}= -\frac{2}{\sqrt{-g}} \frac{\delta S}{\delta g^{\mu\nu}}
\end{equation} 
\footnote{The variation in this equation is defined as follows. We change 
the metric on the membrane world-volume by changing the background solution 
of Einstein's equations with which we work. Note that regular solutions 
of Einstein's equations are completely determined - and therefore 
parametrized - by the induced metric on a bounding surface, which in this 
case is taken to be the world volume of the membrane.}
It is easily verified that the stress tensor \eqref{mst} agrees with \eqref{stf} 
evaluated on the equilibrium solution \eqref{velf}, \eqref{memeo}. In other words the 
offshell action \eqref{memact} generates the equations of motion 
for the shape of stationary solutions, while variation of the value of the onshell action 
w.r.t. the background metric reproduces the conserved stress tensor 
of this solution. \footnote{It may be useful to emphasize a potentially confusing point. The action \eqref{memact} is 
defined as an integral over the full world volume of the membrane - and so is well defined 
also for time dependent membrane shape configurations. In this paper, however, we are interested 
in \eqref{memact} only for stationary membrane configurations. All variations of the action 
\eqref{memact} are performed within the space of stationary membrane shapes - and with respect to killing 
metrics.}

We will now uncover the thermodynamical significance of the action \eqref{memact}. Let 
$t$ be any `time coordinate' that obeys 
\begin{equation} \label{deftime}
k.dt = k_t=1
\end{equation}  
Consider the two time slices of the bulk space time $t=t_1$ and $t=t_2$. Let $\beta= t_1-t_2$.
Let $dB$ represent that part of the membrane world volume that lies between these two times
and let $B$ denote the part of the bulk spacetime enclosed by the membrane between these two time slices. 
In the main text we show that provided \eqref{velf} (but not necessarily \eqref{som} ) holds, the 
membrane energy and entropy is given by

\begin{equation}\label{memenstat} \begin{split}
 E&= \frac{1}{16 \pi \beta}  \left( \int_{dB}\sqrt{-g} {\mathcal K}
- (D-1) \lambda \int_{B} \sqrt{-G} \right) \\
S_{ent} &= \frac{1}{4 \beta} \int_{dB} \frac{\sqrt{-g}}{\sqrt{{-k.k}}} 
\end{split}
\end{equation}
Note that the  second term on the 
RHS of \eqref{memenstat} is proportional to the volume of spacetime 
enclosed by the membrane world volume and the two time slices. The contribution 
of this term vanishes at  $\lambda=0$.

Comparing \eqref{memact} with \eqref{memenstat} 
it follows that the action in \eqref{memact} may be rewritten as 
\begin{equation}\label{actwr}
S= \beta \left( E- T_0 S_{ent} \right)
\end{equation}
where $\beta$ is the `length' of the time coordinate. In Euclidean space $\beta = \frac{1}{T}$ 
where $T$ is the temperature of our system. It follows that the Euclidean action \eqref{actwr} is proportional to the 
logarithm of the partition function (as expected on general grounds)
\begin{equation}\label{pfo}
S=-\ln Z= \frac{E}{T_0} -S_{ent}
\end{equation}
provided we identify
\begin{equation} \label{identt}
T=T_0.
\end{equation}
In other words the arbitrary constant $T_0$ that appears in the action \eqref{memact}
- which we have already identified with the integration constant in \eqref{memeo} 
- is the temperature of the stationary membrane configuration. 

It then follows from \eqref{pfo} that, on shell, \footnote{Naively the action 
\eqref{pfo} changes when we vary the temperature for two reasons. First because 
\eqref{pfo} explicitly depends on $\beta$. Second because the equilibrium 
membrane solution - hence its energy and entropy - depends explicitly on $T=T_0$. 
However the second variation actually vanishes, as the onshell action is stationary 
w.r.t. an arbitrary variation of the membrane configuration.}
\begin{equation}\label{pds}
\partial_\beta S= E,
\end{equation} 
confirming our identification the action 
\begin{equation}
S=-\ln Z
\end{equation}
\footnote{In order to find the partition function of our system we first had to extremize the action w.r.t the membrane 
shape and then evaluate this extremized action. The situation is 
more closely analogous to that of the superfluid partition function (see  \cite{Bhattacharya:2011eea} ) than the 
ordinary fluid partition function of, e.g. \cite{Banerjee:2012iz}.}

Recall that stationary solutions of the membrane equations extremize the action \eqref{pfo}. Viewing 
$\beta=\frac{1}{T_0}$ as a Lagrange multiplier, it follows from \eqref{pfo} that stationary membrane solutions extremize 
membrane entropy at fixed membrane energy. This is satisfying as we expect, on physical grounds, that the equilibrium 
configurations in the  microcanonical ensemble extremize their entropy. 

It follows in particular from \eqref{memeo} that the temperature of a static spherical 
membrane in flat space is given by $T=\frac{\mathcal{{\tilde K}}}{4 \pi}$. In a more general configuration 
that is not necessarily in equilibrium, we simple define the local membrane temperature to be given 
by 
\begin{equation}\label{locts}
T(x)=\frac{\mathcal{{\tilde K}}(x)}{4 \pi},
\end{equation}

We emphasize that the formula \eqref{locts} defines the local temperature of the membrane in any dynamical configuration. The local 
temperature \eqref{locts} is, in general, a function of position and is distinct from the temperature $T_0$ of a stationary 
solution of the membrane equations. In a stationary solution the relationship between $T_0$ and the local membrane temperature 
$T$ follows from \eqref{memeo} and takes the form 
\begin{equation}\label{lgt}
 T(x) = \frac{T_0}{\sqrt{-k.k}}
\end{equation}
In words, the local temperature in equilibrium is given by the global temperature $T_0$ times the effective red shift factor 
$\frac{1}{\sqrt{-k.k}}$. See \cite{Bhattacharyya:2007vs} for a very similar discussion in the context of hydrodynamics 
on a fixed background manifold.

The simplest stationary membrane solutions are those dual to Schwarzschild type black holes of arbitrary size in global 
$AdS$ and global $dS$ spaces \footnote{Schwarzschild black holes in flat space and black branes in $AdS$ space 
can be regarded as special limits of these solutions.}. Quite remarkably we will find below that  the membrane formalism described 
in this subsection reproduces the thermodynamics of the dual black holes exactly - rather than only in the 
large $D$ limit.

\subsection{Fluid Gravity from Membranes}

We now focus on the study of Einstein's equations with a negative cosmological constant, i.e. solutions of the equation \eqref{eeq} 
with $\lambda=1$. A simple solution of these equations is unit radius $AdS_D$ space in Poincare coordinates, i.e. the space 
\begin{equation}\label{adsintro}
 ds^2= \frac{dz^2 + dx^\mu dx_\mu}{z^2}
\end{equation}
where $\mu=0, \ldots D-2$ and $\mu$ indices are raised and lowered using the metric $\eta_{\mu\nu}$. A simple solution of the membrane 
equations is the configuration 
\begin{equation}\label{zsol}
z=\frac{D-1}{4 \pi T_{bb}}, ~~~u^\mu = z v^\mu, ~~~~v^\mu={\rm const}, ~~~ \eta_{\mu\nu} v^\mu v^\nu= -1
\end{equation}
where $T_{bb}$ is the temperature $T_0$ of the membrane configuration. This solution is dual to uniform black brane of temperature $T_{bb}$. By treating the membrane stress tensor 
as a linearized source for Einstein's equations, it is easy to compute the resultant backreaction. For $z <\frac{D-1}{4 \pi T_{bb}}$ 
the resultant spacetime is a linearized normalizable perturbation about $AdS$ space, and the ($AdS/CFT$) boundary stress tensor induced 
by this fluctuation is easily computed. It turns out that this boundary stress tensor agrees precisely (at finite $D$) with the exact 
boundary stress tensor of a uniform black brane of temperature $T_{bb}$ and moving at a uniform velocity $v^\mu$. The membrane entropy density also exactly matches the 
entropy density of the uniform black brane. 

Now consider a membrane whose shape and velocity field take the form listed in \eqref{zsol} with $T_{bb}$ and $v^\mu$ slowly varying functions of the membrane coordinates $x^\mu$. In an expansion in derivatives it is, once again, not difficult to 
solve the `dynamical' linearized Einstein equations to compute the linearized gravitational fluctuations sourced by such a 
membrane. \footnote{We compute the fluctuation fields 
with the boundary conditions that they die off (i.e. are normalizable) towards the boundary of $AdS$, and also that they 
do not blow up as we approach the Poincare horizon.} As in the previous paragraph one can now compute the boundary 
stress tensor induced by this linearized fluctuation. The fact that the boundary stress tensor is conserved 
follows from Einstein's constraint equations evaluated on the boundary. On the other hand the membrane equations follow 
from the constraint equations evaluated `outside' the membrane (the constraint equations are identically obeyed `inside' the 
membrane). 

Given a solution to the dynamical Einstein equations, it is well known that the constraint equations on any slice 
imply the constraint equation on any other slice. It follows that the condition of conservation of the  boundary 
stress tensor is equivalent to the requirement of conservation of membrane stress tensor. At the algebraic level,
the procedure described earlier in this subsection (coupling the membrane to linearized gravity fluctuations) 
allows us to find a linear map from the membrane world volume stress tensor to the boundary stress tensor. The fact allows us to regard 
the boundary stress tensor as a linear functional of the membrane stress tensor (the precise form of this functional 
depends on the membrane shape in a nonlinear way). This functional has the property that it ensures that the 
boundary stress tensor is conserved whenever the membrane stress tensor it is obtained from is also conserved.

The procedure outlined in the previous paragraph yields an expression for the boundary stress tensor in terms of membrane 
stress tensor, and so in terms of membrane variables (membrane shape and velocity field). 
It is possible, however, to perform a field redefinition to a local boundary 
temperature and a local boundary fluid velocity, and rewrite the boundary stress tensor in terms of these new variables. 
In these variables the boundary stress tensor takes the standard form for the stress tensor of a conformal fluid in the 
derivative expansion. Below we have evaluated this expansion to second order in the derivative expansion, and compared our 
results with literature on the fluid gravity correspondence in which the same expansion of the boundary stress tensor as 
a function of the boundary velocity and temperature has been computed in every dimension by an exact direct analysis of 
Einstein's equations. We find that two results (the results of this paper and the exact results of 
the fluid gravity correspondence) are in perfect agreement at zero and first order in the derivative expansion even at 
finite $D$, but deviate from each other (at finite $D$) at second order in the derivative expansion. 

This discussion of the last paragraph implies, 
in particular, that the spectrum of the lightest quasinormal modes around a black brane in an arbitrary number of dimensions 
agrees at finite $D$ and upto first subleading order in $k$, with the corresponding spectrum around the uniform planar 
membrane solution \eqref{zsol}. On the other hand these two spectra deviate at order $k^3$ and at finite $D$. We have 
independently verified that these predictions are borne out. 

Note that traditional hydrodynamics (and so, in the gravitational 
context, fluid gravity) and our large $D$ expansion are distinct expansions of bulk black brane dynamics. 
Fluid gravity functions order by order in an expansion in derivatives; however the coefficients of this expansion 
are computed exactly as functions of $D$. On the other hand the large $D$ membrane equations are constructed 
order by order in $\frac{1}{D}$. At any given order in $\frac{1}{D}$, however, the resultant equations are exact 
in derivatives, and so have terms of all orders in the derivative expansion. 

We have already pointed out that the leading order membrane equations presented in this paper accurately 
reproduces the black brane Navier Stokes equations. In addition the membrane equations capture the 
contribution of infinite number of arbitrarily high derivative corrections to Navier Stokes. 
The membrane equations retain only the contribution of those terms that survive in the improved large $D$ limit. 
From the viewpoint of a boundary observer the truncation to these terms does not appear to help much; 
outside the long wavelength limit the equations for boundary hydrodynamics appear to continue to be a 
nonlocal mess. The miracle 
is that there exists a field redefinition (namely the redefinition that maps boundary to the membrane 
world volume) that turns this nonlocal mess into local - and so tractable - hydrodynamical equations. 
Note that these $D-1$ dimensional equations are local only when formulated on the membrane world volume, 
itself a dynamical $D-1$ dimensional submanifold of the $D$ dimensional bulk $AdS$ space. 

The fact that the membrane equations remain local even outside the traditional boundary derivative expansion 
potentially allows them to capture qualitatively new phenomena. If, for example, the the membrane were to fold 
on itself then the parametrization $z(x^\mu)$, and so the map to boundary fluid variables becomes singular. 
It is, however, manifest from the bulk membrane viewpoint that this singularity is a fake, an artefact of the 
incorrect choice of dynamical variables. We leave a serious investigation of this and other issues to future work.

\section{Details of the formalism}

As explained in the introduction, in this paper we study a membrane that resides on a codimension one submanifold of any background spacetime that obeys 
the Einstein equation \eqref{eeq}. For mathematical purposes it is sometimes convenient to parametrize the membrane 
world volume by the solutions to the equation 
$$\rho-1=0$$ 
where $\rho$ is a suitably chosen scalar function that takes values on the background manifold. Let 
$$|\partial \rho| = \sqrt{\partial_M \rho G^{MN}\partial_N \rho}, ~~~n_A = \frac{\partial_A \rho}{|\partial \rho|}.$$ 
Note that $n_A$ is normal to the membrane world volume and that $n_MG^{MN}n_N=1$.

Our membrane has a stress tensor, ${\cal T}_{MN}$, living on its world volume. The stress tensor has the form  
\begin{equation}\label{sts}
{\cal T}_{MN} = |\partial \rho| \delta(\rho-1)T_{MN}, ~~~n^M {\cal T}_{MN}= n^N T_{M N}=0
\end{equation}
Let $T_{\mu\nu}$
\footnote{In the rest of this paper we 
	will use the indices $M, N  \ldots $ to denote spacetime coordinates and 
	Greek indices $\mu \ldots $ to denote membrane world volume coordinates. 
	$G_{MN}$ denotes the metric of spacetime, while $g_{\mu\nu}$ is the metric 
	on the world volume of the membrane. } denote the pull back of $T_{MN}$ onto the membrane world volume. The equation 
$T_{MN}n^M=0$ ensures that there is as much information in $T_{\mu\nu}$ as $T_{MN}$; knowledge of 
one is sufficient to reconstruct the other. As explained in the introduction, 
 the world volume stress tensor, $T_{\mu\nu}$ for the membrane studied in this paper is taken to be given by the form \eqref{stf} 
 where the membrane shear and extrinsic curvature are defined by 
\begin{equation}\label{sigdef}
\sigma_{\mu\nu} = \frac{1}{2}{\cal P}_\mu^\alpha {\cal P}_\nu^\beta 
\left( \nabla_\alpha u_\beta +\nabla_\beta u_\alpha -{\cal P}_{\alpha \beta} \frac{2\nabla.u}{D-2} \right) 
\end{equation}
\begin{equation}\label{kdef}
K_{\mu\nu} = \left(\frac{\nabla_A n_B + \nabla_B n_A}{2}\right) \frac{\partial X^A}{\partial x^\mu} \frac{\partial X^B}{\partial x^\nu}
\end{equation}
where, $X^M$ are coordinates on the full spacetime and $x^\mu$ are the coordinates on the membrane.

\subsection{Membrane Stress Tensor and equations of motion}

As explained in the introduction, our membrane stress tensor is a sum of two terms, $T_{\mu\nu}^{BY}$
(see \eqref{bysty}) and $T_{\mu\nu}^{fluid}$ (see \eqref{flst}). $T_{\mu\nu}^{BY}$ is identically conserved 
on the membrane world volume (provided it propagates in a background satisfying Einstein equations)
\begin{equation}\label {constd}
\nabla^\mu T^{BY}_{\mu\nu}=0
\end{equation}
The non-trivial part of the membrane equation of motion is the conservation of the fluid stress tensor  
\begin{equation} \label{lome} \begin{split}
		&\nabla^\mu T^{fluid}_{\mu\nu}=0\\
		16 \pi T^{fluid}_{\mu\nu}&= \tilde{\mathcal{K}}~ {\mathcal P}_{\mu\nu} - 2 \sigma_{\mu\nu}
	\end{split}
\end{equation}
It is  useful to decompose the membrane equations of motion  into their components in the direction of and 
orthogonal to $u^\mu$, i.e.  
\begin{equation}\label{north}
u^\nu \nabla^\mu T_{\mu\nu}=0, ~~~ {\cal P}^\nu_\alpha\nabla^\mu T_{\mu\nu}=0, ~~~{\cal P}^\nu_\alpha = \delta^\nu_\alpha + u^\nu
u_\alpha
\end{equation}
Using 
\begin{eqnarray} \label{entprod}
	8\pi \nabla_{\mu}T^{\mu\nu}u_{\nu}&=&-\frac{\tilde{\mathcal{K}}}{2}\nabla\cdot u+\mathcal{P}^{\mu\alpha}\left(\frac{\nabla_\alpha u_\beta+\nabla_\beta u_\alpha}{2}\right)\nabla_\mu u^\beta-\frac{(\nabla\cdot u)^2}{D-2}\nonumber\\
	&=&-\frac{\tilde{\mathcal{K}}}{2}\nabla\cdot u+\mathcal{P}^{\mu\alpha}\left(\frac{\nabla_\alpha u_\beta+\nabla_\beta u_\alpha}{2}\right)\mathcal{P}^{\beta\theta}\nabla_\mu u_\theta-\frac{(\nabla\cdot u)^2}{D-2}\nonumber\\
	&=&-\frac{\tilde{\mathcal{K}}}{2}\nabla\cdot u+\mathcal{P}^{\mu\alpha}\left(\frac{\nabla_\alpha u_\beta+\nabla_\beta u_\alpha}{2}\right)\mathcal{P}^{\beta\theta}\left(\frac{\nabla_\mu u_\theta+\nabla_\theta u_\mu}{2}\right)-\frac{(\nabla\cdot u)^2}{D-2}\nonumber\\
	&=&-\frac{\tilde{\mathcal{K}}}{2}\nabla\cdot u+\sigma_{\alpha\beta}\sigma^{\alpha\beta}\nonumber\\
	\end{eqnarray}
it follows that the first equation in \eqref{north} can be rewritten in the form \eqref{entcur}, and is a statement of 
a local form of the second law of thermodynamics provided ${\tilde {\mathcal K}}$ is 
everywhere positive. 
	
On the other hand, the stress tensor conservation equation projected orthogonal to $u^\mu$ takes the form
\begin{equation}\label{transeq}
\begin{split}
& 16 \pi~ {\cal P}^\nu_\alpha\nabla^\mu T_{\mu\nu} = \left( \tilde {\mathcal K}~u.\nabla u_\nu + \nabla_\nu \tilde {\mathcal K} -2 \nabla^\mu \sigma_{\mu\nu}\right) {\cal P}^\nu_\alpha \\
\end{split}
\end{equation}

In order to explicitly verify that \eqref{transeq} reduces to the membrane equations of motion presented in 
\cite{Bhattacharyya:2015dva,Bhattacharyya:2015fdk,Bhattacharyya:2017hpj} we manipulate \eqref{transeq} as follows.
Let $X^A$ denote any space time coordinates and 
let $\bar R_{ABCD}$ be the background spacetime Riemann tensor. Let 
$x^\alpha$ denote an arbitrary set of coordinates on the membrane world volume and let $e^A_\alpha = \frac{\partial X^A}{\partial x^\alpha}$. Using the Gauss Codacci relationship
\begin{equation}\label{gcr}
R_{\mu\nu}=\mathcal{K}K_{\mu\nu}-K^{\beta}_{\mu}K_{\nu\beta} + \bar R_{ABCD} e^A_\sigma e^B_\nu e^C_\gamma e^D_\mu g^{\sigma\gamma}
\end{equation}
it is not difficult to show that 
\begin{equation} \begin{split} \label{comp}
&16 \pi~ {\cal P}^\nu_\alpha\nabla^\mu T_{\mu\nu}
= \left( \tilde {\mathcal K}~u.\nabla u_\nu + \nabla_\nu \tilde {\mathcal K}\right) {\cal P}^\nu_\alpha + \bigg[ -2 u.\nabla u^\beta \nabla_\beta u_\nu - (\nabla.u) u.\nabla u_\nu \\ & -\nabla^2 u_\nu -u^\gamma u^\mu \nabla_\gamma \nabla_\mu u_\nu  - {\mathcal K} ~u^\mu K_{\mu\nu} + u_\mu K^{\mu\gamma}K_{\gamma\nu} - \bar R_{ABCD} e^A_\sigma e^B_\nu e^C_\gamma e^D_\mu g^{\sigma\gamma} u^\mu \\ & + \frac{2}{D-2}(\nabla.u)u.\nabla u_\nu + \frac{2}{D-2} \nabla_\nu (\nabla.u)  \bigg] {\cal P}^\nu_\alpha
\end{split}
\end{equation} 

At leading order in the large $D$ limit, \eqref{comp} reduces to
 \begin{equation} \begin{split} \label{compu}
&16 \pi~ {\cal P}^\nu_\alpha\nabla^\mu T_{\mu\nu}
= \left(  {\mathcal K}~u.\nabla u_\nu + \nabla_\nu  {\mathcal K}  -\nabla^2 u_\nu 
- {\mathcal K} ~u^\mu K_{\mu\nu} \right){\cal P}^\nu_\alpha
\end{split}
\end{equation}
in agreement with the membrane equations of motion presented in \cite{Bhattacharyya:2015dva,Bhattacharyya:2015fdk,Bhattacharyya:2017hpj}. 
\footnote{Using the large $D$ counting described in \cite{Bhattacharyya:2015dva,Bhattacharyya:2015fdk,Bhattacharyya:2017hpj}
we find that, at leading order in the large $D$ limit $\tilde {\mathcal K}\rightarrow {\mathcal K}$. Moreover 
$\tilde {\mathcal K}~u.\nabla u_\nu$, $\nabla_\nu \tilde {\mathcal K}$, $\nabla^2 u_\nu$ and ${\mathcal K} ~u^\mu K_{\mu\nu}$ 
are all ${\cal O}(D)$ while $\bar R_{ABCD} e^A_\sigma e^B_\nu e^C_\gamma e^D_\mu g^{\sigma\gamma} u^\mu {\cal P}^\nu_\alpha$ 
are all ${\cal O}(1)$. This conclusion holds for all values of the cosmological constant.  }

\subsection{Regular Stationary solutions of Einstein's equations}

We now turn our attention to the construction of stationary solutions to our membrane equations. 
As explained in the introduction, stationary solutions exist only when the background spacetime 
in which the membrane propagates has a killing direction. In this subsection and the next 
we assume this is the case, and denote the killing vector by $k^A$. We now construct a coordinate 
system for any such background spacetime that is adapted to this killing direction. It is useful to look at \cite{Banerjee:2012iz,Bhattacharyya:2012xi} as the setup and construction is similar in flavour

Consider any spacetime with a timelike  killing vector field $k^A$. 
The spacetime in question can be foliated by the $D-1$ parameter set of 
integral curves of this killing vector field, i.e. by curves that obey the 
equation 
\begin{equation}\label{curve}
\frac{dX^A(t)}{dt}=k^A(X)
\end{equation}
where $X^A$ represents an arbitrary set of coordinates in the bulk spacetime. 
Note that there exists a $D-1$ parameter set of such curves which we choose to label by the $D-1$ parameters $X^a$. 
Making an arbitrary ($X^a$ dependent) choice for the origin of the $t$ coordinate in \eqref{curve}, it follows that the 
background spacetime metric takes the `Kaluza Klein' form 
\begin{equation}\label{kist}
ds^2_{ST}= G_{MN} dX^MdX^N = - e^{2 \Sigma(X^a)} ( dt + A_a(X^a) dX^a)^2 + W_{ab}(X^a) dX^a dX^b
\end{equation}

The fact that $\Sigma$, $A_a$ and $W_{ab}$ are all independent 
of $t$ follows from the condition that $\partial_t$ is a killing direction. 
Note also that an $X^a$ dependent shift of the origin of $t$ preserves the form of the metric \eqref{kist}, 
inducing an effective a `Kaluza Klein gauge transformation' on the `Kaluza Klein gauge field' $A_a$. 

We wish to study stationary membrane configurations. As explained in the introduction, this implies, in particular, 
that the killing field $k^A$ - evaluated at any point on the membrane - is tangent to the membrane at that point. 
This requirement forces the membrane world volume to be given by a shape of the form
\begin{equation} \label{fof}
f(X^a)=0
\end{equation} 
(note that the function $f$ does not depend on the `time' $t$.). It follows that the induced metric on the membrane, in a 
stationary configuration, takes the form 
\begin{equation}\label{mos}
ds^2 = - e^{2 \sigma(x)} ( dt + a_i(x) dx^i)^2 + w_{ij}(x) dx^i dx^j 
\end{equation}
where the variables $x^i$ label the  the $D-2$ parameter set of curves \eqref{curve} that 
obey \eqref{fof} and so lie on the membrane. \footnote{In other words the $D-2$ parameters $x^i$ label
the most general solution of \eqref{fof}. This solution is given by the schematic form $X^a(x^i)$. Recall that while 
$a$ runs over $D-1$ variables, $i$ runs over $D-2$ variables.}

As explained in the introduction, the velocity field configuration for a stationary 
solution takes the form \eqref{velf}. It follows from \eqref{velf} that 
\begin{equation}\label{impvelf}
\begin{split}
u.\nabla u_\mu &= \frac{k^\nu}{\sqrt{-k.k}} \nabla_\nu \left( \frac{k_\mu}{\sqrt{-k.k}} \right) \\
&= \frac{k_\mu}{\sqrt{-k.k}} k^\nu \nabla_\nu \left( \frac{1}{\sqrt{-k.k}} \right) + \frac{k^\nu\nabla_\nu k_\mu}{(-k.k)} \\
&= \frac{1}{2} \frac{\nabla_\mu(-k.k)}{(-k.k)} \\
&= \nabla_\mu \ln \sqrt{-k.k}
\end{split}
\end{equation}
\eqref{impvelf}, together with the identity $\sigma_{\mu\nu}=0$ turns the equation of motion 
\eqref{transeq} into the simpler equation \eqref{som}, which can immediately be integrated to 
\eqref{memeo}. 

We now turn to a derivation of the thermodynamical formulae \eqref{memenstat}. Let us begin with the second of 
\eqref{memenstat}. Recall that the entropy of a stationary configuration of the membrane is obtained by integrating 
the entropy current over any spacelike slice of the membrane. 
Consider a spacelike slice of the membrane given by the equation 
\begin{equation} \label{membraneslice}
t=t_0
\end{equation}
where $t_0$ is a constant. 
\footnote{This special choice of slice entails no loss of generality, as the most general slice of spacetime, 
$t=f(x^i)$, can be recast in the form \eqref{membraneslice} by the`Kaluza Klein' $x^i$ dependent shift of the 
origin of $t$.} 

The normal oneform $t$ to this slice - viewed as a oneform on the membrane world volume - is given by 
\begin{equation}\label{tdef}
q= \frac{dt }{\sqrt{e^{-2\sigma}-a_i w^{ij} a_j}}
\end{equation}
Let $g$  represent the determinant of the metric on the $D-1$ membrane world volume and 
let $h$ represent the determinant of the metric on the $D-2$ dimensional membrane slice \eqref{membraneslice}. It is 
easy to find an expression for $g$ and $h$ in terms $w$, the determinant of the metric $w_{ij}$ (see \eqref{mos}). We 
have 
\begin{equation}\label{determinants} \begin{split}
 & \sqrt{-g} = \sqrt{w}e^\sigma, \\
 &\sqrt{h} = \sqrt{w} \sqrt{1-e^{2\sigma} a_i w^{ij} a_j}
\end{split}
\end{equation}
Finally recall that in the coordinate system of \eqref{mos} the velocity vector field $u$ takes the form 
\begin{equation} \label{vspc}
u=e^{-\sigma} \partial_t
\end{equation}

The entropy of the membrane is given by 
\begin{equation} \label{entfor}
S_{ent} = \int \sqrt{h}~q_\mu J^\mu_S 
\end{equation} 
where the integral is taken over the $D-2$ dimensional slice of the membrane world volume 
\eqref{membraneslice}. 
Using \eqref{vspc}, however, it follows that 
$$ J^\mu_S ~q_\mu = \frac{e^{-\sigma}}{4\sqrt{e^{-2\sigma}-a_i w^{ij}a_j}}$$ 
Using \eqref{determinants} it then follows that 
\begin{equation} \label{entor}
S_{ent} = \frac{1}{4} \int \frac{\sqrt{-g}}{\sqrt{-k.k}}
\end{equation} 
where, once again the integral is taken over the $D-2$ dimensional slice of the membrane world volume 
\eqref{membraneslice} and we have used the fact that $\sqrt{-k.k}=e^{\sigma}$. The LHS and RHS of 
\eqref{entor} are both independent of time. Integrating both sides of that equation from $t=t_1$ to $t=t_2$
we obtain the second of  \eqref{memenstat}. 

We now turn to the derivation of the first of \eqref{memenstat}. The energy of the membrane is given by 
\begin{equation} \label{enf}
16 \pi E = -16\pi \int \sqrt{h}~q^\mu T_{\mu\nu} k^\nu = -\int  \sqrt{h}~q^\mu (K_{\mu\nu}-{\mathcal K}g_{\mu\nu}) k^\nu = \int \sqrt{-g}~({\mathcal K}-K^t_t )
\end{equation}
As above, the integral in \eqref{enf} is taken over the $D-2$ dimensional slice of the membrane world volume 
\eqref{membraneslice}. In going from the middle expression in \eqref{enf} to the RHS we have used the fact that 
$k= \partial_t$ and easily verified 
formulae 
$$ \sqrt{h} k.q = \sqrt{-g}, ~~~ \sqrt{h} q_\mu K^\mu_\nu k^\nu = \sqrt{-g} K_t^t$$
Integrating both sides of this equation from $t=t_2$ to $t=t_1$ we obtain 
\begin{equation} \label{enfm}
16 \pi (t_1-t_2)  E = \int_M \sqrt{-g}~({\mathcal K}-K^t_t )
\end{equation}
where the integral on the RHS of \eqref{enfm} is taken over the part of the membrane world volume that lies between 
$t=t_1$ and $t=t_2$.

We will now complete our derivation of  \eqref{memenstat} by demonstrating that 
\begin{equation}\label{ktt}
\int_M \sqrt{-g}~K^t_t = \left(D-1\right)\lambda \int_V \sqrt{-G} 
\end{equation}
The LHS of \eqref{ktt} is integrated, as in \eqref{enfm}, over the part of the membrane contained between 
times $t_1$ and $t_2$. The RHS of \eqref{ktt}, on the other hand, is integrated over the region of the {\it bulk} $D$ 
dimensional spacetime enclosed by three codimension one surfaces: the membrane world volume, the bulk slices $t=t_1$ and
the bulk slice $t=t_2$. 
If \eqref{ktt} holds then clearly \eqref{memenstat} follows from \eqref{enfm}.

In order to establish \eqref{ktt}, consider 
\begin{equation}\label{qte}
 Q= \int_V \sqrt{-G} \nabla_M \left[ (dt)_N \nabla^N k^M \right]
\end{equation}
where the integral is taken over the bulk region $V$ defined in the previous 
paragraph. We will establish \eqref{ktt} by 
evaluating \eqref{qte} in two separate ways.

Our first evaluation uses an integration by parts to express \eqref{qte} as 
\begin{equation}\label{qtet0}
 Q= \int_M \sqrt{-g} n_M  \left[ (dt)_N \nabla^N k^M \right]
\end{equation}
where $n_M$ is the normal to the membrane and the integral is taken over the region
of the membrane world volume for times $t$ that lie between $t_1$ and $t_2$. \footnote{In addition we have 
similar surface terms on the time slices at $t=t_1$ and $t=t_2$. However it is easily 
verified that the contribution of 
the bulk constant time slice at $t_2$ cancels the analogous contribution at $t_1$. }
Recall that $k^M$ is tangent to the membrane, 
in other words $n_M k^M$ vanishes. It follows that $n_M \nabla^N k^M= - k^M \nabla^N n_M$, so that 
\eqref{qtet0} may be rewritten as  
\begin{equation}\label{qteth} \begin{split}
&  Q= - \int_M \sqrt{-g} (\nabla^N n_M)  (dt)_N  k^M\\
&  = - \int_M \sqrt{-g} K^N_{M}  (dt)_N  k^M\\
    &  = -\int_M \sqrt{-g}~K^t_t\\
\end{split}
\end{equation}
where the integral is, once again, taken over the part of the membrane world volume at times between $t_1$ and $t_2$. 
\footnote{In obtaining the first line in \eqref{qteth} starting from \eqref{qte}
we have integrated by parts and used the fact that $n.k=0$.}. \eqref{qteth} is the final result of our first evaluation of 
$Q$. 

Our second evaluation proceeds by expanding out the integrand in \eqref{qte}. We have 
$$ \nabla_M \left( (dt)_N \nabla^N k^M \right)
= (\nabla_M (dt)_N) \nabla^N k^M + dt^N[\nabla_M, \nabla_N] k^M
+ (dt)^N \nabla_N \nabla_M k^M$$
The first term in this expression vanishes because $(\nabla_M (dt)_N)$ is symmetric
\footnote{This follows from the symmetry of $\Gamma$ matrices in our particular
coordinate system.} whereas $\nabla_N k_M$ is antisymmetric in its indices (recall $k^M$ is 
a killing vector). The third term in this equation vanishes
because $\nabla_M k^M$ vanishes. The second term is non-vanishing and is
easily evaluated to be 
$$ R_{NA} (dt)^N k^A= -(D-1) \lambda $$
where in the final equality we have used the bulk Einstein equation \eqref{eeq}. It follows that
\begin{equation}\label{qtef}
 Q= -(D-1) \lambda  \int_V \sqrt{-G} 
\end{equation}
\eqref{qtef} and \eqref{qteth} together establish \eqref{ktt}.

Note that the last step in our derivation of \eqref{ktt} made
crucial use of the fact that the membrane encloses a {\it regular solution}  of Einstein's
equations \eqref{eeq}. Our derivation does not apply to a membrane propagating 
in an arbitrary spacetime, and also does not apply to the membrane 
propagating about a solution of Einstein's equations if that solution encloses either 
a singularity or (secretly) a second asymptotic region, as is the case for a 
black hole spacetime.

\section{The action and its variations} 

We now demonstrate that equilibrium membrane configurations are governed 
by the action \eqref{memact} and establish some properties of this action.

\subsection{Variation of the action w.r.t. the membrane shape}

Consider a membrane whose world volume is given by a smooth codimension one submanifold of the 
ambient spacetime. Let $x^\mu$ represent a set of coordinates on the membrane. The membrane world volume 
can be described by specifying the spacetime coordinates $X^M$ as functions of the membrane coordinates, i.e. by 
the functions $f^M(x^\mu)$ s.t. 
\begin{equation} \label{oldshape}
X^M= f^M(x^\mu)
\end{equation}
We denote the induced metric on this membrane surface by $g_{\mu\nu}(x)$. The extrinsic curvature of the membrane 
surface is denoted by $K_{\mu\nu}(x)$. 

Now consider the slightly displaced membrane described by 
\begin{equation} \label{newshape}
X^M= f^M(x^\mu) + \delta z(x^\mu) n^M(x^\mu)
\end{equation}
Here $n^M(x^\mu)$ is the normal vector of the membrane surface at the point $x^\mu$ and 
$\delta z(x^\mu)$ is an arbitrary infinitesimal displacement function on the membrane. Let 
the induced metric on the displaced surface \eqref{oldshape} be given by $g_{\mu\nu} + \delta g_{\mu\nu}$, and 
let the extrinsic curvature of the displaced surface be given by $K_{\mu\nu} + \delta K_{\mu\nu}$.
In  Appendix \ref{variation} we demonstrate that, to first order in $\delta z$
\begin{equation}\label{variations}\begin{split}
&\delta g_{\mu\nu}= 2 K_{\mu\nu} \delta z\\
& \delta g^{\mu\nu} = -2 K^{\mu\nu} \delta z\\
& \delta K_{\mu\nu} = \left(R_{\mu\nu} + (D-1)\lambda G_{\mu\nu} + 2 K_{\mu\alpha}K^{\alpha}_{\nu}- {\mathcal K} K_{\mu\nu}\right)\delta z -\nabla_\mu\nabla_\nu \delta z\\
& \delta \sqrt{-g}= \sqrt{-g}~ {\mathcal K}~ \delta z\\
& \delta \mathcal{K}= \left( -K_{\mu\nu}K^{\mu\nu}+(D-1)\lambda\right) \delta z - \nabla^2 \delta z\\
& \delta \gamma= \gamma(u.K.u)\delta z \\
& \delta \int_V \sqrt{-G} = \int_M \sqrt{-g} ~\delta z
\end{split}
\end{equation}
where we have used the notation
\begin{equation}\label{defg} 
\gamma = \frac{1}{\sqrt{-k.k}}, ~~~u= \frac{k}{\sqrt{-k.k}} = \gamma k
\end{equation}
In order to obtain the formula for $\delta \gamma$ reported in \eqref{variations} above we have 
used the fact that, for stationary membrane configurations, $n_A k^A=0$ where $n_A$ is the normal to the 
membrane. All of the other formulae in \eqref{variations} are valid even without making this assumption. 

In the last of \eqref{variations} the volume integral on the LHS is taken over $V$, 
the region of spacetime enclosed by the membrane, whereas the integral on the RHS is taken 
over the $M$, the world volume of the membrane.

Using \eqref{variations} it follows immediately that the variation of the action \eqref{memact} under the 
operation \eqref{newshape} is given by \footnote{The variation of this action w.r.t its shape can be more systematically computed using the general formalism developed in \cite{Armas:2013hsa,Armas:2013goa,Armas:2014rva,Armas:2015ssd,Armas:2016mes,Armas:2016xxg,Armas:2017pvj}, and yields the same results as those presented below. 
 We thank J. Armas and J. Bhattacharya for discussions on this point. }
\begin{equation}\label{varact}
\delta S= \frac{1}{16 \pi}  \int_{M} \sqrt{-g}~\left({\mathcal K}^2-K_{\mu\nu}K^{\mu\nu}- \frac{4\pi T_0}{\sqrt{-k.k}}  
\left( {\mathcal K}+u.K.u \right) \right)\delta z
\end{equation}
It follows that the action \eqref{memact} is stationary under shape variations provided that 
\begin{equation}\label{statco} 
\frac{ {\mathcal K}^2-K_{\mu\nu}K^{\mu\nu} }{{\mathcal K}+u.K.u}= \frac{4\pi T_0}{\sqrt{-k.k}}
\end{equation}
In the stationary situation under consideration $\sigma_{\mu\nu}=0$ and so the LHS of \eqref{statco}
equals $\mathcal{{\tilde K}}$ (see \eqref{km})  and \eqref{statco} is the same as \eqref{memeo}. We have 
thus demonstrated that \eqref{memeo} follows as the condition for stationarity of the membrane action 
\eqref{memact}.

\subsection{Variation of the action w.r.t. the metric}

In this subsection we study the change in the membrane action as a response to a variation of the induced metric on the 
membrane world volume - rather than the membrane shape as in the previous subsection. We pause to explain precisely 
what this means. 

Consider a spacetime with a boundary $S$. Consider the Einstein Hilbert action for the spacetime contained within
$S$, supplemented by the Gibbon's Hawking term on the boundary $S$. It is well known that Einstein's equations in 
the interior of $S$ follow from the variation of this functional, subject to the boundary conditions that 
the induced metric on $S$ is a specified metric $g_{\mu\nu}$. Moreover it is expected to be generically true 
that there are at most discretely many solutions to Einstein's equations for any given boundary metric 
$g_{\mu\nu}$. In other words the boundary metric,  on any surface surrounding a region of spacetime, labels
solutions of Einstein's equations in its interior upto discrete ambiguities.  
\footnote{These expectations are best motivated in Euclidean space - and so are expected to apply 
well to the equilibrium spacetimes under study in this section.}

Now the membrane action \eqref{memact} is a functional of both the induced metric on the membrane as well 
as the extrinsic curvature of the membrane. As the extrinsic curvature depends on the normal derivative 
of the spacetime metric away from the membrane, \eqref{memact} would appear to be a functional of both the 
induced metric on the membrane as well as its first normal derivative inwards. However the spacetimes 
on which the membrane propagates are not arbitrary - they are solutions to Einstein's equations. And 
we have just argued in the previous paragraphs that the entire solution to the interior of the membrane 
- hence the normal derivative of the boundary metric on the membrane - and hence the extrinsic curvature 
of the membrane - are all determined by the induced metric 
on the world volume of the membrane. It follows that the variation the extrinsic curvature $K_{\mu\nu}$ (and 
so the membrane action)  w.r.t. the boundary membrane metric is well defined. We define the membrane stress tensor 
in equilibrium via the equation 
\footnote{We emphasize that the variation in \eqref{stdef} is performed onshell. The initial membrane configuration 
in \eqref{stdef} is assumed to be onshell w.r.t shape variations of the membrane. Logically speaking, the final 
membrane configuration in \eqref{stdef} should also be taken onshell, but for the purposes of computing the 
stress tensor \eqref{stdef} this condition is unimportant and can be dropped. The reason for this is simply that 
the variation of the membrane action - due to a change in shape of the membrane - vanishes when taken around
a solution to the membrane equations of motion. }
\begin{equation} \label{stdef}
\delta S= - \frac{1}{2} \int_M \sqrt{-g} T^{\mu\nu} \delta g_{\mu\nu}
\end{equation}
The variation in \eqref{stdef} is performed within the space of stationary membrane metrics (i.e. membrane metrics 
that admit a killing direction). The variation in \eqref{stdef} can be taken to be performed with $k^\mu$ held fixed. 
Though we will not need this for calculational purposes, at the conceptual level it is sometimes useful to work in 
the coordinate system \eqref{mos}. In this coordinate system stationary variations of $g_{\mu\nu}$ are a consequence of 
varying $w_{ij}$, $a_i$ and $\sigma$. Note that we have enough variations to define every component of the stress tensor; 
note also that, with this coordinate choice, all variations are performed holding $k^\mu\partial_\mu= \partial_t$ 
fixed. 

Although the stress tensor \eqref{stdef} is well defined, there is a catch. The variation of the extrinsic 
curvature w.r.t the induced metric on the membrane is, in general, 
a highly nonlocal function of the induced metric on the membrane. \footnote{This is analogous to the fact - familiar 
from the study of electrostatics - that the `normal component of the electric field', $n.\nabla \phi$ 
at a point $x$ just outside a conductor is given by an integral of the form $ \int G(x, y) \phi(y)$ where 
he integral is taken over the boundary of the conductor and $G$ is a Greens function. In this analogy the boundary 
value of the potential $\phi$ plays the role of the induced metric, while the normal component of the 
electric field plays the role of the extrinsic curvature. }. Consequently the variation of a generic action build 
out of Extrinsic curvatures would lead to a highly nonlocal stress tensor \eqref{stdef}. However our membrane action 
\begin{equation}\label{Ft}
S = \frac{1}{16 \pi} \left[  -(D-1)\lambda \int_V \sqrt{-G} + \int_{M} \sqrt{-g}~{\mathcal K} - 4\pi T_0 \int_{M} \sqrt{-g}~\gamma \right]
\end{equation}
is not generic. In particular the sum of the  first two terms in \eqref{Ft} is precisely one half of 
the onshell value of the Einstein action of the region of spacetime enclosed by 
the membrane. \footnote{More precisely, the first term in \eqref{Ft} is half of the onshell value of the bulk part of the action 
$$ \frac{1}{16\pi}\int_V \sqrt{-G} \left( R+\lambda(D-1)(D-2) \right) $$
(this is easily verified by making the the onshell substitution $R=-D(D-1)\lambda$)
while the second term is half of the Gibbons Hawking boundary term 
$$ \frac{1}{8 \pi}  \int_M \sqrt{-g}~ {\mathcal K} $$.} 
It follows that \eqref{Ft} may be rewritten as  
\begin{equation}\label{Ftn}
S = \frac{1}{2} S_{in} - \frac{T_0}{4} \int_{M} \sqrt{-g}~\gamma 
\end{equation}
where 
\begin{equation}\label{sinfi}
S_{in} = \frac{1}{8 \pi} \left[  -(D-1)\lambda \int_V \sqrt{-G} + \int_{M} \sqrt{-g}~{\mathcal K} \right]
\end{equation}
$S_{in}$ is the value of Einstein's action of the spacetime to the interior of our membrane; this can be made
more explicit by using the bulk Einstein equation to rewrite \eqref{sinfi} as 
\begin{equation}\label{sinff}
S_{in} = \frac{1}{16\pi}\int_V \sqrt{-G} \left( R+\lambda(D-1)(D-2) \right)
+ \frac{1}{8 \pi} \int_{M} \sqrt{-g}~{\mathcal K} 
\end{equation}

The only dependence of \eqref{Ftn} on the extrinsic curvature comes from the fact that $S_{in}$ depends 
on $K_{\mu\nu}$. However this dependence is very special. In particular it follows from the Hamilton Jacobi equations 
applied to Einstein gravity that 
\begin{equation}\label{Sinn}
 \delta S_{in} = -\frac{1}{16 \pi}  \int_M \sqrt{-g} \delta g_{\mu\nu}
\left(K^{\mu\nu} - {\cal K} g^{\mu\nu} \right) 
\end{equation}
\footnote{Logically speaking the variation in \eqref{Sinn} is performed in a completely onshell manner 
in the bulk - i.e. from one solution of Einstein's equations parametrized by an induced boundary metric to another solution of Einstein's equations 
parametrized by a slightly varied boundary metric. At the formal level, however, one could ask the following 
question. Suppose we start in a solution of the bulk Einstein equation, but let the bulk metric variation 
be offshell. This offshell variation is an arbitrary function of the full bulk, not just the boundary. 
We could then define the appropriately normalized coefficient of the bulk metric variation to be the 
bulk spacetime stress tensor resulting from the action \eqref{Sinn}. J. Armas has pointed out to us that 
provided we use \eqref{sinff} to define $S_{in}$  then - as follows from standard textbook derivations 
of Einstein's equations - the bulk stress tensor that follows has the form \eqref{sts} where the restriction 
of $T_{MN}$ to the membrane is given by \eqref{Sinn}. Note, in 
particular, that with this choice of $S_{in}$ our bulk stress tensor has no terms proportional to 
$\delta'(\rho-1)$. The offshell stress tensor that follows from the full action \eqref{Ftn} 
also has the form \eqref{sts} where the restriction of $T_{MN}$ to the membrane is 
given by \eqref{ona}. It is important that, in the language of \cite{Armas:2013hsa,Armas:2013goa,Armas:2014rva,Armas:2015ssd,Armas:2016mes,Armas:2016xxg,Armas:2017pvj}, 
we find ${\hat T}_2^{MNO}=0$ (this is equivalent to the fact that the `bulk' stress tensor has
no $\delta'$ pieces). From the point of view of 
\cite{Armas:2013hsa,Armas:2013goa,Armas:2014rva,Armas:2015ssd,Armas:2016mes,Armas:2016xxg,Armas:2017pvj}
it is this fact that allows - for example - the energy current of our membrane to take the simple form 
\eqref{cce} rather than the more complicated form it would have taken had the bulk stress tensor 
also had $\delta'$ pieces. We thank J. Armas and J. Bhattacharya detailed discussions and explanations 
on this topic.}

Consequently the variation of the first two terms in \eqref{Ft} leads to a contribution 
to the membrane stress tensor equal to half of the Brown York stress tensor, 
i.e. 
\begin{equation} \label{byeq}
\delta \left( \frac{S_{in}}{2} \right)= - \frac{1}{2} \int_M \sqrt{-g} 
\left(  \frac{1}{16 \pi} \left(K^{\mu\nu} - {\cal K} g^{\mu\nu} \right) \right) \delta g_{\mu\nu}
 \end{equation}
 and is completely local. 
 
 The third term in \eqref{Ft} is a manifestly local functional of the induced metric on the membrane,
and so its variation w.r.t. the induced metric results in an manifestly local contribution to the stress tensor. 
 \begin{equation}\label{vga}
 \begin{split}
  \delta \left( \sqrt{-g}~\gamma \right) &= \frac{1}{2} \sqrt{-g}~\gamma~g^{\mu\nu}~\delta g_{\mu\nu} + \frac{1}{2}\sqrt{-g}~\gamma^3~k^\mu k^\nu \delta g_{\mu\nu} \\
  &= \frac{1}{2} \sqrt{-g}~\gamma \left( g^{\mu\nu}+u^\mu u^\nu \right)\delta g_{\mu\nu} \\
  &= \frac{1}{2} \sqrt{-g}~\gamma {\cal P}^{\mu\nu} \delta g_{\mu\nu}
 \end{split}
 \end{equation}
 Consequently it follows that 
 \begin{equation}\label{vart}
 \delta \left( - \frac{T_0}{4} \int_{M} \sqrt{-g}~\gamma  \right)
 = - \frac{1}{2} \int_M \sqrt{-g} \left( \frac{T_0}{4} \gamma {\cal P}^{\mu\nu} \right) \delta g_{\mu\nu}
 \end{equation}
 
 Adding \eqref{byeq} and \eqref{vart}, it follows from \eqref{Sinn} that 
 \begin{equation} \label{byeqf}
\delta S = - \frac{1}{2} \int_M \sqrt{-g} 
\left(  \frac{1}{16 \pi} \left(K^{\mu\nu} - K g^{\mu\nu} \right)  + \frac{T_0}{4} \gamma {\cal P}^{\mu\nu} 
\right) \delta g_{\mu\nu}
 \end{equation}
 Comparing \eqref{byeqf} with \eqref{stdef} we conclude that
 \begin{equation} \label{stit}
 16 \pi~T^{\mu\nu} = \frac{4\pi T_0}{\sqrt{-k.k}}~ {\cal P}^{\mu\nu} + (K^{\mu\nu}-{\mathcal K}g^{\mu\nu})
 \end{equation} 
 
 Recall that \eqref{stit} applies only for stationary membranes that obey the onshell condition \eqref{memeo}. 
 Using \eqref{memeo} it follows that \eqref{stit} may be rewritten as 
 \begin{equation} \label{ona}
 16 \pi~T^{\mu\nu} = \tilde {\mathcal K}~ {\cal P}^{\mu\nu} + (K^{\mu\nu}-{\mathcal K}g^{\mu\nu})
 \end{equation} 
 in perfect agreement with \eqref{stf} in the stationary case. In summary, we have demonstrated that the stress tensor that 
 follows from the variation of our membrane action agrees with the general fluid stress tensor \eqref{stf} 
 evaluated on equilibrium configurations. 

As \eqref{ona} is a special case of \eqref{stf}, it follows that it obeys the condition \eqref{KT}. 
We end this subsection with a brief logical explanation (i.e. one that does not rely on algebraic verification) 
that this had to be the case.

Consider a membrane propagating in a given background 
solutions of Einstein's equations. There is one vary easy way to vary the induced metric on the membrane while 
ensuring that the spacetime inside the membrane continues to solve Einstein's equations. One can do this by simply 
infinitesimally displacing the membrane a little bit within the given background solution of Einstein's equations.  
Even though this process does not modify the background metric, it changes the induced metric on the membrane. 
As explained in \eqref{variations}, the change in the induced membrane metric produced by such a manoeuvre 
is equal to $2 K_{\mu\nu} \delta z$ where $\delta z$ is arbitrary.
 As explained in the previous subsection, however, the onshell membrane action is 
stationary under arbitrary variations of the membrane volume, and so we find from \eqref{stdef} that 
\begin{equation} \label{stdefn}
0= - \frac{1}{2} \int_M \sqrt{-g} T^{\mu\nu} (2 K_{\mu\nu} \delta z)
\end{equation}
As \eqref{stdefn} is true for any choice of the function $\delta z$ it follows that 
$$T^{\mu\nu} K_{\mu\nu}=0.$$
In other words the stress tensor defined by varying the action using \eqref{stdef} 
automatically obeys the equation \eqref{KT}.

\section{Simple Static Membrane Configurations and their Thermodynamics}

In this section we study simple static solutions of the membrane equations and compare their thermodynamics with 
that of the dual black holes. The solutions we study are Schwarzschild black holes in flat space, global 
$AdS$ space and de Sitter space. 

\subsection{Coordinates and Conventions}

In this section we study the maximally symmetric backgrounds 
\begin{equation} \label{maxsym}
ds^2 = -f(r) dt^2 +\frac{dr^2}{f(r)} + r^2d\Omega_{D-2}^2, \quad f(r) = 1 + \lambda~r^2.
\end{equation}
Of course \eqref{maxsym} are exact solutions to the Einstein equations \eqref{eeq}. Before proceeding 
with our analysis we pause to describe the coordinates employed in  \eqref{maxsym}. 

When $\lambda=0$ \eqref{maxsym} is just flat space in polar coordinates, and this case requires no further elaboration. 
When $\lambda >0$ the spacetime \eqref{maxsym} is Anti de Sitter space of squared radius $\frac{1}{\lambda}$ in global coordinates. 
Notice that the function $f(r)$ never vanishes in this case. For $r^2 \ll \frac{1}{\lambda}$ this spacetime is approximately flat; 
for $r^2 \gg \frac{1}{\lambda}$ the spacetime approximates Poincare Patch $AdS$ space (i.e. $AdS$ space with planar sections). 
According to the AdS/CFT correspondence, this is the spacetime dual to the vacuum of ${\cal N}=4$ Yang Mills theory. 
Finally when $\lambda<0$, the part of \eqref{maxsym} with $r^2|\lambda|<1$ is the static patch of de Sitter spacetime. 
Recall that the static patch is the causal past of a static observer in global de Sitter spacetime. The submanifold 
$r^2|\lambda|=1$ is the future horizon of the causal patch. Points with $r^2|\lambda|>1$ lie outside the static patch. 
While the killing vector $\partial_t$ is timelike within the causal patch, it is spacelike outside the causal patch. 
As we have explained above, our construction of stationary membranes is based on a timelike killing vector field, which 
we will chose to be $\partial_t$ in the case of the backgrounds \eqref{maxsym}. When $\lambda<0$ the requirement 
that our killing vector field be timelike forces us to restrict our attention to within the static patch. At any rate
below we will focus our attention principally on $\lambda=0$ or $\lambda$ positive. 

\subsection{Exact Black Hole solutions and their Thermodynamics}

It is well known that  following metrics are exact solutions of the Einstein equations 
\eqref{eeq} 
\begin{equation} \label{spbh}
ds^2 = -f(r) dt^2 +\frac{dr^2}{f(r)} + r^2 d\Omega_{D-2}^2, \quad f(r) = 1-\frac{r_0^{D-3}}{r^{D-3}} + \lambda~r^2
\end{equation}
The metrics \eqref{spbh} reduce to \eqref{maxsym} at large $r$. They also possess an event horizon (or in the case of 
$\lambda<0$ an additional event horizon) and so represent Schwarzschild black holes in flat, global AdS and static patch de Sitter 
space respectively. 

The (additional) event horizon of the metric \eqref{spbh} is located at $r=r_H$ determined by the condition that $f(r_H)=0$, i.e. 
$r_H$ obeys the equation 
\begin{equation}\label{rhro}
r_0^{D-3} = \left( 1+\lambda r_H^2 \right)r_H^{D-3}
\end{equation}
At least for $\lambda=0$ and $\lambda>0$ - the cases to which we restrict attention in most of the rest of this subsection - 
the mass, entropy and temperature of the black hole solutions are unambiguously 
well defined; specializing to this case, the mass and entropy of the black holes were listed in, e.g., \cite{Witten:1998zw}
and are given by;  \footnote{In the case $\lambda=0$ the black hole mass is its usual ADM energy. 
In the case $\lambda>0$ the black hole mass is given by integrating the boundary stress tensor (Brown York stress tensor plus
suitable counterterms) over the boundary sphere, and coincides with the energy of the dual field theory on $S^{D-2}$.}
\begin{equation}\label{bhth}
\begin{split}
& M_{bh} = \frac{(D-2)\left( 1+\lambda r_H^2 \right)r_H^{D-3}\Omega_{D-2}}{16\pi}, \quad S_{bh} = \frac{r_H^{D-2}\Omega_{D-2}}{4}, \\ & T_{bh} =\frac{1}{4\pi r_H}\left[ (D-3) + (D-1)\lambda r_H^2 \right]
\end{split}
\end{equation}

\subsection{Membrane solutions and their thermodynamics}

We now study stationary membrane solutions in the background \eqref{maxsym}. We use the formalism developed 
in earlier sections, base our construction on the killing vector $k^\mu \partial_\mu = \partial_t$. It is easily 
verified that the spherical membranes  $r={\tilde r}_H$ are solutions of the stationary membrane equations
\eqref{memeo}. As we will see below, the thermodynamics of spherical membranes at 
$r={\tilde r}_H$ exactly matches \eqref{bhth} provided we make the identification 
${\tilde r}_H=r_H$, forcing us to identify ${\tilde r}_H$ with $r_H$. We use this foreknowledge to lighten 
the notation of this subsection by simply dropping the tilde on $r_H$ in all the formulae that follow. 

For the membrane shape under consideration it is not difficult to verify that 
\begin{equation}\label{smqu}
\begin{split}
& q_{\mu}= \left(1+\lambda r_H^2\right)^{1/2} (dt)_{\mu}, \quad k^\mu = (\partial_t)^\mu,\quad  u^\mu = \left(1+\lambda r_H^2\right)^{-1/2}(\partial_t)^\mu \\ &\quad K_{tt} = -\lambda r_H (1+\lambda r_H^2)^{1/2},\quad K_{ta} = 0, \quad K_{ab} = r_H (1+\lambda r_H^2)^{1/2}\Omega_{ab}
\end{split}
\end{equation}
and that 
\begin{equation}\label{moreqt}
\begin{split}
&\mathcal{K}= r_H\lambda (1+\lambda r_H^2)^{-1/2} + \frac{D-2}{r_H}(1+\lambda r_H^2)^{1/2},\\ & K_{\mu\nu} K^{\mu\nu}=\lambda^2r_H^2(1+\lambda r_H^2)^{-1}+\frac{D-2}{r_H^2}(1+\lambda r_H^2), ~~~u.K.u=-\lambda r_H(1+\lambda r_H^2)^{-1/2}, \\ &\tilde{ \mathcal{K}}=(1+\lambda r_H^2)^{-1/2}\left[\frac{D-3}{r_H}+(D-1)\lambda r_H\right],  ~~~\sqrt{-k.k}= (1+\lambda r_H^2)^{1/2}
\end{split}
\end{equation}

It follows from the second last and the last of \eqref{moreqt} and from \eqref{memeo} that the temperature, $T_0$, of this membrane 
configuration is given by 
\begin{equation}\label{tempmem}
T_0 = \frac{1}{4\pi r_H} \left[ (D-3)+(D-1)\lambda r_H^2 \right]
\end{equation}
in perfect agreement with \eqref{bhth}.

The energy of our membrane is given by 
\begin{equation}\label{sme}
 E = -\int \sqrt{h}~q^\mu T_{\mu\nu} k^\nu = -\frac{1}{16\pi}\int \sqrt{h}~q^\mu (K_{\mu\nu}-{\mathcal K}g_{\mu\nu}) k^\nu
\end{equation}
and its entropy by
\begin{equation}\label{sms}
S_{ent} = \int \sqrt{h}~q_\mu J^\mu_S = \frac{1}{4}  \int \sqrt{-g}~\gamma
\end{equation} 
Substituting \eqref{smqu} into \eqref{sme} and \eqref{sms} we find the explicit results
\begin{equation} \label{enentmem}
\begin{split}
E = \frac{(D-2)r_H^{D-3}(1+\lambda r_H^2)\Omega_{D-2}}{16\pi}, \quad S_{ent} = \frac{r_H^{D-2}\Omega_{D-2}}{4}
\end{split}
\end{equation}
Once again \eqref{enentmem} is in perfect agreement with \eqref{bhth}.

As a check we note that 
\begin{equation}\label{tcheck}
\frac{\partial E}{\partial S_{ent}} = \frac{\partial E}{\partial r_H} \left(\frac{\partial S_{ent}}{\partial r_H}\right)^{-1} = \frac{1}{4\pi r_H}\left[ (D-3) + (D-1)\lambda r_H^2 \right] =T_0
\end{equation}
(where we have used \eqref{tempmem} in the last equality). We conclude that the thermodynamical temperature 
or our system is, indeed, $T_0$. 

Finally, it is not difficult to evaluate the $S=-\ln Z$ of our spherical membrane solutions. Using \eqref{memact} we find 
\begin{equation}\label{sfm}
-\ln Z = S= \frac{r_H^{D-3}\Omega_{D-2}}{16 \pi T_0}\left[ (D-2)\left(1+\lambda r_H^2\right) -4\pi T_0 r_H \right]
\end{equation} 
The partition function \eqref{sfm} has been presented as a function of $r_H$; however one can, in principle, 
invert \eqref{tempmem} to obtain $r_H$ as a function of temperature and so view $\ln Z$ as a function of temperature. 

As a check, it is not difficult to use \eqref{sfm},  together with the thermodynamical relations 
\begin{equation}\label{thermrel}
E= \partial_\beta S, ~~~S_{ent} = \partial_T(TS)
\end{equation}
together with the explicit formula for the temperature \eqref{tempmem} to reproduce the relations 
\eqref{enentmem}. 

In this section we study only the membrane duals of static Schwarzschild type black holes. We largely leave the 
generalization of this discussion to rotating Kerr type black holes to future work. However see Appendix 
\ref{rot} for preliminary work in this direction.

\subsection{The membrane and boundary stress tensors}

It may be verified that the induced metric on the membrane and its world volume stress tensor, 
evaluated on the equilibrium configurations of this section are given by 
\begin{equation}\label{memst} \begin{split}
    & ds^2= -(1+\lambda r_H^2)dt^2 + r_H^2 d \Omega_{D-2}^2 \\
    & 16\pi T_{tt} = \frac{D-2}{r_H}(1+\lambda r_H^2)^{3/2} , ~~~~ T_{ta} =0, ~~~~16 \pi T_{ab}= (1+\lambda r_H^2)^{-1/2}\lambda r_H^3\Omega_{ab}
\end{split}
\end{equation}  
where $\Omega_{ab}$ is the metric on the world volume of a unit sphere. As a check on this formula it may be verified that 
\begin{equation}\label{enform} 
  \int \frac{\sqrt{-g}}{-g_{tt}} T_{tt} =M_{bh}
\end{equation}
where $g$ represents the metric on the world volume of the membrane.

It is interesting to specialize \eqref{memst} to the case $\lambda=1$ (in which case our solution is a spherical membrane in 
a unit radius $AdS$ space) and compare \eqref{memst} with the boundary stress tensor of the dual gravitational black hole 
\eqref{spbh}. The stress tensor lives on the manifold on which the field theory is defined, i.e. 
\begin{equation}\label{ftm} 
ds^2= -dt^2 + d \Omega_{D-2}^2
\end{equation}
Its form may be read off, for instance, from section 5.3 of \cite{Bhattacharyya:2007vs} and is given by 
\begin{equation}\label{bdyst} \begin{split}
    & 16\pi T^B_{tt}= 2m(D-2) = (D-2)r_H^{D-3}(1+r_H^2), ~~~~T^B_{ta} =0, \\  ~~~~& 16 \pi T^B_{ab}= 2m\Omega_{ab} = r_H^{D-3}(1+ r_H^2)\Omega_{ab}
\end{split}
\end{equation}
It is also  easily verified that 
\begin{equation}\label{enformft} 
  \int \frac{\sqrt{-g}}{-g_{tt}} T^B_{tt} =M_{bh}
\end{equation}
where, here  $g$ represents the metric \eqref{ftm}.

The fact that \eqref{enformft} and \eqref{enform} are both true of course means that 
the membrane and field theory stress tensors are related to each other. However the 
precise relationship, while easy to state,
\footnote{The requirement that \eqref{enformft} and \eqref{enform} be simultaneously valid 
determines the ratio of the membrane and boundary $T_{tt}$ components. The requirement that 
the boundary stress tensor is traceless, while the membrane stress tensor obeys 
\eqref{introkt} then also determines the ratio of energy density to pressure, 
on both the boundary and the membrane.}
is not visually transparent.

 In the large $r_H$ limit,
on the other hand, \eqref{memst} simplifies to 
\begin{equation}\label{memext}
\begin{split}
    & ds^2 \equiv \mathbf{g}_{\alpha \beta} dx^\alpha dx^\beta = r_H^2(-dt^2 + d \Omega_{D-2}^2) \\
    & 16 \pi \mathbf{T}_{tt}=(D-2)r_H^2 , ~~~~16 \pi \mathbf{T}_{ab}= r_H^2\Omega_{ab}
\end{split}
\end{equation}
while \eqref{bdyst} simplifies to 
\begin{equation}\label{bdyext} \begin{split}
    & ds^2 \equiv \mathbb{G}_{\alpha \beta} dx^\alpha dx^\beta = -dt^2 + d \Omega_{D-2}^2 \\
    & 16 \pi \mathbb{T}_{tt}= (D-2)r_H^{D-1}  , ~~~~16 \pi \mathbb{T}_{ab}= r_H^{D-1} \Omega_{ab}
\end{split}
\end{equation}
In this case \eqref{memext} and \eqref{bdyext} are simply related by the
Weyl scaling
$$ \mathbf{g}_{\alpha \beta} = r_H^2 \mathbb{G}_{\alpha \beta}, ~~~
\mathbf{T}^\alpha_\beta= \frac{1}{r_H^{D-1}} \mathbb{T}^\alpha_\beta. $$
where the Field Theory metric and Stress tensor are written with `hollow' letters ($\mathbb{G}_{\alpha \beta},~ \mathbb{T}_{\mu\nu}$) and the Membrane metric and Stress tensor are written with `thick bold' letters ($\mathbf{g}_{\alpha \beta},~ \mathbf{T}_{\mu\nu}$).

\subsection{Spectrum of linearized Excitations}

In Appendix \ref{lin} we have linearized our membrane equations around the spherical 
membranes dual to Schwarzschild black holes in flat space. The final results for 
the spectrum of linearized fluctuations is presented in 
\eqref{qem}, \eqref{qemt}, \eqref{qvm}. It is easily verified that 
this spectrum agrees with the results of Emparan, Suzuki and Tanabe \cite{Emparan:2014aba}
at leading order in the large $d$ limit. It should be straightforward to generalize this fluctuation 
analysis to static membranes at non-zero $\lambda$, however we have not done this 
calculation in this paper.

\section{Fluid Gravity from Membrane Dynamics}

In this section we specialize to the study of the motion of a membrane in planar $AdS$ space, and ask ourselves 
the following question: what does the membrane dynamics look like from the perspective of a boundary observer. 

In principle this question is easily answered in the following manner. The membrane is a source for linearized gravitational 
fluctuations about pure $AdS$ space. The precise form of these fluctuations may be obtained by convoluting 
the membrane stress tensor with the appropriate Green's function. The Green's function may be obtained along the lines 
of the analysis of \cite{Bhattacharyya:2016nhn} (in which the same problem was solved about flat space). 
This Green's function may be used to construct a linear map from the membrane to the boundary stress tensor 
of the schematic form 
\begin{equation} \label{bulkbdry}
\mathbb{T}(x)=  \int H(x,y) \mathbf{T} (y)
\end{equation}
for some kernel function $H(x,y)$ (all indices have been omitted in the highly schematic equation
\eqref{bulkbdry}). As gravitational fluctuations 
can only be consistently sourced by a conserved bulk stress tensor, the map \eqref{bulkbdry} is well defined only when 
$\mathbf{T} (y)$ is conserved in the bulk. Whenever this is the case, $\mathbb{T}(x)$ is well defined - and is automatically 
conserved and traceless on the boundary. In other words  \eqref{bulkbdry} maps a membrane stress tensor that is conserved in the bulk 
to a boundary stress tensor conserved on the boundary. We will see below that the relationship between these two conservation 
equations is very tight - at the algebraic level the map \eqref{bulkbdry} converts the membrane world volume stress tensor 
conservation equation \eqref{sttr} into a conservation equation for the boundary stress tensor, while the equation 
\eqref{introkt} is mapped  to the condition that the boundary stress tensor is traceless. 

Restated, the membrane equations - which we have so far viewed as conservation equations on the world volume of the 
membrane - may be recast as conservation equations in the flat boundary spacetime $R^{d-1,1}$. If we adopt this presentation 
then it is very unnatural to use the membrane velocity and height function as our dynamical variables, as these variables 
do not naturally live on the boundary. Instead, as we explain below, it is natural 
for the boundary observer to use a boundary velocity field $v^\mu(x)$  and a local boundary temperature field $T(x)$ to study dynamics (we will provide precise definitions of these variables in terms of the boundary stress tensor 
below). \footnote{In the long wavelength limit this choice of 
boundary variables is standard in the study of hydrodynamics. We emphasize, however, that these variables are well defined, 
and so can be utilized, even outside this limit.} Note that the boundary velocity field has the same number 
of components as the bulk membrane velocity field while the `location' variable of the bulk membrane is traded for 
the boundary temperature. Using the map \eqref{bulkbdry}, the explicit form of the membrane stress tensor
\eqref{stf} yields precise expressions for the boundary variables in terms of the bulk variables. These expressions 
take the schematic form 
\begin{equation} \label{frd}
v^\mu= v^\mu(u^\mu, z),~~~~T= T(z, u^\mu)
\end{equation}
where $z$ denotes the location of the membrane in the radial $AdS$ direction (see below) and $u^\mu$ is the membrane 
world volume velocity field. The relations \eqref{frd} may be inverted, and may be regarded as a field redefinition from membrane 
to boundary variables. The boundary stress tensor may now be re expressed in terms of $v^\mu$ and $T$, and the condition that 
the boundary stress tensor is conserved yields a set of boundary equations of motion for these natural boundary variables. In the long wavelength 
limit - which we will now focus on - these are simply the equations of boundary hydrodynamics. 

In general the expressions \eqref{frd} are highly non-local; as a consequence the boundary dynamical system for the variables 
$v^\mu$ and $T$ is, in general, highly non-local Consider however a limit in which the membrane is `nearly flat' (see below 
for what this means)  and varies slowly in the `field theory directions'. In this limit it turns out that the map between 
membrane and boundary variables is approximately local. The boundary stress tensor is, thus, also an approximately local 
functional of boundary variables in this limit - and takes the form of a hydrodynamical stress tensor that is expressed 
in terms of the boundary temperature and velocity field by a set of constitutive relations that may be obtained and 
presented, order by order, in a derivative expansion. In this section we focus in this limit and  work out the resultant boundary
hydrodynamical constitutive relations at upto second order in the derivative expansion. 

Of course the exact finite $D$ expressions for the constitutive relations of the boundary stress tensor are known upto 
second order in the derivative expansion - the determination of these coefficients was achieved as part of the 
programme of the fluid gravity 
correspondence. Comparing our results with those of fluid gravity we find - perhaps unexpectedly - that our membrane 
induced constitutive relations agree exactly -at finite $D$ - with the results of fluid gravity at zeroth and first order 
in the derivative expansion. At second order in derivatives, however, the membrane constitutive relations agree with 
the exact results of fluid gravity only at large $D$ and deviate from the exact results in a power series in $\frac{1}{D}$. 

The papers \cite{Emparan:2016sjk,Herzog:2016hob} have previously demonstrated that the equations of `scaled black brane 
dynamics' reduce - under an appropriate field redefinition - to the equations of boundary hydrodynamics at large 
$D$ in an appropriate scaling limit. The analysis of this subsection generalizes the discussions of  
\cite{Emparan:2016sjk,Herzog:2016hob} in several ways. First, in this paper we map the full nonlinear membrane equations of 
motion to full nonlinear equations of boundary hydrodynamics, and do not work in a particular scaling limit. 
Next, the starting point of our analysis is the equations for probe membrane dynamics. Our probe membrane is 
defined by the improved stress tensor \eqref{stf} and its motion is well defined at finite $D$. In this section we 
map our finite $D$ probe membrane dynamics to the equations of finite $D$ boundary hydrodynamics, and obtain results that 
agree exactly with those of fluid gravity at zero and first order in derivatives even at finite $D$. Finally, the method we employ
in our analysis utilizes the linearized backreaction of the membrane on gravity, and - in our opinion - conceptually 
clarifies the relationship between membrane dynamics and boundary hydrodynamics. 

Let us end these introductory comments by re emphasizing that the improved membrane stress tensor
\eqref{stf} yields the exact zero and first order constitutive relations of hydrodynamics even at finite $D$. 
Recall that our improved stress tensor \eqref{stf} represents the sum over of what - from other points of view - would 
be regarded as a very particularly chosen infinite set of corrections to the leading large $D$ stress tensor \eqref{std}. The fact 
that precisely this infinite class of terms was sufficient to obtain exact results for zero and first order 
fluid coefficients suggests that improved membrane equations presented in this paper represents a 
useful resummation of $\frac{1}{D}$ perturbation theory.

\subsection{Equilibrium}

\subsubsection{Black Branes}

In the previous section we studied the membrane solutions dual to static black holes of radius $r_H$ in $AdS_D$ 
spacetime. In the limit that $r_H \to \infty$, black hole reduce (locally) to black branes and their dual 
spherical membranes in global AdS space reduce locally to planar membranes in Poincare patch AdS space. We use notation
\begin{equation}\label{ddef}
 d=D-1
\end{equation}
Recall that a black brane in AdS space is defined by the metric
\begin{equation} \label{bbme}
ds^2 = \frac{1}{\rho^2} \left[ -\left(1-\frac{\rho^d}{\textbf{z}^d}\right)dt^2 +\frac{d\rho^2}{\left( 1 -\frac{\rho^d}{\textbf{z}^d} \right)} + \delta_{ij} dx^i dx^j \right]
\end{equation}
In \eqref{bbme}  $\rho=\textbf{z}$ is the event horizon of the black brane. Now we rewrite the metric \eqref{bbme} in Fefferman-Graham coordinates by change of $\rho$ variable to $z$ \footnote{Fefferman-Graham coordinates are defined by the requirement  $G_{zz}=\frac{1}{z^2}$ and $G_{z\mu}=0$.}. The metric becomes
\begin{equation}\label{fgbb}
ds^2 = \frac{1}{z^2}\left[ -\frac{\left(1-\frac{z^d}{4\textbf{z}^d}\right)^2}{\left(1+\frac{z^d}{4\textbf{z}^d}\right)^{2-4/d}} dt^2 +dz^2 + \left( 1+\frac{z^d}{4\textbf{z}^d} \right)^{4/d} \delta_{ij} dx^i dx^j \right]
\end{equation}
Throughout this section we employ the Fefferman-Graham coordinate choice. 
Now expanding the metric \eqref{fgbb} in power series in $\frac{z^d}{\textbf{z}^d}$ and retaining terms upto the first 
subleading order in this expansion we get
\begin{equation}\label{bbe}
ds^2 =  -\left[1-\left(\frac{d-1}{d}\right)\frac{z^{d-2}}{\textbf{z}^d}\right]dt^2 + \frac{dz^2}{z^2} + \left[1+\left(\frac{1}{d}\right)\frac{z^{d-2}}{\textbf{z}^d}\right]\delta_{ij} dx^i dx^j 
\end{equation}
Now we can recover the boundary stress tensor corresponding to the black brane solution \eqref{bbe} by the prescription
\begin{equation}\label{bbs}
 \mathbb{T}^{\mu}_{\nu} = -\frac{1}{8\pi} \lim\limits_{z\rightarrow0} \frac{K^\mu_\nu-\delta^\mu_\nu}{z^d}
 \end{equation}
Note that to use the prescription \eqref{bbs}, it is sufficient to use the metric expanded to linear order in 
$\frac{z^d}{4\textbf{z}^d}$, i.e. \eqref{bbe}, rather than full metric \eqref{fgbb}. Indeed it is easily verified that the boundary stress tensor 
corresponding to this solution is just the coefficient of $z^{d-2}$ in the metric \eqref{bbe}, and the boundary stress tensor \footnote{
For this subsection, the boundary fluid-gravity metric, Stress tensor and Entropy current are written with `hollow'
letters ($\mathbb{G}_{\alpha \beta},~ \mathbb{T}_{\mu\nu},~ \mathbb{J}_S^\mu$) and the Membrane metric, 
Stress tensor and Entropy current are written with `thick bold' letters ($\mathbf{g}_{\alpha \beta},~ \mathbf{T}_{\mu\nu}~ \mathbf{J}_S^\mu$).} 
dual to the black brane solution is given by 
\begin{equation}\label{bbs0}
16 \pi \mathbb{T}_{\mu\nu} = \left(\frac{4\pi T_{bb}}{d}\right)^d \left(\eta_{\mu\nu}+dv_\mu v_\nu\right)
\end{equation}
where 
 $v^\mu= (1, 0, 0, 0 ...)$ and the temperature of the black brane $T_{bb}$ is given by 
\begin{equation}\label{tempobb}
T_{bb} = \frac{d}{4\pi\textbf{z}}, 
\end{equation}
The boundary entropy current corresponding to the 
black brane is given by 
\begin{equation}\label{bec}
\mathbb{J}_S^\mu= \frac{1}{4}\left(\frac{4\pi T_{bb}}{d}\right)^{d-1} v^\mu
\end{equation}

\subsubsection{Flat Membranes}

The membrane configuration dual to this black brane is given by the submanifold 
$z={\mathbf z}$ of the pure AdS metric 
\begin{equation} \label{pame}
ds^2 = \frac{dz^2 + \eta_{\mu\nu} dx^\mu dx^\nu}{z^2} = \frac{dz^2 - dt^2 + dx^i dx_i}{z^2} 
\end{equation}
The membrane induced metric and stress tensor are given by
\begin{equation}\label{bmst}
\begin{split} 
{\bf g}_{\mu\nu} &= \frac{\eta_{\mu\nu}}{{\bf z}^2} \\
{\cal T}_{MN}& = z \delta(z-{\mathbf z}) T_{MN}\\
T_{z z}&= T_{\mu z}=0,~~~T_{M=\mu, N=\nu} ={\rm {\bf T}_{\mu\nu}=independent ~of}~ x^\mu
\end{split}
\end{equation}

\footnote{The factor of $z$ on the RHS of \eqref{bmst} is the factor of $|\partial \rho|$ in \eqref{sts}. 
In the case at hand $\rho=-z+\textbf{z}+1$.}
Until the very end of this subsection we will use no property of ${\bf T}_{\mu\nu}$ other than the fact that it is constant.

Let us now regard the stress tensor \eqref{bmst} as a source for gravitational fluctuations 
about AdS space \eqref{pame} and compute the resultant linearized gravitational response. 
We consider the most general linearized correction to the background metric of the form
\begin{equation}\label{fbm}
ds^2 \equiv \mathfrak{G}_{MN} dX^M dX^N = \left( G_{MN} + h_{MN} \right)dX^M dX^N = \frac{dz^2 + \eta_{\mu\nu} dx^\mu dx^\nu}{z^2} + h_{MN} dX^M dX^N 
\end{equation}
We adopt the Fefferman-Graham coordinate choice and so set 
\begin{equation}
h_{zM} = 0   
\end{equation} 
The linearized Einstein equations evaluate to
\begin{equation}\label{fgee}
\begin{split}
E_{zz} &= -\frac{(d-1)}{2}z \partial_z h - (d-1)h + \frac{z^2}{2} \partial^2 h - \frac{z^2}{2} \partial_\alpha \partial_\beta h^{\alpha\beta} \\
E_{z\mu} &= \left( \frac{z^2}{2}\partial_z + z \right)\left( \partial_\alpha {h^{\alpha}}_{\mu} - \partial_\mu h \right) \\
E_{\mu\nu} &= \frac{z^2}{2}\left( \partial_\nu \partial_\alpha h{^\alpha}_\mu +\partial_\mu \partial_\alpha {h^\alpha}_\nu - \partial^2 h_{\mu\nu} \right) -\frac{z^2}{2}\partial_z^2 h_{\mu\nu}-\frac{z^2}{2}\partial_\mu\partial_\nu h+\left(\frac{d-5}{2}\right)z\partial_z h_{\mu\nu}\\&+(d-2)h_{\mu\nu} + \left[ \frac{z^2}{2} \partial_z^2 h + \frac{z^2}{2}\partial^2 h-\frac{z^2}{2}\partial_\alpha\partial_\beta h^{\alpha\beta}-\frac{(d-5)}{2}z\partial_z h -(d-2)h \right]\eta_{\mu\nu}
\end{split}
\end{equation} 
In \eqref{fgee}, the $\mu,\nu$ indices are raised with $\eta^{\mu\nu}$ and $h\equiv h_{\mu\nu}\eta^{\mu\nu}$.
Now in this case, clearly the resultant response inherits the translational invariance in the $x^\mu$ directions of the source \eqref{bmst}. Away from $z=\mathbf{z}$ the response is thus a translationally invariant solution to the linearized Einstein equations about $AdS$ space. 
In the Fefferman-Graham gauge it is easily verified that the most general linearized solution 
of this form is given by 
\begin{equation} \label{hmn}
h_{\mu\nu}= \frac{1}{z^2}\left(A^{(out)}_{\mu\nu} z^{d} + A^{(in)}_{\mu\nu} \right)
\end{equation}
The requirement that our fluctuation is normalizable ensures $A^{(in)}_{\mu\nu}=0$ outside the membrane , i.e. for $z<\textbf{z}$. On the 
other hand the requirement that the fluctuation remain bounded on the Poincare horizon forces $A^{(out)}_{\mu\nu}=0$ inside the 
membrane i.e. for $z>\textbf{z}$. The requirement that the fluctuation $h_{\mu\nu}$ is continuous across the membrane 
implies that $$A^{(in)}_{\mu\nu} = \textbf{z}^d A^{(out)}_{\mu\nu} \equiv \textbf{z}^d A_{\mu\nu}$$ So we have
\begin{equation}\label{hmnd}
   h_{\mu\nu} =
   \begin{cases}
     A_{\mu\nu} \textbf{z}^{d} z^{-2}, & \text{for}~~ z \ge \textbf{z} \\
     A_{\mu\nu} z^{d-2}, & \text{for}~~ z \le \textbf{z}
   \end{cases}
\end{equation} 
Finally, the junction matching condition on the membrane (refer \cite{Bhattacharyya:2016nhn} for similar calculation) relates the discontinuity of the Extrinsic curvature (as calculated in the linearized metric \eqref{fbm}) to membrane stress tensor as   
 \begin{equation}\label{byd}
 \textbf{T}_{\mu\nu} = -\frac{1}{8\pi} \left( \left[K^{(out)}_{\mu\nu}-K^{(in)}_{\mu\nu}\right] - \left[{\mathcal K}^{(out)}-{\mathcal K}^{(in)}\right] \bf{g}_{\mu\nu} \right)
 \end{equation}
 We find the answers for Extrinsic curvature of the membrane seen from the inside and outside as
 \begin{equation}\label{kio}
 K_{\mu\nu} =
    \begin{cases}
      \frac{\eta_{\mu\nu}}{\textbf{z}^2}+\textbf{z}^{d-2} A_{\mu\nu}, & \text{for}~~ z \ge \textbf{z} \\
      \frac{\eta_{\mu\nu}}{\textbf{z}^2}-\frac{d-2}{2}\textbf{z}^{d-2} A_{\mu\nu}, & \text{for}~~ z \le \textbf{z}
      \end{cases}
 \end{equation}
 Using \eqref{kio} in \eqref{byd} we get
 \begin{equation}\label{meten}
 A_{\mu\nu} = \frac{16\pi}{d} \frac{\textbf{T}_{\mu\nu}}{\textbf{z}^{d-2}}
 \end{equation}
It follows that the backreaction of the probe membrane
modifies the metric of $AdS$ spacetime to 
\begin{equation}\label{brm1}
 \mathfrak{G}_{MN} dX^M dX^N =
   \begin{cases}
     \frac{1}{z^2}\left(dz^2 + \eta_{\mu\nu} dx^\mu dx^\nu + A_{\mu\nu} \textbf{z}^{d}  dx^\mu dx^\nu\right) , & \text{for}~~ z \ge \textbf{z} \\
     \frac{1}{z^2} \left(dz^2 + \eta_{\mu\nu} dx^\mu dx^\nu + A_{\mu\nu} z^{d} dx^\mu dx^\nu \right) , & \text{for}~~ z \le \textbf{z}
   \end{cases}
\end{equation} 
where $A_{\mu\nu}$ is given by \eqref{meten}. As we have explained above, the  boundary Stress tensor is defined by the limit
 \begin{equation}\label{pbs}
 \mathbb{T}^{\mu}_{\nu} = -\frac{1}{8\pi} \lim\limits_{z\rightarrow0} \frac{K^\mu_\nu-\delta^\mu_\nu}{z^d}
 \end{equation}
 Using \eqref{kio} in \eqref{pbs} we get the answer for boundary Stress tensor as
 \begin{equation}\label{bst1}
 \mathbb{T}_{\mu\nu} = \frac{\mathbf{T}_{\mu\nu}}{\textbf{z}^{d-2}} 
 \end{equation}
 \eqref{bst1} is the main result of this subsection. The analogous relationship between membrane and boundary metrics and energy currents 
 takes the form 
\begin{equation} \begin{split}\label{relbetst}
\mathbb{G_{\mu\nu}}&= \mathbf{z}^2\mathbf{g}_{\mu\nu} \\
\mathbb{J}_E^\mu &\equiv \mathbb{T^{\mu}_{\nu}} k^\nu =  \frac{\mathbf{T^{\mu}_{\nu}} k^\nu}{\mathbf{z}^d} =  \frac{\mathbf{J}_E^\mu}{\mathbf{z}^d}\\
\end{split}
\end{equation}
Note that \eqref{relbetst} are meaningful equations because we have used the `same' $x^\mu$ coordinate on the membrane 
and on the boundary of spacetime. Note that \eqref{relbetst} simply expresses the condition that the membrane velocity field and stress 
tensor are Weyl equivalent to the boundary velocity field and stress tensor in the case of stationary black branes. Note in particular 
that the boundary energy (charge carried by the current $\mathbb{J}_E^\mu$) contained in a part of a spacelike slice of the  
boundary is given by 
\begin{equation} \label{intfg} 
\int \sqrt{\mathbb{G}_{ind}}~ \mathbb{J}_E^\mu~ r_\mu 
\end{equation} 
where $\sqrt{\mathbb{G}_{ind}}$ is the boundary (fluid gravity) metric induced on the spatial slice, and $r_\mu$ is the unit normal 
to this slice. 
Noting that 
$$\sqrt{\mathbb{G}_{ind}}= \textbf{z}^{d-1} \sqrt{\mathbf{g}_{ind}}. ~~~r_\mu = \textbf{z}~ q_\mu$$
and using \eqref{relbetst} it follows that \eqref{intfg} can be rewritten as  
\begin{equation} \label{intm} 
\int \sqrt{\mathbf{g}_{ind}}~ \mathbf{J}_E^\mu~ q_\mu 
\end{equation} 
Identical comments apply to the entropy. In summary, the energy/entropy contained in any part of the boundary, computed 
using boundary currents, is identical to the energy/entropy of the `same' region of the membrane, computed using 
membrane currents. In particular the formulae \eqref{bhth} in the planar black brane limit are easily reproduced directly from the membrane side.

Finally, to end this subsection we plug the explicit form of the membrane stress tensor $T_{MN}$ into the formulae above and obtain 
explicit formulae for the linearized metric perturbation, the boundary stress tensor and the boundary entropy current 
dual to our flat membrane configuration. In order to obtain these quantities we note that the induced metric and extrinsic curvature 
of the membrane located at $z={\bf z}$ are given by 
\begin{equation}\label{pmqu}
\begin{split}
ds^2_{ind} \equiv \mathbf{g}_{\mu\nu}dx^\mu dx^\nu = \frac{\eta_{\mu\nu}}{\mathbf{z}^2}dx^\mu dx^\nu = \frac{-dt^2 + dx^i dx_i}{{\mathbf z}^2} , \quad  K_{\mu\nu} = \mathbf{g}_{\mu\nu}
\end{split}
\end{equation}
It follows that the temperature of the flat membrane configuration is given by 
\begin{equation}\label{pmt}
T_m = \frac{\tilde {\mathcal K}\sqrt{-k.k}}{4\pi} = \frac{d}{4\pi\textbf{z}}
\end{equation}
and that the world volume membrane stress tensor is given by 
  \begin{equation}\label{eb0}
  16 \pi {\bf T}_{\mu\nu} =  \frac{\eta_{\mu\nu}}{\textbf{z}^2} + d~u_\mu u_\nu
  \end{equation}
Putting \eqref{eb0} in \eqref{bst1} it follows that the boundary stress tensor induced by the flat membrane of this subsubsection agrees exactly - 
at finite $D$ - with the boundary stress tensor of the exact black brane solution \eqref{bbs0} provided we make the identification 
(see  \eqref{tempobb}, \eqref{pmt})
\begin{equation} \label{rufd}
 T_{bb} = T_m = \frac{d}{4 \pi~ \mathbf{z}}, ~~~v^\mu = \frac{u^\mu}{\mathbf{z}}
\end{equation}
With these identifications the relations between the membrane and boundary entropy currents is given by 
\begin{equation}\label{mbec}
\mathbb{J}_S^\mu= \frac{\mathbf{J}_S^\mu}{\mathbf{z}^{d}} 
\end{equation}

We can get the explicit value of the total linearized metric outside the membrane using explicit value of the Stress tensor \eqref{eb0}. 
Setting $v^\mu=(1,0, 0 ...0)$ we find 
  \begin{equation}\label{mme}
  h_{tt} = \left(\frac{d-1}{d}\right)\left(\frac{4\pi T_m}{d}\right)^d z^{d-2}, \quad h_{ij} = \left(\frac{1}{d}\right)\left(\frac{4\pi T_m}{d}\right)^dz^{d-2}~\delta_{ij}
  \end{equation}
  Note that \eqref{mme} matches exactly with the expansion at linear order in $\frac{z^d}{\textbf{z}^d}$ 
  of the Fefferman Graham form of exact black brane metric i.e. \eqref{bbe}. It follows immediately that the boundary stress tensor 
  dual to our flat membrane configuration exactly matches the boundary stress tensor of a black brane once we use the identifications 
  \eqref{rufd}.

\subsection{The boundary stress tensor in the derivative expansion}

In the previous subsection we computed the linearized metric fluctuation (and thereby the boundary stress tensor) sourced by a membrane 
like stress tensor \eqref{bmst} that was localized at a constant value of $z$ ($z={\bf z}$) and was also uniform in space. Our final result was 
presented in \eqref{brm1}. In this subsection will generalize the computation of the previous subsection in the following manner. 
We compute the linearized metric fluctuation sourced by the stress tensor
\begin{equation}\label{bmstm}
\begin{split} 
{\cal T}^{MN}& = z \sqrt{1+ \partial_\mu {\bf z} \partial^\mu{\bf z}} ~\delta(z-{\mathbf z (x)})~ T^{MN}\\
T^{z z}&= \partial_\mu {\bf z}~ \partial_\nu {\bf z}~ {\bf T}^{\mu\nu}, ~~~
T^{\nu z}= \partial_\mu {\bf z}~ {\bf T}^{\mu \nu},~~~T^{M=\mu, N=\nu} = {\bf T}^{\mu\nu}
\end{split}
\end{equation}
where ${\bf z}$ and ${\bf T}^{\mu\nu}$ are no longer constants but are slowly varying functions of $x^\mu$. \footnote{In the equation 
\eqref{bmstm} $\mu$ indices have been raised using the induced metric on membrane ${\bf g}^{\mu\nu}$. }. In what follows we view ${\bf T}^{\mu\nu}$ as
a tensor valued field on the membrane world volume, and use the induced metric on the membrane to raise and lower its indices. 

We will take advantage of the slowly varying nature 
of the functions ${\bf z}$ and ${\bf T}_{\mu\nu}$ to perform our computation to first nontrivial order (which turns out to be the 
second order) in an expansion of the derivatives of these fields.

At leading (zero) order in the derivative expansion the metric fluctuation sourced by \eqref{bmstm} is simply given by the local 
form of \eqref{brm1}, i.e.
\begin{equation}\label{metform}
 \mathfrak{G}_{MN}dX^M dX^N =\frac{dz^2 + \eta_{\mu\nu} dx^\mu dx^\nu}{z^2}+h_{\mu\nu} dx^\mu dx^\nu
\end{equation}
where 
\begin{equation}\label{brm2}
h_{\mu\nu} =
   \begin{cases}
     A_{\mu\nu}(x) \textbf{z}(x)^{d} z^{-2}, & \text{for}~~ z \ge \textbf{z}(x) \\
     A_{\mu\nu}(x) z^{d-2}, & \text{for}~~ z \le \textbf{z}(x)
   \end{cases}
\end{equation} 
with
\begin{equation}\label{meten2}
 A_{\mu\nu}(x) = \frac{16\pi}{d} \frac{\textbf{T}_{\mu\nu}(x)}{\textbf{z}(x)^{d-2}}
 \end{equation}

Of course the metric \eqref{brm2} does not exactly solve Einstein's equations linearized about $AdS$ space; 
when we plug \eqref{brm2} into Einstein's equations \eqref{eeq} with $\lambda=1$, the LHS of these equations evaluates 
to an expression that is not zero, but turns out to be of second order in `field theory' (i.e. $x^\mu$) derivatives. 
In order to find a solution to the linearized Einstein equations valid to second order in derivatives we replace $h_{\mu\nu}$ in  \eqref{brm2} by  
\begin{equation}\label{brm3}
h_{\mu\nu} =
   \begin{cases}
     A_{\mu\nu}(x) \textbf{z}(x)^{d} z^{-2} +\delta h_{\mu\nu}(z,x) , & \text{for}~~ z \ge \textbf{z}(x) \\
     A_{\mu\nu}(x) z^{d-2} +\delta h_{\mu\nu}(z,x) , & \text{for}~~ z \le \textbf{z}(x)
   \end{cases}
\end{equation} 
where $\delta h_{\mu\nu}$ is an as yet unknown correction. We then plug \eqref{brm3} into 
the Einstein equation \eqref{eeq}. We assume that $\delta h_{\mu\nu}$ is of second order 
in derivatives, and work consistently to this order (i.e. we ignore all terms in the equation that are of third or higher order). The LHS of \eqref{eeq} now has terms of two sorts. First we have the `source' terms, independent of $\delta h_{\mu\nu}$ that we 
have already encountered earlier in this paragraph. In addition we have new terms proportional to $\delta h_{\mu\nu}$. Setting the sum of these terms to zero in the 
dynamical Einstein equations (dynamical w.r.t. evolution in $z$) yields an equations of 
the schematic form 
\begin{equation}\label{sfeq}
H \delta h_{\mu\nu}= s_{\mu\nu}
\end{equation}
where $s_{\mu\nu}$ are source terms and $H$ is a differential operator of second order 
in $z$ derivatives. Note that the differential operator has no derivatives in the 
$x^\mu$ directions - $x^\mu$ derivatives on $\delta h_{\mu\nu}$ result in expressions that 
are of third or higher order in derivatives and so are ignored at the order at which 
we work. 

In order to obtain a unique solution to the equations \eqref{sfeq} we impose the following
boundary conditions. First we demand that the `outside' solution is normalizable. Second 
we demand that the `inside' solution does not blow up at $z=\infty$. Third we require 
that $\delta h_{\mu\nu}$ is continuous across the membrane located at $z={\bf z}(x)$. 
Fourth we require the solution to obey the appropriate junction matching condition 
across the membrane (see below). These four conditions allow us to determine the four 
integration constants (two for the outside solution and two for the inside solution)
that appear in the most general solution of \eqref{sfeq} and thereby obtain a 
unique solution for $\delta h_{\mu\nu}$. The algebra involved in our work out is 
straightforward and we simply present our final results. 
\begin{equation}\label{hmnd2}
 \delta  h_{\mu\nu}(z,x) =
   \begin{cases}
    C^{(in)}_{\mu\nu}  z^{-2} + B^{(in)}_{\mu\nu}, & \text{for}~~ z \ge \textbf{z}(x) \\
     C^{(out)}_{\mu\nu} z^{d-2} + B^{(out)}_{\mu\nu} z^{d}, & \text{for}~~ z \le \textbf{z}(x)
   \end{cases}
\end{equation}
where
\begin{equation}\label{bva}
\begin{split}
B^{(out)}_{\mu\nu} &= -\frac{\partial^2A^{(out)}_{\mu\nu}}{2(d+2)} \\
B^{(in)}_{\mu\nu}  &= -\frac{1}{2(d-2)}\left( \partial_\nu\partial^\alpha A^{(in)}_{\alpha\mu} + \partial_\mu\partial^\alpha A^{(in)}_{\alpha\nu}-\partial^2A^{(in)}_{\mu\nu} - \frac{\partial^\alpha \partial^\beta A^{(in)}_{\alpha\beta}}{d-1}\eta_{\mu\nu} \right) \\
C^{(out)}_{\mu\nu}  &= \frac{\textbf{z}^2}{2d}\partial^2A^{(out)}_{\mu\nu} - \frac{1}{d(d-2)}\left( \partial_\nu \partial^\alpha A^{(in)}_{\alpha\mu} + \partial_\mu \partial^\alpha A^{(in)}_{\alpha\nu} - \partial^2A^{(in)}_{\mu\nu}  \right) \\& -\frac{\partial^\alpha \textbf{z} \partial_\alpha \textbf{z}}{2} A^{(out)}_{\mu\nu} + \left(\frac{\partial^\alpha\partial^\beta A^{(in)}_{\alpha\beta}}{d(d-2)\textbf{z}^{d-2}}-\partial_\alpha \textbf{z} \partial_\beta \textbf{z}~ A^{(out)}_{\alpha\beta}\right)\eta_{\mu\nu} \\
C^{(in)}_{\mu\nu} &= C^{(out)}_{\mu\nu}\textbf{z}^d + B^{(out)}_{\mu\nu} \textbf{z}^{d+2} - \textbf{z}^2 B^{(in)}_{\mu\nu} \\
A^{(in)}_{\mu\nu} &= \textbf{z}^d A^{(out)}_{\mu\nu} \equiv \textbf{z}^d A_{\mu\nu}
\end{split}
\end{equation}
In order to obtain the results listed above we have used the fact that the extrinsic 
curvature of the slice $z={\bf z}(x)$ is given up to linear order in $h_{\mu\nu}$ and second order in field theory derivatives by 
\begin{equation}\label{exf}
\begin{split}
K_{\mu\nu} &= \bigg[ \frac{\eta_{\mu\nu}}{z^2} - \frac{z}{2}\partial_z h_{\mu\nu} + \frac{\partial_\mu \partial_\nu \textbf{z}}{z} + \frac{\partial_\mu \textbf{z}\partial_\nu \textbf{z}}{z^2} + \frac{z}{4}\partial_z h_{\mu\nu}\partial^\alpha \textbf{z} \partial_\alpha \textbf{z} \\ &- \frac{z}{2}\partial^\alpha \textbf{z} \left( \partial_\mu h_{\alpha\nu} + \partial_\nu h_{\alpha\mu} - \partial_\alpha h_{\mu\nu} \right)+\frac{1}{2z^2}\left( -\partial^\alpha \textbf{z}\partial_\alpha \textbf{z} + \textbf{z}^2 h^{\alpha\beta}\partial_\alpha{\bf z}\partial_\beta \textbf{z} \right)\eta_{\mu\nu}
\\& -\left(\frac{z}{2}\partial_z +1\right) \left( h_{\mu\alpha}\partial^\alpha \textbf{z} \partial_\nu \textbf{z} + h_{\nu\alpha}\partial^\alpha \textbf{z} \partial_\mu \textbf{z}\right) \bigg]_{z\rightarrow\textbf{z}}
\end{split}
\end{equation} 
Note that the expression \eqref{exf} depends on $\partial_z h_{\mu\nu}$. As this quantity
jumps across the membrane, the extrinsic curvature `above' the membrane is discontinuously
different from the same quantity `below' the membrane. The difference between these 
two quantities is governed by the `junction condition' mentioned above
 \begin{equation}\label{byda}
 \textbf{T}_{\mu\nu} = -\frac{1}{8\pi} \left( \left[K^{(out)}_{\mu\nu}-K^{(in)}_{\mu\nu}\right] - \left[{\mathcal K}^{(out)}-{\mathcal K}^{(in)}\right] \bf{g}_{\mu\nu} \right)
 \end{equation}
It is not difficult to evaluate the boundary stress tensor dual to the solution 
presented above using the definition \eqref{pbs}; we find 
\begin{equation}\label{abst}
\begin{split}
{\mathbb T}_{\mu\nu} &= \frac{{\mathbf T}_{\mu\nu}}{{\mathbf z}^{d-2}} + \frac{{\mathbf z}^2}{2d}~ \partial^2 \left(\frac{{\mathbf T}_{\mu\nu}}{{\mathbf z}^{d-2}}\right)  - \frac{\partial^\alpha {\mathbf z} \partial_\alpha {\mathbf z}}{2~{\mathbf z}^{d-2}} {\mathbf T}_{\mu\nu} +\frac{\partial^\alpha \partial^\beta\left( {\mathbf z}^2 {\mathbf T}_{\alpha\beta} \right) }{d(d-2){\mathbf z}^{d-2}}\eta_{\mu\nu} - \frac{\partial^\alpha{\mathbf z} \partial^\beta{\mathbf z}{\mathbf T}_{\alpha\beta}}{{\mathbf z}^{d-2}}\eta_{\mu\nu} \\& - \frac{1}{d(d-2){\mathbf z}^{d-2}}\left[ \partial_\nu \partial^\alpha \left({\mathbf z}^2 {\mathbf T}_{\alpha\mu}\right) + \partial_\mu \partial^\alpha \left({\mathbf z}^2 {\mathbf T}_{\alpha\nu}\right) - \partial^2 \left({\mathbf z}^2 {\mathbf T}_{\mu\nu}\right) \right]
\end{split}
\end{equation}

The results \eqref{hmnd2}, \eqref{bva} were obtained by solving the 
dynamical Einstein equations. The Einstein constraint equations (for evolution along the $z$ direction) remain to be solved. 
The situation with these equations is closely analogous to that encountered in section 4.3 of \cite{Bhattacharyya:2016nhn} 
in a distinct but related context.

 Let us first recall the following general property of Einstein's equations: provided the dynamical equations are solved everywhere, the constraint equations are automatically solved everywhere if they are solved on a single slice. As we have already dealt with the dynamical equations, it remains only to solve the 
constraint equations on any one slice on the outside and on any other slice on the `inside'. It is convenient 
to choose these slices to be the membrane world volume, approached either from the outside or from the inside. 

Let us recall that the constraint equations are of two sorts; the momentum constraint equations and the `Hamiltonian' constraint
equations. Let us first deal with the momentum constraint equations. These equations are simply the statement that the 
Brown York tensor of the full metric (background plus fluctuation) is conserved on our slice. 
Now as in section 4.3 of \cite{Bhattacharyya:2016nhn}, it turns out that this condition is automatic for the inside solution
(this is suggested by the general argument of section 4.3 of \cite{Bhattacharyya:2016nhn} and we have explicitly algebraically 
verified that it is the case for the explicit solution presented above). On the other hand the Brown York stress tensor
is not identically conserved just outside the membrane. However it follows from \eqref{byda} that the difference between 
the conservation of the BY tensor outside and the BY tensor inside the membrane is simply the condition that the membrane 
stress tensor is conserved on its world volume. As the inside BY tensor is identically conserved, it follows that the 
outside BY tensor is conserved - and hence the outside Einstein constraint equation obeyed - if and only if the membrane 
stress tensor is conserved on its world volume. 

We have already mentioned above, once the membrane stress tensor is conserved on the membrane world volume, this 
automatically ensures that the momentum constraint equations are solved everywhere. The momentum constraint equations are particularly interesting when evaluated on the boundary of $AdS$, where they assert the conservation of the boundary 
stress tensor \eqref{abst}. It follows, in other words, that conservation of the membrane stress tensor and the 
boundary stress tensor must be algebraically equivalent statements: one must imply the other. It is easy to directly 
verify that this is the case. In particular we have algebraically verified, using \eqref{abst}, that \eqref{sttr} is 
algebraically equivalent to the condition 
\begin{equation} \label{algidenteo}
\nabla^\mu {\mathbb T}_{\mu\nu} =0
\end{equation}
(where $\nabla^\mu$ in \eqref{algidenteo} is the boundary field theory covariant derivative -i.e. the raised partial derivative 
in flat space. \footnote{Note that, as in discussions of the fluid gravity correspondence, the equation \eqref{algidenteo} has 
an explicit derivative. It follows that the constraint equation \eqref{algidenteo} at $(n+1)^{th}$ order is completely determined
by knowledge of the stress tensor at $n^{th}$ order.}

In a similar manner, the Hamilton constraint equations are automatically (identically) obeyed for the inside solution. 
The condition that they are also obeyed on the outside solution follows provided that \eqref{introkt} holds 
(see around 4.25 of \cite{Bhattacharyya:2016nhn} for a proof).
At the boundary of $AdS$, on the other hand, this constraint equation simply reduces to the condition that the 
boundary stress tensor is traceless. It follows, in other words, that the tracelessness of the boundary stress tensor
\begin{equation} \label{algident}
{\mathbb T}_{\mu\nu} \eta^{\mu\nu}=0
\end{equation}
must be algebraically identical to the condition \eqref{introkt} for the membrane stress tensor. Using the explicit
result \eqref{abst} we have directly verified that this is the case.

In summary, the solution \eqref{hmnd2}, \eqref{bva} solves all Einstein momentum 
constraint equations in addition to the Einstein dynamical equations if and only if the membrane stress tensor is 
conserved on its world volume and also obeys the equation \eqref{introkt}. The resultant boundary stress tensor 
\eqref{abst} is then automatically conserved and traceless.

\subsection{Boundary stress tensor in terms of fluid variables}

Plugging the explicit form of the membrane stress tensor, \eqref{stf} into the general formula \eqref{abst}, we find that the boundary stress tensor dual to our membrane -accurate to second order in derivatives - is given by 
 \begin{equation}\label{sobs}
 \begin{split}
  \mathbb{T}_{\mu\nu} &=  t^{(0)}_{\mu\nu} + t^{(1)}_{\mu\nu} +t^{(2)}_{\mu\nu} \\
 t^{(0)}_{\mu\nu} &=\frac{1}{\textbf{z}^d}\left(\eta_{\mu\nu} + d~v_\mu v_\nu\right) \\
 t^{(1)}_{\mu\nu} &= - \frac{2}{\textbf{z}^{d-1}} \sigma_{\mu\nu} \\
 t^{(2)}_{\mu\nu} &=   \frac{1}{\textbf{z}^{d-2}}\Bigg[ \left[-\left(\frac{d}{2}\right) \frac{\partial^\alpha\textbf{z}\partial_\alpha\textbf{z}}{\textbf{z}^2} + \left(\frac{d-2}{d-1}\right)\frac{\partial^\alpha\partial_\alpha \textbf{z}}{\textbf{z}} -\left(\frac{d}{d-1}\right)\frac{v^\alpha v^\beta\partial_\alpha\partial_\beta\textbf{z}}{\textbf{z}}\right]P_{\mu\nu} \\& + \left[\left(\frac{d-1}{2}\right)\frac{\partial^\alpha\textbf{z}\partial_\alpha\textbf{z}}{\textbf{z}^2} - \frac{\partial^\alpha\partial_\alpha \textbf{z}}{\textbf{z}} \right]\eta_{\mu\nu} + \left[ \frac{\partial_\mu\textbf{z}\partial_\nu\textbf{z}}{\textbf{z}^2} + \frac{\partial_\mu\partial_\nu\textbf{z}}{\textbf{z}} \right] - d \frac{\partial^\alpha\textbf{z}}{{\bf z}}\left( v_\mu \partial_\alpha v_\nu + v_\nu \partial_\alpha v_\mu \right) \\& +\frac{1}{2}\left(\frac{d}{d-2}\right)\left(v_\mu\partial^2 v_\nu+v_\nu\partial^2 v_\mu + 2 \partial^\alpha v_\mu \partial_\alpha v_\nu \right) - \frac{1}{d-2}\big[ \left(v_\mu \partial_\nu + v_\nu \partial_\mu \right)(\partial.v) \\&+ (\partial.v)\left(\partial_\mu v_\nu+\partial_\nu v_\mu\right) + \left(\partial_\mu v^\alpha \partial_\alpha v_\nu + \partial_\nu v^\alpha \partial_\alpha v_\mu\right) + v.\partial\left( \partial_\mu v_\nu + \partial_\nu v_\mu \right) \big] \\&+ \left[d\frac{\partial^\alpha \textbf{z}\partial_\alpha \textbf{z}}{2\textbf{z}^2}-\frac{\partial^\alpha \partial_\alpha \textbf{z}}{2\textbf{z}}\right]\left(\eta_{\mu\nu} + d v_\mu v_\nu\right)-\left[ \frac{\partial^\alpha\textbf{z}\partial_\alpha\textbf{z}}{\textbf{z}^2} + d\frac{(v.\partial \textbf{z})^2}{\textbf{z}^2} \right]\eta_{\mu\nu} \\&+ \frac{1}{d-2}\left[ (\partial.v)^2 + 2 v.\partial(\partial.v) + \partial^\alpha v^\beta \partial_\beta v_\alpha \right]\eta_{\mu\nu} - d \frac{\left(v.\partial {\bf z}\right)^2}{{\bf z}^2}v_\mu v_\nu \Bigg]
 \end{split}
 \end{equation} 
 where,
 \begin{equation}
 v^\mu = \frac{u^\mu}{\mathbf{z}},\quad P_{\mu\nu} = \eta_{\mu\nu} + v_\mu v_\nu ,\quad \sigma_{\mu\nu} = \left(\frac{\partial_\alpha v_\beta+\partial_\beta v_\alpha }{2}\right)P^\alpha_\mu P^\beta_\nu-\left(\frac{\partial.v}{d-1}\right)P_{\mu\nu}
 \end{equation}
 
 The expression for $t^{(2)}_{\mu\nu}$ above can be simplified by recalling that we are 
 interested only in onshell configurations of our boundary fluid. At zero order in derivatives, the conservation of $t^{(0)}_{\mu\nu}$ yields
  \begin{equation}\label{foc}
 \partial^\mu t^{(0)}_{\mu\nu} = 0, \quad \partial_\alpha \partial^\mu t^{(0)}_{\mu\nu} = 0
 \end{equation} 
From \eqref{foc} we get
\begin{equation}\label{dcon}
\begin{split}
\frac{\partial_\mu\textbf{z}}{\textbf{z}} &= v.\partial v_\mu - \frac{\partial.v}{d-1}v_\mu
\\ \frac{\partial_\mu \partial_\nu\textbf{z}}{\textbf{z}} &= \left(v.\partial v_\mu - \frac{\partial.v}{d-1}v_\mu\right)\left(v.\partial v_\nu - \frac{\partial.v}{d-1}v_\nu\right) - \frac{1}{d-1}\partial_\mu(\partial.v)v_\nu \\&- \frac{\partial.v}{d-1}\partial_\mu v_\nu + \partial_\mu v^\lambda \partial_\lambda v_\nu + v.\partial (\partial_\mu v_\nu)
\end{split}
\end{equation}
Of course the object that is really conserved is the full stress tensor rather than simply $t^{(0)}_{\mu\nu}$.  This means that the RHS of \eqref{dcon} has corrections that are of higher order in derivatives. We  will now use the equations \eqref{dcon} to 
simplify $t^{(2)}_{\mu\nu}$; the corrections to \eqref{dcon} yield terms of third or higher 
order in derivatives and so can be ignored. We thus proceed to simplify 
\eqref{sobs} by using \eqref{dcon} to replace occurrence of a term in  $t^{(2)}_{\mu\nu}$ involving derivatives of ${\bf z}$ with the expressions on the 
RHS of \eqref{dcon}. The resultant expression for $t^{(2)}_{\mu\nu}$ is a sum of two derivative terms with all derivatives acting on the velocity field $v_\mu$. 
The final expression for the resulting expression is somewhat cumbersome and we do not explicitly list it here. 

We will now perform a field redefinition from the natural membrane variables 
$v_\mu$ and ${\bf z}$ to more natural - and more standard - boundary variables. 
Let us define the  Landau Frame boundary velocity field ${\mathsf v}_\mu$ and
and the boundary temperature  ${\mathsf T}$ by the conditions  
\begin{equation}\label{vtdef}
{\mathbb{T}^\mu}_\nu {\mathsf v}^\nu = -(d-1)\left(\frac{4\pi \mathsf{T}}{d}\right)^{d}~ {\mathsf v}^\mu
\end{equation} 
In other words ${\mathsf v}^\mu$ is the unique timelike eigenvector of the boundary 
stress tensor (normalized to be a boundary velocity field) and  $\mathsf{T}$ is simply defined in terms of its eigenvalue.  
It is not difficult to solve for ${\mathsf v}_\mu$ and $\mathsf{T}$ in terms of $v_\mu$ 
and ${\bf z}$, order by order in the derivative expansion. At zero order in derivatives 
we work with the simple stress tensor $t^{(0)}_{\mu\nu}$ ; it is easily verified that
\begin{equation}\label{zoi}
 {\mathsf T} = T= \frac{d}{4 \pi~ \mathbf{z}}, ~~~{\mathsf v}^\mu = v^\mu =\frac{u^\mu}{\mathbf{z}}
\end{equation}
Note that, at this order, \eqref{zoi} agrees with \eqref{rufd} as we might have anticipated on general grounds.

The relation 
$$t^{(1)}_{\mu\nu} v^\mu=0$$
immediately implies that the solution \eqref{zoi} continues to hold at first order 
in derivatives. The situation is more complicated at second order. At this order 
\eqref{zoi} is corrected to 
\begin{equation}\label{rdf}
 {\mathsf T} =  T(1+\delta T), \quad {\mathsf v}_\mu = v_\mu + \delta v_\mu
\end{equation} 
where,
\begin{equation}\label{tred}
\begin{split}
\delta T &= \frac{1}{d(d-1)}\left(\frac{d}{4\pi T}\right)^2\Bigg[ -\frac{1}{2}\left(\frac{d^2-7d+8}{d-2}\right)\sigma_{\alpha\beta}\sigma^{\alpha\beta} + \frac{1}{2}\left(\frac{d^2-3d+8}{d-2}\right)\omega_{\alpha\beta}\omega^{\alpha\beta} \\& -\frac{(d-4)}{2}v.\partial(\partial.v) + \frac{(d-1)(d-2)}{2}v.\partial v_\lambda v.\partial v^\lambda - \frac{(d-1)(d-2)}{2}\left(\frac{\partial.v}{d-1}\right)^2\Bigg]
\end{split}
\end{equation}
and
\begin{equation}\label{vred}
\begin{split}
\delta v_\mu &= \frac{P^\lambda_\mu}{d}\left(\frac{d}{4\pi T}\right)^2 \Bigg[ -\frac{1}{2}\left(\frac{2d^2-5d+4}{(d-1)(d-2)}\right)(\partial.v)v.\partial v_\lambda + \frac{1}{2}\left(\frac{3d-4}{(d-1)(d-2)}\right)\partial_\lambda (\partial.v) \\& +
\frac{(d-4)}{2(d-2)}v.\partial(v.\partial v_\lambda)-\frac{d}{2(d-2)}\partial^2 v_\lambda + (d)~ v.\partial v^\alpha \partial_\alpha v_\lambda -\frac{(d-4)}{2(d-2)}\partial_\lambda v_\alpha v.\partial v^\alpha \Bigg]  
\end{split}
\end{equation}
Plugging \eqref{tred} and \eqref{vred} into \eqref{sobs} we obtain our final expression 
for the boundary stress tensor expressed in terms of boundary Landau frame temperature 
and velocity fields
\begin{equation}\label{bst}
\begin{split}
\mathbb{T}_{\mu\nu} &= p \left(\eta_{\mu\nu}+d ~{\mathsf v}_\mu {\mathsf v}_\nu\right)-2\eta {\mathsf \sigma}_{\mu\nu}\\& +2\eta \left(\frac{d}{4\pi {\mathsf T}}\right)\Bigg[ \left( \sigma_\mu^\lambda\sigma_{\lambda\nu} - \frac{\sigma_{\alpha\beta}\sigma^{\alpha\beta}}{d-1}P_{\mu\nu} \right) - \frac{2}{d-2}\left( \omega_\mu^\lambda\omega_{\lambda\nu} + \frac{\omega_{\alpha\beta}\omega^{\alpha\beta}}{d-1}P_{\mu\nu} \right) \\& - \frac{1}{2}\left(\frac{d}{d-2}\right)\left(\omega_\mu^\lambda\sigma_{\lambda\nu}+\omega_\nu^\lambda\sigma_{\lambda\mu}\right)+\frac{1}{2}\left(\frac{d-4}{d-2}\right)\left({\mathsf v}.D \sigma_{\mu\nu}\right) \Bigg]
\end{split}
\end{equation}
Where we have 
\begin{equation}\label{qbst}
\begin{split}
 \quad p = \frac{1}{16\pi} \left(\frac{4\pi \mathsf{T}}{d}\right)^d, \quad \eta = \frac{1}{16\pi} \left(\frac{4\pi \mathsf{T}}{d}\right)^{d-1}, \quad {\mathsf v}.D\sigma_{\mu\nu} = P^\alpha_\mu P^\beta_\nu~{\mathsf v}.\partial\sigma_{\alpha\beta} + \frac{\partial.{\mathsf v}}{d-1}\sigma_{\mu\nu}
\end{split}
\end{equation}
and the quantities $\sigma_{\mu\nu},~ \omega_{\mu\nu},~ P_{\mu\nu}$ are constructed from ${\mathsf v}$. As a nontrivial check of the algebra leading up to \eqref{bst} we 
note that the stress tensor \eqref{bst} is Weyl covariant (see \cite{Bhattacharyya:2008mz,Haack:2008cp}). 

Let us now compare the second order hydrodynamical stress tensor \eqref{bst} with 
the corresponding object obtained from the fluid gravity map listed in \cite{Bhattacharyya:2008jc,Bhattacharyya:2008mz,Haack:2008cp}. In the current paper 
we have worked with a flat boundary metric, and so should set the boundary Weyl tensor 
in the fluid gravity papers listed above to zero. In this case the results of \cite{Bhattacharyya:2008jc,Bhattacharyya:2008mz,Haack:2008cp} are 
\begin{equation}\label{fgst}
\begin{split}
\mathbb{T}^{(fg)}_{\mu\nu} &= p \left(\eta_{\mu\nu}+d ~{\mathsf v}_\mu {\mathsf v}_\nu\right)-2\eta {\mathsf \sigma}_{\mu\nu}\\& - 2 \eta\tau_{\omega}\left[ {\mathsf v}.D \sigma_{\mu\nu} +\omega_\mu^\lambda\sigma_{\lambda\nu}+\omega_\nu^\lambda\sigma_{\lambda\mu} \right] 
+2 \eta \left(\frac{d}{4\pi {\mathsf T}}\right) \left[ {\mathsf v}.D \sigma_{\mu\nu} +\sigma_\mu^\lambda\sigma_{\lambda\nu} - \frac{\sigma_{\alpha\beta}\sigma^{\alpha\beta}}{d-1}P_{\mu\nu} \right] \\& \text{where,}~~ \tau_{\omega} = 
\left(\frac{d}{4\pi {\mathsf T}}\right) \int_{1}^{\infty} \frac{y^{d-2}-1}{y(y^d-1)} dy = 
\left(\frac{d}{4\pi {\mathsf T}}\right) \left( \frac{1}{2}-\frac{\pi^2}{3d^2}+ {\cal O}(\frac{1}{d^3}) \right) 
\end{split}
\end{equation}
The quantities $p$ and $\eta$ in \eqref{fgst} were listed in \eqref{qbst}. 

Clearly \eqref{bst} agrees exactly (at finite $d$) with \eqref{fgst} at zero and 
first order in the derivative expansion. At second order the two stress tensors 
have the same tensor structures. The coefficients of individual tensor structures 
match perfectly at leading order in the large $d$ limit, but deviate from each other 
at subleading orders in this expansion. 

The `flow' from membrane hydrodynamics to boundary hydrodynamics derived in this section has 
some similarities with the analysis of \cite{Brattan:2011my}. It might be interesting to explore this connection in greater 
detail in the future.

\subsection{Quasinormal modes from membrane stress tensor about uniform planar membrane in AdS}
In the previous subsection we demonstrated that the nonlinear equations that govern the motion of a membrane in planar AdS space reduce, 
in the derivative expansion, to the equations of boundary hydrodynamics. The boundary stress tensor is given in terms of the local boundary 
fluid velocity and temperature by a constitutive relation that agrees on the dot with the finite $D$ fluid gravity constitutive relation 
at first order in the derivative expansion, but deviates (at finite $D$) from fluid gravity at second and higher orders in this expansion. 

In this subsection we will explore related physics by performing a related but distinct computation - we use the membrane equations to compute 
the spectrum of small fluctuations about an exactly planar membrane in Poincare patch $AdS$ space, and compare our results with the spectrum of 
quasinormal modes about the dual black brane in $AdS$ space. Once again we find that the spectrum computed using our membrane equations 
perfectly reproduces black brane quasinormal mode spectrum to leading and first subleading order in $k$, but reproduces higher order corrections 
only in the large $D$ limit. 

We consider background spacetime $AdS_D$ with $\lambda=\frac{1}{L^2}=1$
\begin{equation}
ds^2 = -r^2 dt^2 + \frac{dr^2}{r^2} + r^2 (dx^a dx_a)
\end{equation}
Let the planar membrane be located at $r=r_0$. For convenience we choose $r_0=1$; it is easy to reinstate factors of $r_0$ in the final 
answer. In this section we closely follow the method used in \cite{Bhattacharyya:2017hpj}; we refer the reader interested in details to that 
paper and report only key results. 

Consider the membrane configuration
 \begin{equation}\label{pfl}
  \begin{split}
   r &= 1 + \epsilon \delta r(t,a) \\
   u &= -(1 + \epsilon \delta r) dt + \epsilon \delta u_a(t,a) dx^{a} 
  \end{split}
 \end{equation}
(the $\delta r$ dependence in the velocity fluctuation is dictated by the requirement that $u^2=-1$). 
The induced metric on membrane is
 \begin{equation}\label{pim}
  ds^2 = g_{\mu\nu} dx^\mu dx^\nu = -(1 + 2 \epsilon \delta r) dt^2 + (1 + 2 \epsilon \delta r) (dx^adx_a)
 \end{equation}
The projector orthogonal to the fluid velocity is easily evaluated; we find 
\begin{equation}
  \mathcal{P}^a_b = \delta^a_b, \quad \mathcal{P}^t_t = 0, \quad \mathcal{P}^t_a = \epsilon \delta u_a, \quad  \mathcal{P}^a_t = -\epsilon \delta u^a
\end{equation}
We have the membrane equation
\begin{equation}\label{meq} \begin{split}
& \nabla.u=0\\
&16 \pi~ {\cal P}^\nu_\alpha\nabla^\mu T_{\mu\nu} = \left( \tilde {\mathcal K}~u.\nabla u_\nu + \nabla_\nu \tilde {\mathcal K} -2 \nabla^\mu \sigma_{\mu\nu}\right) {\cal P}^\nu_\alpha \equiv E_\nu {\cal P}^\nu_\alpha
\end{split}
\end{equation}
To linear order in fluctuations we find 
\begin{equation} \begin{split}
&\sigma_{tt} = 0, \quad \sigma_{ta} = 0,\quad \sigma_{ab} = \epsilon \frac{\partial_a \delta u_b+\partial_b \delta u_a}{2} + \epsilon \partial_t \delta r \delta_{ab}\\
&\tilde {\mathcal K} = (D-1) +2 \epsilon\left( \partial_t^2\delta r -\partial^2\delta r \right) + \left(\frac{D-1}{D-2}\right) \epsilon\partial^2 \delta r \\
&u. \nabla u_t=0, \quad  u. \nabla u_a = \epsilon \partial_t \delta u_a + \epsilon \partial_a \delta r   \\
\end{split}
\end{equation}
Using these results the membrane equations  \eqref{meq} simplify to  
\begin{equation}\label{linpm}
\begin{split}
&\partial_a \delta u^a = -(D-2) \partial_t \delta r\\
&V_a \equiv (D-1) (\partial_t \delta u_a+\partial_a \delta r) + 2 (\partial_a \partial_t^2 -\partial_a \partial^2 \delta r) + \left(\frac{D-1}{D-2}\right) \partial_a \partial^2 \delta r \\&- \left( \partial^2 \delta u_a -(D-2) \partial_a \partial_t \delta r \right) - 2 \partial_a \partial_t \delta r = 0  
\end{split}
\end{equation}
Inserting the plane wave expansion 
\begin{equation} \label{pwe} 
 \delta r = a e^{i k.x -i \omega t}, ~~~\delta u_a = b_a e^{i k.x -i \omega t}
\end{equation}
into \eqref{linpm}, we find that our equations have solutions if and only if $\omega$ obeys either the sound wave 
dispersion relation (recall $d=D-1$ and $k=\sqrt{k.k}$)
\begin{equation}\label{sf}
\omega^s = \pm \left(\frac{k}{\sqrt{d-1}}\right)\left[\frac{\sqrt{d^2(d-1)^2+4(d-1)^2 k^2+2(d-2)k^4}}{d(d-1)+2k^2}\right] - i\left[\frac{(d-2)k^2}{d(d-1)+2k^2}\right]
\end{equation}
or the shear wave dispersion relation
\begin{equation}\label{vf}
\omega^v = -i\frac{k^2}{d}
\end{equation}
Note, in particular, that \eqref{vf} takes an incredibly simple purely imaginary form. 

In order to compare with the spectrum of quasinormal modes about black branes, we expand these results in power series in $k$. We get
\begin{equation}\label{fgc}
\omega^v = -i\frac{k^2}{d} + {\cal O}(k^3), \quad \omega^s = \pm \left(\frac{k}{\sqrt{d-1}}\right) - i\left[\frac{(d-2)k^2}{d(d-1)}\right] + {\cal O}(k^3)
\end{equation}
The results \eqref{fgc} exactly (i.e. at arbitrary values of $D$ and not merely at large $D$) 
match the spectrum of the lightest quasinormal modes expanded around a black brane to the respective orders 
reported in the derivative expansion \cite{Emparan:2015rva} (see equation (6.1) and (6.2) in that paper); as might 
have been anticipated from the fact that our membrane exactly reproduces the fluid gravity stress tensor at zero 
and first order in derivatives even at finite $D$ (see above). It is also, however, easily verified 
that \eqref{sf} and \eqref{vf} do not match the exact finite $D$ gravitational results at higher orders in $k$.
(however the match persists in the large $D$ limit). This could also have been anticipated from the fact that our membrane accurately reproduces the second order 
terms in the hydrodynamical stress tensor only at large $D$ (see above).

Note that the  paper \cite{Emparan:2015rva} directly computed the black brane quasi normal modes within gravity in an expansion in large $D$. 
They obtained results very similar to our \eqref{fgc}; however the effective coefficient functions of the various terms in \eqref{fgc} 
were obtained in \cite{Emparan:2015rva} order by order in an expansion in $\frac{1}{D}$ (upto a particular order see (4.23),(4.24),(4.25)
of that paper). In contrast our membrane 
equations reproduces the reported coefficients exactly. 

We find it very encouraging that the simple membrane equations of this paper reproduce 
some gravitational results exactly as a function of $D$. It appears that the simple membrane equations presented in this paper 
(whose form was dictated by physical consistency requirements) resum an infinite class of corrections of other approaches, and so do a particularly 
good job of reproducing gravitational results to higher accuracy than might have been reasonable to expect. 

\section{Discussion}

In this paper we have made four main points. 
\begin{itemize}
\item At least at leading order, 
it is possible to `improve' the large $D$ perturbative expansion of black hole physics presented in earlier work. The improved leading order equations are 
chosen so that they agree with earlier derived results at leading order in the large $D$ 
limit but also  define consistent probe membrane dynamics at finite $D$ . Even though our improved equations define 
consistent probe dynamics at finite $D$, they do not exactly reproduce black hole physics at finite $D$ in generic situations, even though
they appear to work surprisingly well in some equilibrium and near equilibrium configurations. 
\item The velocity field in stationary solutions of the improved membrane equations is always proportional to a killing vector of the background spacetime 
in which the membrane propagates. The membrane shape
in such configurations obeys a differential equation that follows from extremizing a simple action for the membrane shape. Onshell this action reduces to the 
thermodynamical membrane partition function. 
\item The thermodynamics of static spherical membranes in flat space and global AdS space, obtained via this procedure, agrees exactly with that of their dual 
black holes even at finite $D$. 
\item The motion of a membrane in Poincare Patch AdS space sources linearized 
gravitational fluctuations and so a boundary stress tensor. In the long wavelength
limit the resultant boundary stress tensor is a hydrodynamical stress tensor 
for a conformal boundary fluid. At zero and first order in the derivative expansion, 
this stress tensor exactly reproduces the results of the fluid gravity correspondence even at finite $D$. At second order in derivatives, the
fluid dual to improved probe membrane agrees with the 
second order fluid gravity stress tensor at large $D$, but deviates
from these exact results at finite $D$. 
\end{itemize}

Each of the points listed above throws up several interesting questions and directions for future research. 
One immediate question is whether the improvement of the leading large $D$ membrane equations, 
presented in this paper  can be systematically continued order by order, in large $D$ perturbation theory. 
More precisely the question is the following. Given any positive integer $n$, can we always (in principle) find 
an improved $n^{th}$ order membrane stress tensor with the following 
two properties. First, that the expansion of our improved stress tensor 
to $n^{th}$ order in  $\frac{1}{D}$ agrees with the `naive' $n^{th}$ order 
stress tensor obtained from the naive large $D$ expansion (i.e. by following the algorithm
presented in \cite{Dandekar:2016fvw, Bhattacharyya:2016nhn}). Second, that our improved 
$n^{th}$ order stress tensor autonomously defines consistent probe dynamics at finite $D$. 
\footnote{As a first  calculational check it would be 
useful to obtain explicit results for the improved large $D$ expansion at 
first subleading order in $\frac{1}{D}$. } We suspect that the answer to this 
question is in the affirmative, and that the techniques developed in 
\cite{Bhattacharyya:2013lha} and related subsequent work \cite{Bhattacharyya:2014bha,Haehl:2013hoa,Haehl:2014zda,Haehl:2015pja,Crossley:2015tka,Haehl:2015foa,Crossley:2015evo,Haehl:2015uoc,Haehl:2016pec,Haehl:2016uah,Glorioso:2016gsa,Jensen:2017kzi,Gao:2017bqf,Glorioso:2017fpd,Glorioso:2017lcn,Geracie:2017uku}
will prove useful in demonstrating this issue. In particular, a device adopted in several of the papers \cite{Haehl:2013hoa,Haehl:2014zda,Haehl:2015pja,Crossley:2015tka,Haehl:2015foa,Crossley:2015evo,Haehl:2015uoc,Haehl:2016pec,Haehl:2016uah,Glorioso:2016gsa,Jensen:2017kzi,Gao:2017bqf,Glorioso:2017fpd,Glorioso:2017lcn,Geracie:2017uku}
- namely the use of `diffeomorphisms' as the basic degrees of freedom to describe hydrodynamics - may have a very natural 
generalization to the context of this paper, as a single bulk diffeomorphisms (starting from a prescribed membrane world volume) 
could generate both the most general membrane shape as well as the most general membrane velocity field. 
We hope to return to these questions in the future. \footnote{We thank M. Rangamani for discussions on this topic.}

There are also several interesting open questions relating to the action that 
governs equilibrium membrane configurations. First, as we have explained in the 
main text, we suspect that the very simple general structure of this action 
- namely that it is given by the sum of a Gibbons Hawking term and the 
action for a stationary fluid on the membrane - persists to every 
order in the $\frac{1}{D}$ expansion. It would be useful to explicitly verify this 
expectation, atleast at first subleading order in $\frac{1}{D}$. Second, 
it is natural to wonder whether this structure of the action - that it is the sum of 
a Gibbons Hawking like term plus a fluid action - generalizes to the study 
of an arbitrary higher derivative diffeomorphically invariant theory of gravity. 
Finally, it may be  interesting to investigate whether there is a sense in which the offshell membrane action presented in this paper can be obtained from an offshell gravitational action for an appropriate dual set of configurations. 

In Appendix \ref{rot} we have noted that the exact finite $D$ agreement between 
spherical membranes and their dual Schwarzschild black holes appears 
not to carry over to rotating black holes. It may be possible to construct a further 
improved stress tensor (and correspondingly, improved membrane 
equations of motion and actions) whose rotating membrane solutions exactly 
reproduce the thermodynamics of arbitrary Myers Perry black holes at finite $D$. 
In this context it is encouraging to recall that, in the context of the fluid 
gravity correspondence, it was possible reproduce the 
exact thermodynamics of AdS Kerr black holes using only the second order 
corrected fluid stress tensor \cite{Bhattacharyya:2008mz}. 

Finally, we find it absolutely fascinating that even the leading order large $D$ membrane equations are equivalent to a set of equations of boundary hydrodynamics that 
reproduce the correct fluid constitutive relations at zero and first order in derivatives even at finite $D$, but also automatically resum a very particular infinite class of higher derivative corrections to the Navier Stokes equations - namely those 
that survive at large $D$. It would be interesting to compare this resummation with other partial resummations of 
the hydrodynamical derivative expansion investigated in the hydrodynamics literature (see e.g. \cite{Heller:2015dha,Basar:2015ava,Aniceto:2015mto,Buchel:2016cbj,Spalinski:2017mel}). We also note that some higher derivative corrections 
to the Navier Stokes equations - like the Israel Stewart correction - turn the parabolic Navier Stokes PDEs into hyperbolic 
PDEs. It would be interesting to investigate whether the corrections induced by our membrane also have 
this property (i.e. whether the membrane equations are hyperbolic PDEs). 

We re-emphasize that our improved membrane equations 
define a generalization of the Navier Stokes equations that can be used to study the dynamics of thermal systems outside the
validity of hydrodynamics (i.e. at length scales shorter than thermal length scales) 
atleast in the large $D$ limit. We have already pointed out that the membrane picture 
suggests the possibility of qualitatively new phenomena - like membrane 
folds - that cannot be captured by the variables of hydrodynamics.

It would be useful to generalize the discussion of this paper to the study of improved equations, the
partition function and hydrodynamics of charged membranes (see \cite{Bhattacharyya:2015fdk,Bhattacharyya:2016nhn}).

Apart from all these issues of principle, it would also be interesting to put the formulae presented 
in this paper to practical use.  It would be interesting to use the improved membrane equations presented in this paper as the 
starting point for a `rederivation' of the equations of black fold dynamics \footnote{We thank Mukund Rangamani for
a discussion on this point.} and to compare our results with the exact gravitational results   
 \cite{Emparan:2009cs,Emparan:2009at,Emparan:2011br,Camps:2012hw}. Such a discussion could proceed along 
 the lines of our `rederivation' of boundary hydrodynamics from our improved membrane equations, presented earlier in this paper.

 It is already known that the black hole membranes have a `Gregory-Laflamme like' instability at large $D$. At large $D$, however, 
 this transition is of second order and ends up in a wiggly string. It would be interesting to re-investigate Gregory Laflamme physics 
 using the improved membrane equations presented in this paper. As our probe membranes define consistent dynamics even at 
 finite $D$, it is meaningful to ask whether their Gregory Laflamme like transition switches from  
 second to first order below a critical value of $D$ (recall this is the case for actual black strings; the critical value of 
 $D$ is 13.5 \cite{Suzuki:2015axa}). Assuming this is the case as a related analysis suggests \cite{Suzuki:2015axa}, it would be interesting to 
 investigate whether the equations of membrane hydrodynamics  presented in this paper capture the fascinating dynamics of the `self-similar cascade and pinch off' observed in \cite{Lehner:2010pn}. It is far from clear that this will turn out to be 
 the case \footnote{We thank R. Emparan for emphasizing this to us.}. Nonetheless we find the possibility tantalizing, 
 as it holds out the promise of relating the mysterious process of horizon bifurcation to the more mundane process 
 of hydrodynamical droplet formation in a semi quantitative manner.

Finally it is possible that the formalism developed in this paper 
can be combined with that of 
\cite{Bhattacharyya:2013lha,Bhattacharyya:2014bha,Haehl:2013hoa,Haehl:2014zda,Haehl:2015pja,Crossley:2015tka,Haehl:2015foa,Crossley:2015evo,Haehl:2015uoc,Haehl:2016pec,Haehl:2016uah,Glorioso:2016gsa,Jensen:2017kzi,Gao:2017bqf,Glorioso:2017fpd,Glorioso:2017lcn,Geracie:2017uku} to establish a second law of thermodynamics 
for dynamical event horizons in higher derivative theories of gravity. We hope to return to 
this point in the future. 

\section*{Acknowledgments}
We would like to thank  R. Loganayagam, M. Mandlik, F. Pretorius, S. Thakur for useful discussions. We would especially like to thank J. Armas, J. Bhattacharya, S. Bhattacharyya, M. Rangamani, U. Mehta and U. Sharma for several extremely useful discussions over the course of this project. We would also like to thank J. Armas, J. Bhattacharya, J. de Boer, R. Emparan and M. Rangamani 
for comments on a preliminary draft of this manuscript. Y.D. would like to acknowledge the hospitality of the IIT-Kanpur while this work was in progress. The work of all authors was supported in part by a UGC/ISF Indo Israel grant, and the Infosys Endowment for Research into the Quantum Structure of Spacetime. Finally we would all like to acknowledge our debt to the steady support of the people of India for research in the basic sciences.

\appendix

\section{Shape variations} \label{variation}

In this appendix, we demonstrate the results \eqref{variations}. That is, we calculate variations of various membrane quantities with respect to change in shape of membrane. We find it useful to use the Gaussian normal coordinates for this purpose. 
The form of the spacetime metric in Gaussian normal coordinates is
\begin{equation}\label{gncs}
ds^2 = G_{MN}dx^Mdx^N = dz^2 + g_{\mu\nu}(z,x^\mu)dx^\mu dx^\nu
\end{equation}
 and we take the membrane surface at $z=0$. The induced metric on the membrane worldvolume is $g_{\mu\nu}(0,x^\mu)$. We use overhead bar for the quantities defined in spacetime metric. Unbarred quantities are defined in induced metric on membrane. The normal to membrane is $n=dz$. The Christoffel symbols for the spacetime metric \eqref{gncs} are
\begin{eqnarray}
\bar \Gamma^M_{zz} = 0,\quad  \bar \Gamma^z_{zM} = 0,\quad \bar \Gamma^z_{\mu\nu} = -\frac{1}{2} \partial_z g_{\mu\nu},\quad  \bar \Gamma^\mu_{z\nu} = \frac{1}{2}g^{\mu\alpha}\partial_z g_{\alpha \nu}, \quad \bar \Gamma^\mu_{\nu\rho} = \Gamma^\mu_{\nu\rho} 
\end{eqnarray} 
 $\bar \nabla_Mn_N$ and $\bar P^M_N \equiv \delta^M_N -n^Mn_N$ evaluate to
\begin{equation}
\begin{split}
& \bar\nabla_z n_z = 0, \quad \bar\nabla_z n_\mu = 0, \quad \bar\nabla_\mu n_z = 0, \quad \bar\nabla_\mu n_\nu = \frac{1}{2} \partial_z g_{\mu\nu} \\
& \bar P^z_z = \bar P^z_\mu = \bar P^\mu_z = 0, \quad \bar P^\mu_\nu = \delta^\mu_\nu
\end{split}
\end{equation}
Thus the Extrinsic curvature of the membrane evaluates to
\begin{equation}\label{ec}
K_{\mu\nu} = \frac{1}{2} \partial_z g_{\mu\nu} |_{z=0}, \quad K^{\mu\nu} = -\frac{1}{2} \partial_z g^{\mu\nu} |_{z=0}
\end{equation}
Now we consider a new membrane surface $z=\delta z(x^i)$ (Note that $\delta z$ is not a function of $t$, so $x^i$ are rest of the spacial coordinates).  We work in the linear order in shape perturbations. Using \eqref{ec}, the change in the induced metric on the membrane can be found to be
\begin{equation}\label{varg}
\begin{split}
 g_{\mu\nu}(z=\delta z,x^\mu) &= g_{\mu\nu}(z=0,x^\mu) + \partial_z g_{\mu\nu}(z,x^\mu)|_{z=0} ~\delta z \\ \therefore \delta g_{\mu\nu} &= 2 K_{\mu\nu} \delta z
\end{split}
\end{equation}
and for the inverse metric it is
\begin{equation}\label{vargi}
\delta g^{\mu\nu} = -2 K^{\mu\nu} \delta z
\end{equation}
Using \eqref{varg} we get the variation
\begin{equation}\label{vard}
\delta \sqrt{-g} = \sqrt{-g}~ {\mathcal K}~ \delta z 
\end{equation}
The normal to new surface is $n=dz-\partial_\mu \delta z(x^i)dx^\mu$. For the new surface, $\bar \nabla_Mn_N$ and $\bar P^M_N \equiv \delta^M_N -n^Mn_N$ evaluate to (with $\nabla_\mu$ denotes the covariant derivative on the membrane worldvolume) 
\begin{equation}\label{iss}
\begin{split}
& \bar\nabla_z n_z = 0, \quad \bar\nabla_z n_\mu = \bar\nabla_\mu n_z =\frac{1}{2}\nabla^\rho \delta z~\partial_z g_{\mu\rho} ,\quad \bar\nabla_\mu n_\nu = \frac{1}{2} \partial_z g_{\mu\nu}-\nabla_\mu\nabla_\nu \delta z \\
& \bar P^z_z = 0,\quad \bar P^z_\mu = \nabla_\mu \delta z,\quad \bar P^\mu_z = \nabla^\mu \delta z, \quad \bar P^\mu_\nu = \delta^\mu_\nu
\end{split}
\end{equation}
Using \eqref{ec} and \eqref{iss}, the Extrinsic curvature for the new surface is found to be
\begin{equation}\label{ecn}
K_{\mu\nu}|_{z=\delta z} = \frac{1}{2} \partial_z g_{\mu\nu}|_{z=\delta z} - \nabla_\mu \nabla_\nu \delta z = - \nabla_\mu \nabla_\nu \delta z + K_{\mu\nu}|_{z=0} + \frac{1}{2} \partial^2_z g_{\mu\nu}|_{z=0}~ \delta z
\end{equation}
Hence we get
\begin{equation}\label{vare}
\delta K_{\mu\nu} = - \nabla_\mu \nabla_\nu \delta z + \frac{1}{2} \partial^2_z g_{\mu\nu} \delta z
\end{equation}
 Ricci tensor $\bar R_{MN}$ in spacetime evaluates to
\begin{equation}\label{EE}
\begin{split}
\bar R_{zz} &= -\frac{1}{2}g^{\mu\nu}\partial^2_z g_{\mu\nu} - \frac{1}{4} \partial_z g_{\mu\nu}\partial_z g^{\mu\nu} \\
\bar R_{z\mu} &= \nabla_\nu \bar\Gamma^\nu_{z\mu} - \nabla_\mu \bar\Gamma^\nu_{\nu z} \\
 \bar R_{\mu\nu} &= R_{\mu\nu} - \frac{1}{2}\partial_z^2 g_{\mu\nu}+\frac{1}{2}\partial_z g_{\mu\alpha}g^{\alpha \beta}\partial_z g_{\beta \nu} -\frac{1}{4} \partial_z g_{\mu\nu}(g^{\alpha \beta}\partial_z g_{\alpha \beta}) \\ &= R_{\mu\nu} -\frac{1}{2}\partial_z^2 g_{\mu\nu} + 2 K_{\mu\alpha}K^{\alpha}_{\nu}- {\mathcal K} K_{\mu\nu} 
\end{split}
\end{equation}
Because the spacetime metric solves Einstein equations,
\begin{equation}\label{onco}
\bar R = -D(D-1)\lambda, \quad\bar R_{MN} = -(D-1)\lambda G_{MN}
\end{equation}
Thus using \eqref{onco} and \eqref{EE} in \eqref{vare} we get
\begin{equation}\label{vka}
\delta K_{\mu\nu} = \left(R_{\mu\nu} + (D-1)\lambda G_{\mu\nu} + 2 K_{\mu\alpha}K^{\alpha}_{\nu}- {\mathcal K} K_{\mu\nu}\right)\delta z -\nabla_\mu\nabla_\nu \delta z
\end{equation}
Using \eqref{vargi}, \eqref{vka} and Gauss's identity, 
$R={\mathcal K}^2-K_{\mu\nu}K^{\mu\nu}-(D-1)(D-2)\lambda$  we get
\begin{equation}\label{varkapa}
\delta {\mathcal K} = \delta K_{\mu\nu} g^{\mu\nu} + K_{\mu\nu}\delta g^{\mu\nu} = \left( -K_{\mu\nu}K^{\mu\nu}+(D-1)\lambda\right) \delta z - \nabla^2 \delta z
\end{equation}
Using \eqref{vard} and \eqref{varkapa} we get
\begin{equation}\label{vargk}
\delta (\sqrt{-g}{\mathcal K}) = \sqrt{-g}\left({\mathcal K}^2-K_{\mu\nu}K^{\mu\nu}+(D-1)\lambda-\nabla^2 \right)\delta z
\end{equation}
Notice that the term $\sqrt{-g} \nabla^2 \delta z$ in \eqref{vargk} is total derivative.

The variation of the volume term can be seen to be
\begin{equation}
\delta \left[-(D-1)\lambda \int_V \sqrt{-G}\right] = -(D-1)\lambda\int_M \sqrt{-g} ~\delta z
\end{equation}
The variation of $\gamma$ becomes \footnote{This can be seen from the following manipulations 
\begin{equation}
\partial_z\gamma = \frac{1}{2}\gamma^3 n.\nabla (k^Mk_M) = \gamma^3 k^Mn.\nabla k_M = -\gamma^3k^M n^N \nabla_M k_N = \gamma^3 k^M k^N \nabla_M n_N = \gamma (u.K.u)   
\end{equation}
Where we have used the fact that there is a Killing vector $k^M$ in spacetime whose pullback on the membrane is $k^\mu$ (see section \ref{ssol}). In the third step, we use the Killing equation. In the fourth and last step we use the fact that $k^Mn_M=0$ on the membrane.}
(Recall $\gamma=\frac{1}{\sqrt{-k.k}}$) 
\begin{equation}\label{vg}
\delta \gamma = (\partial_z \gamma)~\delta z = \gamma(u.K.u)\delta z  
\end{equation}
Using \eqref{vg} and \eqref{vard} we get
\begin{equation}\label{vS}
\delta (\sqrt{-g}~\gamma ) = \sqrt{-g}~\gamma ({\mathcal K}+u.K.u)~\delta z
\end{equation}
This completes the demonstration of \eqref{variations}.

\section{QNM for spherical membrane in flat spacetime} \label{lin}
In this section, we find the quasinormal mode spectrum for linearized fluctuations about a spherical membrane in arbitrary $D$ dimensional flat spacetime background. Since the calculation is very similar to done e.g. in \cite{Bhattacharyya:2015fdk,Bhattacharyya:2017hpj} we present only key steps. For details, \cite{Bhattacharyya:2015fdk,Bhattacharyya:2017hpj} can be referred.
We consider the background spacetime metric
\begin{equation}\label{smbg}
ds^2_{ST} = -dt^2+ dr^2 +r^2 d\Omega_{D-2}^2
\end{equation}
We consider the shape and velocity fluctuations about a uniform spherical membrane, so we consider the shape and the velocity field of the following form
\begin{equation}\label{smfl}
r=1+\epsilon \delta r (t,\theta^a), \quad u =  -dt + \epsilon \delta u_a(t,\theta^a) d\theta^a 
\end{equation}
We will always work in linear order in $\epsilon$. Putting \eqref{smfl} in \eqref{smbg} we get the induced metric on the membrane 
\begin{equation}
ds^2 = -dt^2 + (1+2\epsilon \delta r)d\Omega_{D-2}^2
\end{equation}  
We have the membrane equations
\begin{equation}\label{msq} \begin{split}
& \nabla.u=0\\
&16 \pi~ {\cal P}^\nu_\alpha\nabla^\mu T_{\mu\nu} = \left( \tilde {\mathcal K}~u.\nabla u_\nu + \nabla_\nu \tilde {\mathcal K} -2 \nabla^\mu \sigma_{\mu\nu}\right) {\cal P}^\nu_\alpha \equiv E_\nu {\cal P}^\nu_\alpha
\end{split}
\end{equation}
We use the notation that $\Omega_{ab}$ denotes the metric on the unit sphere, $\nabla_a$ denotes the covariant derivative on the unit sphere, and $\nabla^2\equiv\nabla^a\nabla_a$. To linear order in fluctuations we calculate the quantities present in \eqref{msq}
\begin{equation} \begin{split}
& {\cal P}^t_t=0,\hspace{5mm}{\cal P}^a_t=-\epsilon \delta u^a,\hspace{5mm}{\cal P}^t_a=\epsilon \delta u_a,\hspace{5mm}{\cal P}^a_b=\delta^a_b \\
&\sigma_{tt} = 0, \quad \sigma_{ta} = 0,\quad \sigma_{ab} = \frac{\epsilon}{2}\bigg(\nabla_a\delta u_b+\nabla_b\delta u_a\bigg)+\epsilon\partial_t\delta r\Omega_{ab}\\
&\tilde {\mathcal K} = (D-3)-(D-3)\epsilon\delta r+2\epsilon\partial^2_t\delta r-\epsilon\bigg(\frac{D-3}{D-2}\bigg)\nabla^2\delta r \\
&u. \nabla u_t=0, \quad  u. \nabla u_a = \epsilon \partial_t \delta u_a  \\
\end{split}
\end{equation}
Hence the membrane equations  \eqref{msq} simplify to  
\begin{equation}\label{linsm}
\begin{split}
&\nabla^a\delta u_a+(D-2)\partial_t\delta r=0\\
& V_a \equiv -(D-3)\nabla_a\delta r+2\partial_t^2\nabla_a\delta r-\bigg(\frac{D-3}{D-2}\bigg)\nabla_a\nabla^2\delta r+(D-3)\delta_t\delta u_a\\ &-\nabla^b\nabla_a\delta u_b
-\nabla^2\delta u_a-2\partial_t\nabla_a\delta r=0
\end{split}
\end{equation}
We write the velocity field as  
\begin{equation}\label{vde}
\delta u_a = \delta v_a +\nabla_a \Phi,\quad  \text{with} \quad \nabla^a \delta v_a = 0
\end{equation}
Putting \eqref{vde} into the first equation in \eqref{linsm} we get
\begin{equation}\label{pheq}
\nabla^2 \Phi = -(D-2) \partial_t \delta r
\end{equation}
We expand the fluctuations in the Spherical Harmonic basis as
\begin{equation}\label{smf}
\delta r = \sum_{l,m} a_{l,m}Y_{l,m}e^{-i\omega^s_l t}, \quad \delta v_a = \sum_{l,m} b_{l,m} Y^{l,m}_a e^{-i \omega^v_l t}
\end{equation}
Recall that for the Spherical Harmonics
\begin{equation}\label{sid}
\nabla^2 Y_{l,m} = -l(D-3+l) Y_{l,m}, \quad \nabla^2 Y^{l,m}_a = -\left(-l(D-3+l)-1\right) Y^{l,m}_a
\end{equation}
We take the divergence of the second equation in \eqref{linsm} i.e. $\nabla^a V_a$. We then eliminate the terms containing $\Phi$ using \eqref{pheq}. We put the basis \eqref{smf} and use \eqref{sid}, to get the scalar QNM frequencies, which are found to be
\begin{equation}\label{qem}
\omega^s_l = \pm \frac{\sqrt{-b^2-4ac}}{2a} -i \frac{b}{2a}
\end{equation}
where,
\begin{equation}\label{qemt}
\begin{split}
 a &=l\big(l+D-3\big)+\frac{\big(D-3\big)\big(D-2\big)}{2}\\
b &=\big(D-3\big)\bigg[l\big(l+D-3\big)-\big(D-2\big)\bigg]\\
c &=l\big(l+D-3\big)\bigg(\frac{D-3}{2}\bigg)\bigg[1-\frac{l\big(l+D-3\big)}{D-2}\bigg]
\end{split}
\end{equation}
Using the fact that $\delta r$ solves the equation $\nabla^aV_a=0$ the second equation in \eqref{linsm} reduces to the equation only for the variable $\delta v_a$. Putting \eqref{smf} into this equation and using \eqref{sid} we find the vector QNM frequencies
\begin{equation}\label{qvm}
\omega^{(v)}_l=-i\bigg[\frac{l\big(l+D-3\big)-1}{D-3}-1\bigg]
\end{equation}
Expanding the answers \eqref{qem} and \eqref{qvm} in a power series in $1/D$, we get 
\begin{equation}\label{ldm}
\begin{split}
\omega^s_l &= \pm \sqrt{l-1} -i(l-1) \pm \frac{l\sqrt{l-1}(2l-3)}{2D}-i\frac{l(l-1)}{D} + {\cal O}(D^{-2})\\
\omega^v_l &= -i(l-1)- \frac{i(l^2-1)}{D}+ {\cal O}(D^{-2})
\end{split}
\end{equation}
Whereas the actual answers found from gravity analysis in \cite{Emparan:2014aba} and from Membrane paradigm approach in \cite{Dandekar:2016fvw} are 
\begin{equation}\label{mpm}
\begin{split}
\omega^s_l &= \pm \sqrt{l-1}-i(l-1) \pm \frac{\sqrt{l-1}(3l-4)}{2D} -i\frac{(l-1)(l-2)}{D} + {\cal O}(D^{-2})\\
\omega^v_l &= -i(l-1)- \frac{i(l-1)^2}{D}+ {\cal O}(D^{-2})
\end{split}
\end{equation}
 Note that the answers of \eqref{ldm} and \eqref{mpm} match at leading order but differ at the subleading orders in $1/D$.

\section{Membrane Energy and Bulk Hamiltonian} \label{ham}

In the main text we have demonstrated that the first two terms in the action 
\eqref{Ft} have a simple bulk interpretation - they are equal to half the action of the bulk 
region enclosed by the membrane. We will now present an alternative - but equivalent 
- reinterpretation of the same two terms  in \eqref{Ft} in terms of the Hamiltonian 
of the region of spacetime enclosed by the membrane. 

In order to do this we first rewrite the stationary spacetime \eqref{kist} 
in the standard ADM form 
\begin{equation} \label{admform}
ds^2_{ST} = G_{MN} dX^M dX^N = -N^2 dt^2 +q_{ab} (dX^a +N^a dt)(dX^b +N^b dt)
\end{equation} 
where, the various metric coefficients are related to \eqref{kist} by the relations
\begin{equation}
q_{ab} N^b = -e^{2\Sigma} A_a, \quad -N^2 + q_{ab}N^a N^b = -e^{2\Sigma}, \quad q_{ab} = -e^{2\Sigma}A_a A_b +W_{ab}
\end{equation}
Notice that
\begin{equation}
k^M = (\partial_t)^M, \quad q_M = -N (dt)_M, \quad k^M = N q^M + N^a e^M_a \quad \text{with}\quad  e^M_a = \left(\frac{\partial X^M}{\partial X^a}\right)_{t} 
\end{equation} 
where $k^M$ is the killing vector field as usual, $q_M$ is the unit normalized normal 
vector orthogonal to slices of constant time $t$.
As is well known, the offshell action of a region of spacetime can be rewritten 
in terms of the Hamiltonian of general relativity (see e.g. section 4.2 of \cite{poisson_2004})
\begin{equation}\label{hfe}
\begin{split}
{\cal S_G} &= \frac{1}{16 \pi} \bigg[ \int_{V}~\sqrt{-G}~(\bar R-2\Lambda)~d^DX + 2\int_{M} ~\sqrt{-g}~{\mathcal K}~d^{D-1}x  \\& ~~~~+ 2 \int_{\Sigma_{t_1}} \sqrt{q}~ C_{ab} ~d^{D-1}X - 2 \int_{\Sigma_{t_2}} \sqrt{q}~ C_{ab} ~d^{D-1}X  \bigg] \\ &= \int dt \left( \int_{\Sigma_t} p^{ab}\dot q_{ab} -H_G \right) \\
 \text{where} &\quad p^{ab} \equiv \frac{\sqrt{q}}{16\pi} (C^{ab}-C q^{ab}), \quad \dot q_{ab} \equiv {\mathcal L}_t q_{ab} = ({\mathcal L}_k G_{MN})e^M_ae^N_b 
\end{split}
\end{equation} 
In \eqref{hfe}, $\Sigma_t$ is the spacelike slice of spacetime at time $t$. $C_{ab}, C$ are the extrinsic curvature and its trace of the spacelike slice of the spacetime as embedded in the spacetime. $\Sigma_{t_2},~\Sigma_{t_1}$ are respectively the initial and final spacelike slices.  Focusing on the special case of the stationary solutions of 
interest to us we have 
$$ \dot q_{ab} =0$$
Moreover, onshell, the Hamiltonian of spacetime is given by the ADM formula 
(see Equation (4.80) of \cite{poisson_2004})
\begin{equation} \label{oham}
H_G = -\frac{1}{8\pi} \int_{S_t} \sqrt{-g} \left( {\mathcal K} + q^M \nabla_M n_N q^N  -\frac{N^a}{N} (C_{ab}-Cq_{ab})n^b \right)\\
\end{equation}
In the special case at hand \eqref{oham} can be further simplified. 
\begin{equation}
\begin{split}
N^a q_{ab} n^b &= \left(k^M-N q^M\right) G_{MN} n^N = 0 \\
N^a C_{ab} n^b &= \left(k^M-N q^M\right) \nabla_M q_N n^N = - k^M {K_M}^N q_N + N q^M \nabla_M n_N q^N \\&= N K^t_t + N q^M \nabla_M n_N q^N
\end{split}
\end{equation}

Hence, we get 
\begin{equation}
H_G = -\frac{1}{8 \pi}  \int_{S_t} \sqrt{-g}~({\mathcal K}-K^t_t)
\end{equation}
Where $S_t$ is the boundary of the $\Sigma_t$, that is the timeslice of the membrane worldvolume at time $t$. 

It follows that the first two terms in the action \eqref{Ft} are equal both to
the `length of time' (equal to $\beta$ in Euclidean space) times 
\begin{itemize} 
\item[1.] Half of the General Relativistic Hamiltonian (i.e. ADM energy) of the 
region of spacetime enclosed by the membrane
\item[2.] The actual energy $E$ of the membrane 
\end{itemize}

The discussion in this Appendix provides an alternate derivation of the equation \eqref{pfo}.

\section{Rotating membranes in 4 dimensions} \label{rot}

It would be interesting to find the exact solutions corresponding to rotating membrane solutions at all values of $D$. The problem we need to solve 
is the following. Specializing to the case of even $D$, consider flat space in the coordinates
\begin{equation}\label{fsm}
 ds^2=-dt^2 + dz^2 + \sum_{i=1}^{[D/2]} dr_i^2 + r_i^2  d \phi_i^2 
\end{equation}
Consider the killing vector 
\begin{equation}\label{kkv}
 k=   \partial_t + \sum_i \omega_i \partial_{\phi_i} 
\end{equation}
With this choice of $k$ we need to find the membrane shape that obeys the equation \eqref{memeo}. 
\footnote{For other studies of reliable fluid descriptions of localized black holes see e.g. 
\cite{Lahiri:2007ae, Bhattacharya:2009gm, Armas:2015ssd, Armas:2016xxg}.} 

We postpone the general consideration of this problem to future work. For the present, we focus our attention on a simple 
special example, namely $D=4$. In this case the most general velocity field is characterized by a single rotational velocity $\omega$, 
and the construction of the membrane shape - dual to the Kerr black hole - turns out to be particularly easy. The trick turns out to be
a good choice of coordinates; in this case the zero mass  Boyer-Lindquist coordinates.

As our starting point consider the flat space metric in Minkowski coordinates
\begin{equation}\label{fsmu}
 ds^2=-dt^2 + dz^2 + dx^2 + dy^2
\end{equation}
Then perform the coordinate change to the zero mass Boyer-Lindquist coordinates 
\begin{equation}\label{blc}
 z=r \cos \theta, ~~~x= \sqrt{r^2 + a^2} \sin \theta \cos \phi, ~~~ y= \sqrt{r^2 + a^2} \sin \theta \sin \phi
\end{equation}
Under which \eqref{fsmu} becomes
\begin{equation}
ds^2 = G_{MN} dx^M dx^N = -dt^2 + \frac{r^2+a^2 \cos^2\theta}{r^2+a^2}dr^2 + \left(r^2+a^2\cos^2\theta \right)d\theta^2  + (r^2+a^2)\sin^2\theta d\phi^2 
\end{equation}
Under this coordinate change the killing vector \eqref{kkv} retains its form 
\begin{equation}\label{kform}
 k= \partial_t + \omega \partial_\phi
\end{equation}
We will find it useful to define a new constant $a$ of dimension length by the equation 
\begin{equation} \label{omega} 
 \omega= \frac{a}{r_H^2+a^2}
\end{equation}
Working in the coordinate system \eqref{blc} we will now demonstrate that the surface 
\begin{equation} \label{sursol}
r=r_H
\end{equation}
(together with the choice of $k$ listed in \eqref{kform} and \eqref{omega}) solve 
\eqref{memeo} with 
\begin{equation}\label{ktem}
4 \pi T_0 = \frac{r_H}{a^2 + r_H^2}
\end{equation}

In order to see this we note that the velocity field corresponding to the killing vector \eqref{omega}, \eqref{kform} 
is given by 
\begin{equation}
u^M = \gamma k^M, \quad \gamma = \frac{1}{\sqrt{-k^MG_{MN}k^N}} = \left(1-\frac{a^2 \sin^2\theta}{r_H^2+a^2}\right)^{-1/2}  
\end{equation} 
For the surface \eqref{sursol} we find 
\begin{equation} \label{knn} \begin{split}
                              &{\cal K} = \frac{r_H}{\sqrt{r_H^2+a^2}}\frac{2r_H^2+a^2(1+\cos^2\theta)}{\left(r_H^2+a^2\cos^2\theta\right)^{3/2}} \\
                              &K_{MN}K^{MN}= \frac{r_H^2}{r_H^2+a^2}\frac{\left(r_H^2+a^2\right)^2+\left(r_H^2+a^2\cos^2\theta\right)^2}{\left(r_H^2+a^2\cos^2\theta\right)^3} \\
                              & u.K.u= \frac{r_H }{\sqrt{r_H^2+a^2}}\frac{a^2\sin^2\theta}{\left(r_H^2+a^2\cos^2 \theta\right)^{3/2}}\\
                             \end{split}
\end{equation}
It follows that
\begin{equation}\label{ser}
\frac{\tilde{\cal K}}{\gamma}=\frac{{\cal K}^2-K_{MN}K^{MN}}{\gamma({\cal K}+u.K.u)} = \frac{r_H}{r_H^2+a^2} = 4 \pi T_0 
\end{equation}
demonstrating that the surface $r=r_H$ solves the equations \eqref{memeo} with $T_0$ given in \eqref{ktem}.

Let us emphasize that the quantities $\gamma$, ${\cal K}$, $K_{MN}$, $u.K.u$  - which went into the LHS of \eqref{ser} - all depend on 
$\theta$ in a nontrivial manner. Interestingly however, the $\theta$ dependences of the combination of these 
quantities that appears in $\tilde{\cal K}$ cancel out, allowing the configuration 
$r=r_H$ to solve \eqref{memeo}.

Inverting \eqref{ktem} to solve for the parameter $r_H$ in terms of $a$ and $T_0$ we find 
\begin{equation}
r_H = m \pm \sqrt{m^2-a^2}, \quad \text{where}~~m\equiv\frac{1}{8\pi T_0}
\end{equation}

It is not difficult to determine the thermodynamical charges of our solution. The entropy is given by
\begin{equation}\label{kms}
S_{ent} = \int_{sM} \sqrt{h}~q_\mu J^\mu_S = \frac{1}{4}  \int_{sM} \sqrt{-g}~\gamma = \pi (r_H^2+a^2)
\end{equation}
Where, $sM$ denotes integration over the spacelike slice of the membrane. $h$ is the determinant of the metric on this slice. $g$ is the determinant of the metric on the membrane worldvolume.
    
The mass of the membrane is given by 
\begin{equation} \label{massform}
\begin{split}
M &= -\int_{sM} \sqrt{h}~q^\mu T_{\mu\nu} (\partial_t)^\nu = \frac{-1}{16\pi}\int_{sM} \sqrt{h}~q^\mu (\tilde {\mathcal K}P_{\mu\nu} + K_{\mu\nu}-{\mathcal K}g_{\mu\nu}) (\partial_t)^\nu \\&= \frac{r_H^2+a^2}{2a}\tan^{-1}\left( \frac{a}{r_H} \right) 
\end{split}
\end{equation} 
 and the angular momentum
 \begin{equation}\label{angform}
 \begin{split}
 J &= \int_{sM} \sqrt{h}~q^\mu T_{\mu\nu} (\partial_\phi)^\nu = \frac{1}{16\pi}\int_{sM} \sqrt{h}~q^\mu (\tilde {\mathcal K}P_{\mu\nu} + K_{\mu\nu}-{\mathcal K}g_{\mu\nu}) (\partial_\phi)^\nu \\ &= \frac{r_H^2+a^2}{4a}\left[ -r_H+\frac{r_H^2+a^2}{a}\tan^{-1}\left(\frac{a}{r_H}\right) \right]
 \end{split}
 \end{equation} 
 It is easily verified that our results obey the first law of thermodynamics
\begin{equation} \label{fl}
dM = T_0 dS_{ent} +\omega dJ
\end{equation}
 
The `energy' of the membrane - i.e. conserved charge $E = M-\omega J$ of membrane dual to the killing vector $k$ is given by 
\begin{equation}\label{kme}
\begin{split}
E &= -\int_{sM} \sqrt{h}~q^\mu T_{\mu\nu} k^\nu = \frac{-1}{16\pi}\int_{sM} \sqrt{h}~q^\mu (K_{\mu\nu}-{\mathcal K}g_{\mu\nu}) k^\nu \\&= \frac{r_H}{4} \left[ 1+\left(\frac{a}{r_H}+\frac{r_H}{a}\right)\tan^{-1}\left(\frac{a}{r_H}\right) \right]
\end{split}
\end{equation}
Provided we restrict attention to those variations that keep $\omega$ fixed we have (from \eqref{fl})
\begin{equation}
 dE= T_0 dS_{ent}
\end{equation}
in agreement with the general analysis presented earlier in this paper (recall that it was assumed - for the 
purpose of that analysis - that the killing vector $k_\mu$ - and hence $\omega$ of this subsection - 
is kept constant while taking all variations). 

The Partition function for the rotating membrane in 4D flat spacetime, written in terms of chemical potentials becomes 
\begin{equation}\label{rmpf}
\ln Z = \frac{-1}{4 T_0 \omega} \tan^{-1} \left(\frac{\omega}{4\pi T_0}\right)
\end{equation} 
Whereas, the partition function for actual Kerr black hole (see \cite{Gibbons:1976ue}) is (with $M$ as the mass of black hole)
\begin{equation}\label{kbpf}
\ln Z = -\frac{M}{2T_0} = -\frac{1}{8\pi T_0^2 + 4T_0\sqrt{4\pi^2 T_0^2 +\omega^2}}
\end{equation}
Note that for $\omega\rightarrow0$ we have both the partition functions reduce to $-\frac{1}{16\pi T_0^2}$ 
in agreement with \eqref{sfm} and \eqref{tcheck} at $D=4$. 
It is easy to check that the partition functions \eqref{rmpf} and \eqref{kbpf} satisfy the thermodynamic relations
\begin{equation}
J = T_0 \frac{\partial \ln Z}{\partial \omega}, \quad -T_0^2 \frac{\partial \ln Z}{\partial T_0} = -M + \omega J, \quad S_{ent} = \ln Z + T_0 \frac{\partial \ln Z}{\partial T_0}
\end{equation}
It is also easy to check that the thermodynamical charges that we have computed for our  4D rotating membrane 
above obey the `Smarr relation'
\begin{equation} \label{smarr}
M = 2 \omega J +2 T_0 S_{ent}
\end{equation} 
(of course the exact thermodynamical charges for the Kerr black hole - see below -  also obey \eqref{smarr}). 

It is natural to interpret the rotating membrane solution presented in this paper as the dual to the Kerr black hole 
solution given, for instance, pages (221, 222) of \cite{poisson_2004} with electromagnetic charge $Q$ of 
\cite{poisson_2004} set to zero  and the parameter $r_+$ of \cite{poisson_2004} identified with  $r_H$ of this subsection and $a$ and $M$ of \cite{poisson_2004} identified with  $a$ and $M$ of this subsection. 
With these identifications, the entropy of the Kerr black hole agrees exactly with the \eqref{kms}. 
However the mass of the Kerr black hole does {\it not} agree exactly with \eqref{massform}; indeed we find the 
correct gravitational results for the Kerr black hole mass only once we make the replacement 
$$ \frac{1}{a} \tan^{-1} \frac{a}{r_H} \rightarrow \frac{1}{r_H}$$. 
This replacement is exact in the limit $a \to 0$, and so at $\omega=0$. However the two expressions above 
differ already at ${\cal O}(\omega^2)$. 

The match between our membrane's angular momentum and the angular momentum of the Kerr black hole is even worse. 
The equation \eqref{angform} reduces to the formula for {\it half of} the angular momentum of the Kerr black hole 
under the replacement 
$$ \frac{1}{a} \tan^{-1} \frac{a}{r_H} \rightarrow \frac{1}{r_H}$$
The surprise here is the additional factor of half which means that the membrane description fails 
to quantitatively reproduce the even the leading order - order $\omega$. \footnote{Recall that the energy and angular 
momentum enter thermodynamical relations in the combination $E-\omega J$. The mismatch of $J$ at order $\omega$ 
is, therefore, connected to the mismatch of $E$ at order $\omega^2$ noted above. }

Of course the discrepancies of this subsection all occur at $D=4$ - which is as far from the large $D$ limit 
as we can be. Consequently the thermodynamical mismatches described above do not contradict any clearly established
expectation. Nonetheless  - given the fact 
that our membrane worked so remarkably well for static black holes, we find them disappointing. Given the 
fact that the second order fluid gravity correspondence was able to exactly reproduce the thermodynamics of Kerr-AdS black holes, it seems likely to us that the membrane stress tensor \eqref{stf}, will turn out to admit 
an additional improvement term that is irrelevant at large $D$ and in static situations, but will allow us to reproduce the thermodynamics of rotating black hole solutions exactly at finite $D$. We postpone the study of this possibility to future work.

\bibliography{larged}

\providecommand{\href}[2]{#2}\begingroup\raggedright\begin{thebibliography}{10}

\bibitem{Emparan:2013moa}
R.~Emparan, R.~Suzuki and K.~Tanabe, \emph{{The large D limit of General
  Relativity}}, \href{http://dx.doi.org/10.1007/JHEP06(2013)009}{\emph{JHEP}
  {\bfseries 1306} (2013) 009},
  [\href{https://arxiv.org/abs/1302.6382}{{\ttfamily 1302.6382}}].

\bibitem{Emparan:2013xia}
R.~Emparan, D.~Grumiller and K.~Tanabe, \emph{{Large-D gravity and low-D
  strings}},
  \href{http://dx.doi.org/10.1103/PhysRevLett.110.251102}{\emph{Phys.Rev.Lett.}
  {\bfseries 110} (2013) 251102},
  [\href{https://arxiv.org/abs/1303.1995}{{\ttfamily 1303.1995}}].

\bibitem{Emparan:2014cia}
R.~Emparan and K.~Tanabe, \emph{{Universal quasinormal modes of large D black
  holes}}, \href{http://dx.doi.org/10.1103/PhysRevD.89.064028}{\emph{Phys.Rev.}
  {\bfseries D89} (2014) 064028},
  [\href{https://arxiv.org/abs/1401.1957}{{\ttfamily 1401.1957}}].

\bibitem{Emparan:2014aba}
R.~Emparan, R.~Suzuki and K.~Tanabe, \emph{{Decoupling and non-decoupling
  dynamics of large D black holes}},
  \href{http://dx.doi.org/10.1007/JHEP07(2014)113}{\emph{JHEP} {\bfseries 07}
  (2014) 113}, [\href{https://arxiv.org/abs/1406.1258}{{\ttfamily 1406.1258}}].

\bibitem{Giribet:2013wia}
G.~Giribet, \emph{{Large D limit of dimensionally continued gravity}},
  \href{http://dx.doi.org/10.1103/PhysRevD.87.107504}{\emph{Phys. Rev.}
  {\bfseries D87} (2013) 107504},
  [\href{https://arxiv.org/abs/1303.1982}{{\ttfamily 1303.1982}}].

\bibitem{Prester:2013gxa}
P.~D. Prester, \emph{{Small black holes in the large D limit}},
  \href{http://dx.doi.org/10.1007/JHEP06(2013)070}{\emph{JHEP} {\bfseries 06}
  (2013) 070}, [\href{https://arxiv.org/abs/1304.7288}{{\ttfamily 1304.7288}}].

\bibitem{Emparan:2013oza}
R.~Emparan and K.~Tanabe, \emph{{Holographic superconductivity in the large D
  expansion}}, \href{http://dx.doi.org/10.1007/JHEP01(2014)145}{\emph{JHEP}
  {\bfseries 1401} (2014) 145},
  [\href{https://arxiv.org/abs/1312.1108}{{\ttfamily 1312.1108}}].

\bibitem{Bhattacharyya:2015dva}
S.~Bhattacharyya, A.~De, S.~Minwalla, R.~Mohan and A.~Saha, \emph{{A membrane
  paradigm at large D}},
  \href{http://dx.doi.org/10.1007/JHEP04(2016)076}{\emph{JHEP} {\bfseries 04}
  (2016) 076}, [\href{https://arxiv.org/abs/1504.06613}{{\ttfamily
  1504.06613}}].

\bibitem{Bhattacharyya:2015fdk}
S.~Bhattacharyya, M.~Mandlik, S.~Minwalla and S.~Thakur, \emph{{A Charged
  Membrane Paradigm at Large D}},
  \href{http://dx.doi.org/10.1007/JHEP04(2016)128}{\emph{JHEP} {\bfseries 04}
  (2016) 128}, [\href{https://arxiv.org/abs/1511.03432}{{\ttfamily
  1511.03432}}].

\bibitem{Dandekar:2016fvw}
Y.~Dandekar, A.~De, S.~Mazumdar, S.~Minwalla and A.~Saha, \emph{{The large D
  black hole Membrane Paradigm at first subleading order}},
  \href{http://dx.doi.org/10.1007/JHEP12(2016)113}{\emph{JHEP} {\bfseries 12}
  (2016) 113}, [\href{https://arxiv.org/abs/1607.06475}{{\ttfamily
  1607.06475}}].

\bibitem{Dandekar:2016jrp}
Y.~Dandekar, S.~Mazumdar, S.~Minwalla and A.~Saha, \emph{{Unstable `black
  branes' from scaled membranes at large $D$}},
  \href{http://dx.doi.org/10.1007/JHEP12(2016)140}{\emph{JHEP} {\bfseries 12}
  (2016) 140}, [\href{https://arxiv.org/abs/1609.02912}{{\ttfamily
  1609.02912}}].

\bibitem{Bhattacharyya:2016nhn}
S.~Bhattacharyya, A.~K. Mandal, M.~Mandlik, U.~Mehta, S.~Minwalla, U.~Sharma
  et~al., \emph{{Currents and Radiation from the large $D$ Black Hole
  Membrane}}, \href{http://dx.doi.org/10.1007/JHEP05(2017)098}{\emph{JHEP}
  {\bfseries 05} (2017) 098},
  [\href{https://arxiv.org/abs/1611.09310}{{\ttfamily 1611.09310}}].

\bibitem{Bhattacharyya:2017hpj}
S.~Bhattacharyya, P.~Biswas, B.~Chakrabarty, Y.~Dandekar and A.~Dinda,
  \emph{{The large D black hole dynamics in AdS/dS backgrounds}},
  \href{https://arxiv.org/abs/1704.06076}{{\ttfamily 1704.06076}}.

\bibitem{Emparan:2014jca}
R.~Emparan, R.~Suzuki and K.~Tanabe, \emph{{Instability of rotating black
  holes: large D analysis}},
  \href{http://dx.doi.org/10.1007/JHEP06(2014)106}{\emph{JHEP} {\bfseries 1406}
  (2014) 106}, [\href{https://arxiv.org/abs/1402.6215}{{\ttfamily 1402.6215}}].

\bibitem{Emparan:2015rva}
R.~Emparan, R.~Suzuki and K.~Tanabe, \emph{{Quasinormal modes of (Anti-)de
  Sitter black holes in the 1/D expansion}},
  \href{http://dx.doi.org/10.1007/JHEP04(2015)085}{\emph{JHEP} {\bfseries 04}
  (2015) 085}, [\href{https://arxiv.org/abs/1502.02820}{{\ttfamily
  1502.02820}}].

\bibitem{Emparan:2015hwa}
R.~Emparan, T.~Shiromizu, R.~Suzuki, K.~Tanabe and T.~Tanaka, \emph{{Effective
  theory of Black Holes in the 1/D expansion}},
  \href{http://dx.doi.org/10.1007/JHEP06(2015)159}{\emph{JHEP} {\bfseries 06}
  (2015) 159}, [\href{https://arxiv.org/abs/1504.06489}{{\ttfamily
  1504.06489}}].

\bibitem{Suzuki:2015iha}
R.~Suzuki and K.~Tanabe, \emph{{Stationary black holes: Large $D$ analysis}},
  \href{http://dx.doi.org/10.1007/JHEP09(2015)193}{\emph{JHEP} {\bfseries 09}
  (2015) 193}, [\href{https://arxiv.org/abs/1505.01282}{{\ttfamily
  1505.01282}}].

\bibitem{Suzuki:2015axa}
R.~Suzuki and K.~Tanabe, \emph{{Non-uniform black strings and the critical
  dimension in the $1/D$ expansion}},
  \href{http://dx.doi.org/10.1007/JHEP10(2015)107}{\emph{JHEP} {\bfseries 10}
  (2015) 107}, [\href{https://arxiv.org/abs/1506.01890}{{\ttfamily
  1506.01890}}].

\bibitem{Emparan:2015gva}
R.~Emparan, R.~Suzuki and K.~Tanabe, \emph{{Evolution and endpoint of the black
  string instability: Large D solution}},
  \href{http://dx.doi.org/10.1103/PhysRevLett.115.091102}{\emph{Phys. Rev.
  Lett.} {\bfseries 115} (2015) 091102},
  [\href{https://arxiv.org/abs/1506.06772}{{\ttfamily 1506.06772}}].

\bibitem{Tanabe:2015hda}
K.~Tanabe, \emph{{Black rings at large D}},
  \href{http://dx.doi.org/10.1007/JHEP02(2016)151}{\emph{JHEP} {\bfseries 02}
  (2016) 151}, [\href{https://arxiv.org/abs/1510.02200}{{\ttfamily
  1510.02200}}].

\bibitem{Tanabe:2015isb}
K.~Tanabe, \emph{{Instability of the de Sitter Reissner–Nordstrom black hole
  in the $1/D$ expansion}},
  \href{http://dx.doi.org/10.1088/0264-9381/33/12/125016}{\emph{Class. Quant.
  Grav.} {\bfseries 33} (2016) 125016},
  [\href{https://arxiv.org/abs/1511.06059}{{\ttfamily 1511.06059}}].

\bibitem{Chen:2015fuf}
B.~Chen, Z.-Y. Fan, P.~Li and W.~Ye, \emph{{Quasinormal modes of Gauss-Bonnet
  black holes at large D}},
  \href{http://dx.doi.org/10.1007/JHEP01(2016)085}{\emph{JHEP} {\bfseries 01}
  (2016) 085}, [\href{https://arxiv.org/abs/1511.08706}{{\ttfamily
  1511.08706}}].

\bibitem{Emparan:2016sjk}
R.~Emparan, K.~Izumi, R.~Luna, R.~Suzuki and K.~Tanabe, \emph{{Hydro-elastic
  Complementarity in Black Branes at large D}},
  \href{http://dx.doi.org/10.1007/JHEP06(2016)117}{\emph{JHEP} {\bfseries 06}
  (2016) 117}, [\href{https://arxiv.org/abs/1602.05752}{{\ttfamily
  1602.05752}}].

\bibitem{Sadhu:2016ynd}
A.~Sadhu and V.~Suneeta, \emph{{Nonspherically symmetric black string
  perturbations in the large dimension limit}},
  \href{http://dx.doi.org/10.1103/PhysRevD.93.124002}{\emph{Phys. Rev.}
  {\bfseries D93} (2016) 124002},
  [\href{https://arxiv.org/abs/1604.00595}{{\ttfamily 1604.00595}}].

\bibitem{Herzog:2016hob}
C.~P. Herzog, M.~Spillane and A.~Yarom, \emph{{The holographic dual of a
  Riemann problem in a large number of dimensions}},
  \href{http://dx.doi.org/10.1007/JHEP08(2016)120}{\emph{JHEP} {\bfseries 08}
  (2016) 120}, [\href{https://arxiv.org/abs/1605.01404}{{\ttfamily
  1605.01404}}].

\bibitem{Tanabe:2016pjr}
K.~Tanabe, \emph{{Elastic instability of black rings at large D}},
  \href{https://arxiv.org/abs/1605.08116}{{\ttfamily 1605.08116}}.

\bibitem{Tanabe:2016opw}
K.~Tanabe, \emph{{Charged rotating black holes at large D}},
  \href{https://arxiv.org/abs/1605.08854}{{\ttfamily 1605.08854}}.

\bibitem{Rozali:2016yhw}
M.~Rozali and A.~Vincart-Emard, \emph{{On Brane Instabilities in the Large $D$
  Limit}}, \href{http://dx.doi.org/10.1007/JHEP08(2016)166}{\emph{JHEP}
  {\bfseries 08} (2016) 166},
  [\href{https://arxiv.org/abs/1607.01747}{{\ttfamily 1607.01747}}].

\bibitem{Chen:2016fuy}
B.~Chen and P.-C. Li, \emph{{Instability of Charged Gauss-Bonnet Black Hole in
  de Sitter Spacetime at Large $D$}},
  \href{https://arxiv.org/abs/1607.04713}{{\ttfamily 1607.04713}}.

\bibitem{Chen:2017wpf}
B.~Chen, P.-C. Li and Z.-z. Wang, \emph{{Charged Black Rings at large D}},
  \href{http://dx.doi.org/10.1007/JHEP04(2017)167}{\emph{JHEP} {\bfseries 04}
  (2017) 167}, [\href{https://arxiv.org/abs/1702.00886}{{\ttfamily
  1702.00886}}].

\bibitem{Chen:2017hwm}
B.~Chen and P.-C. Li, \emph{{Static Gauss-Bonnet Black Holes at Large $D$}},
  \href{http://dx.doi.org/10.1007/JHEP05(2017)025}{\emph{JHEP} {\bfseries 05}
  (2017) 025}, [\href{https://arxiv.org/abs/1703.06381}{{\ttfamily
  1703.06381}}].

\bibitem{Rozali:2017bll}
M.~Rozali, E.~Sabag and A.~Yarom, \emph{{Holographic Turbulence in a Large
  Number of Dimensions}},  \href{https://arxiv.org/abs/1707.08973}{{\ttfamily
  1707.08973}}.

\bibitem{Chen:2017rxa}
B.~Chen, P.-C. Li and C.-Y. Zhang, \emph{{Einstein-Gauss-Bonnet Black Strings
  at Large $D$}}, \href{http://dx.doi.org/10.1007/JHEP10(2017)123}{\emph{JHEP}
  {\bfseries 10} (2017) 123},
  [\href{https://arxiv.org/abs/1707.09766}{{\ttfamily 1707.09766}}].

\bibitem{Kovtun:2004de}
P.~Kovtun, D.~T. Son and A.~O. Starinets, \emph{{Viscosity in strongly
  interacting quantum field theories from black hole physics}},
  \href{http://dx.doi.org/10.1103/PhysRevLett.94.111601}{\emph{Phys. Rev.
  Lett.} {\bfseries 94} (2005) 111601},
  [\href{https://arxiv.org/abs/hep-th/0405231}{{\ttfamily hep-th/0405231}}].

\bibitem{Caldarelli:2008mv}
M.~M. Caldarelli, O.~J.~C. Dias, R.~Emparan and D.~Klemm, \emph{{Black Holes as
  Lumps of Fluid}},
  \href{http://dx.doi.org/10.1088/1126-6708/2009/04/024}{\emph{JHEP} {\bfseries
  04} (2009) 024}, [\href{https://arxiv.org/abs/0811.2381}{{\ttfamily
  0811.2381}}].

\bibitem{Bhattacharya:2011eea}
J.~Bhattacharya, S.~Bhattacharyya and S.~Minwalla, \emph{{Dissipative
  Superfluid dynamics from gravity}},
  \href{http://dx.doi.org/10.1007/JHEP04(2011)125}{\emph{JHEP} {\bfseries 04}
  (2011) 125}, [\href{https://arxiv.org/abs/1101.3332}{{\ttfamily 1101.3332}}].

\bibitem{Banerjee:2012iz}
N.~Banerjee, J.~Bhattacharya, S.~Bhattacharyya, S.~Jain, S.~Minwalla and
  T.~Sharma, \emph{{Constraints on Fluid Dynamics from Equilibrium Partition
  Functions}}, \href{http://dx.doi.org/10.1007/JHEP09(2012)046}{\emph{JHEP}
  {\bfseries 09} (2012) 046},
  [\href{https://arxiv.org/abs/1203.3544}{{\ttfamily 1203.3544}}].

\bibitem{Bhattacharyya:2007vs}
S.~Bhattacharyya, S.~Lahiri, R.~Loganayagam and S.~Minwalla, \emph{{Large
  rotating AdS black holes from fluid mechanics}},
  \href{http://dx.doi.org/10.1088/1126-6708/2008/09/054}{\emph{JHEP} {\bfseries
  09} (2008) 054}, [\href{https://arxiv.org/abs/0708.1770}{{\ttfamily
  0708.1770}}].

\bibitem{Bhattacharyya:2012xi}
S.~Bhattacharyya, S.~Jain, S.~Minwalla and T.~Sharma, \emph{{Constraints on
  Superfluid Hydrodynamics from Equilibrium Partition Functions}},
  \href{http://dx.doi.org/10.1007/JHEP01(2013)040}{\emph{JHEP} {\bfseries 01}
  (2013) 040}, [\href{https://arxiv.org/abs/1206.6106}{{\ttfamily 1206.6106}}].

\bibitem{Armas:2013hsa}
J.~Armas, \emph{{How Fluids Bend: the Elastic Expansion for Higher-Dimensional
  Black Holes}}, \href{http://dx.doi.org/10.1007/JHEP09(2013)073}{\emph{JHEP}
  {\bfseries 09} (2013) 073},
  [\href{https://arxiv.org/abs/1304.7773}{{\ttfamily 1304.7773}}].

\bibitem{Armas:2013goa}
J.~Armas, \emph{{(Non)-Dissipative Hydrodynamics on Embedded Surfaces}},
  \href{http://dx.doi.org/10.1007/JHEP09(2014)047}{\emph{JHEP} {\bfseries 09}
  (2014) 047}, [\href{https://arxiv.org/abs/1312.0597}{{\ttfamily 1312.0597}}].

\bibitem{Armas:2014rva}
J.~Armas and T.~Harmark, \emph{{Constraints on the effective fluid theory of
  stationary branes}},
  \href{http://dx.doi.org/10.1007/JHEP10(2014)063}{\emph{JHEP} {\bfseries 10}
  (2014) 063}, [\href{https://arxiv.org/abs/1406.7813}{{\ttfamily 1406.7813}}].

\bibitem{Armas:2015ssd}
J.~Armas, J.~Bhattacharya and N.~Kundu, \emph{{Surface transport in
  plasma-balls}}, \href{http://dx.doi.org/10.1007/JHEP06(2016)015}{\emph{JHEP}
  {\bfseries 06} (2016) 015},
  [\href{https://arxiv.org/abs/1512.08514}{{\ttfamily 1512.08514}}].

\bibitem{Armas:2016mes}
J.~Armas, J.~Gath, V.~Niarchos, N.~A. Obers and A.~V. Pedersen, \emph{{Forced
  Fluid Dynamics from Blackfolds in General Supergravity Backgrounds}},
  \href{http://dx.doi.org/10.1007/JHEP10(2016)154}{\emph{JHEP} {\bfseries 10}
  (2016) 154}, [\href{https://arxiv.org/abs/1606.09644}{{\ttfamily
  1606.09644}}].

\bibitem{Armas:2016xxg}
J.~Armas, J.~Bhattacharya, A.~Jain and N.~Kundu, \emph{{On the surface of
  superfluids}}, \href{http://dx.doi.org/10.1007/JHEP06(2017)090}{\emph{JHEP}
  {\bfseries 06} (2017) 090},
  [\href{https://arxiv.org/abs/1612.08088}{{\ttfamily 1612.08088}}].

\bibitem{Armas:2017pvj}
J.~Armas and J.~Tarrio, \emph{{On actions for (entangling) surfaces and
  DCFTs}},  \href{https://arxiv.org/abs/1709.06766}{{\ttfamily 1709.06766}}.

\bibitem{Witten:1998zw}
E.~Witten, \emph{{Anti-de Sitter space, thermal phase transition, and
  confinement in gauge theories}},
  \href{http://dx.doi.org/10.4310/ATMP.1998.v2.n3.a3}{\emph{Adv. Theor. Math.
  Phys.} {\bfseries 2} (1998) 505--532},
  [\href{https://arxiv.org/abs/hep-th/9803131}{{\ttfamily hep-th/9803131}}].

\bibitem{Bhattacharyya:2008mz}
S.~Bhattacharyya, R.~Loganayagam, I.~Mandal, S.~Minwalla and A.~Sharma,
  \emph{{Conformal Nonlinear Fluid Dynamics from Gravity in Arbitrary
  Dimensions}},
  \href{http://dx.doi.org/10.1088/1126-6708/2008/12/116}{\emph{JHEP} {\bfseries
  12} (2008) 116}, [\href{https://arxiv.org/abs/0809.4272}{{\ttfamily
  0809.4272}}].

\bibitem{Haack:2008cp}
M.~Haack and A.~Yarom, \emph{{Nonlinear viscous hydrodynamics in various
  dimensions using AdS/CFT}},
  \href{http://dx.doi.org/10.1088/1126-6708/2008/10/063}{\emph{JHEP} {\bfseries
  10} (2008) 063}, [\href{https://arxiv.org/abs/0806.4602}{{\ttfamily
  0806.4602}}].

\bibitem{Bhattacharyya:2008jc}
S.~Bhattacharyya, V.~E. Hubeny, S.~Minwalla and M.~Rangamani, \emph{{Nonlinear
  Fluid Dynamics from Gravity}},
  \href{http://dx.doi.org/10.1088/1126-6708/2008/02/045}{\emph{JHEP} {\bfseries
  02} (2008) 045}, [\href{https://arxiv.org/abs/0712.2456}{{\ttfamily
  0712.2456}}].

\bibitem{Brattan:2011my}
D.~Brattan, J.~Camps, R.~Loganayagam and M.~Rangamani, \emph{{CFT dual of the
  AdS Dirichlet problem : Fluid/Gravity on cut-off surfaces}},
  \href{http://dx.doi.org/10.1007/JHEP12(2011)090}{\emph{JHEP} {\bfseries 12}
  (2011) 090}, [\href{https://arxiv.org/abs/1106.2577}{{\ttfamily 1106.2577}}].

\bibitem{Bhattacharyya:2013lha}
S.~Bhattacharyya, \emph{{Entropy current and equilibrium partition function in
  fluid dynamics}},
  \href{http://dx.doi.org/10.1007/JHEP08(2014)165}{\emph{JHEP} {\bfseries 08}
  (2014) 165}, [\href{https://arxiv.org/abs/1312.0220}{{\ttfamily 1312.0220}}].

\bibitem{Bhattacharyya:2014bha}
S.~Bhattacharyya, \emph{{Entropy Current from Partition Function: One
  Example}}, \href{http://dx.doi.org/10.1007/JHEP07(2014)139}{\emph{JHEP}
  {\bfseries 07} (2014) 139},
  [\href{https://arxiv.org/abs/1403.7639}{{\ttfamily 1403.7639}}].

\bibitem{Haehl:2013hoa}
F.~M. Haehl, R.~Loganayagam and M.~Rangamani, \emph{{Effective actions for
  anomalous hydrodynamics}},
  \href{http://dx.doi.org/10.1007/JHEP03(2014)034}{\emph{JHEP} {\bfseries 03}
  (2014) 034}, [\href{https://arxiv.org/abs/1312.0610}{{\ttfamily 1312.0610}}].

\bibitem{Haehl:2014zda}
F.~M. Haehl, R.~Loganayagam and M.~Rangamani, \emph{{The eightfold way to
  dissipation}},
  \href{http://dx.doi.org/10.1103/PhysRevLett.114.201601}{\emph{Phys. Rev.
  Lett.} {\bfseries 114} (2015) 201601},
  [\href{https://arxiv.org/abs/1412.1090}{{\ttfamily 1412.1090}}].

\bibitem{Haehl:2015pja}
F.~M. Haehl, R.~Loganayagam and M.~Rangamani, \emph{{Adiabatic hydrodynamics:
  The eightfold way to dissipation}},
  \href{http://dx.doi.org/10.1007/JHEP05(2015)060}{\emph{JHEP} {\bfseries 05}
  (2015) 060}, [\href{https://arxiv.org/abs/1502.00636}{{\ttfamily
  1502.00636}}].

\bibitem{Crossley:2015tka}
M.~Crossley, P.~Glorioso, H.~Liu and Y.~Wang, \emph{{Off-shell hydrodynamics
  from holography}},
  \href{http://dx.doi.org/10.1007/JHEP02(2016)124}{\emph{JHEP} {\bfseries 02}
  (2016) 124}, [\href{https://arxiv.org/abs/1504.07611}{{\ttfamily
  1504.07611}}].

\bibitem{Haehl:2015foa}
F.~M. Haehl, R.~Loganayagam and M.~Rangamani, \emph{{The Fluid Manifesto:
  Emergent symmetries, hydrodynamics, and black holes}},
  \href{http://dx.doi.org/10.1007/JHEP01(2016)184}{\emph{JHEP} {\bfseries 01}
  (2016) 184}, [\href{https://arxiv.org/abs/1510.02494}{{\ttfamily
  1510.02494}}].

\bibitem{Crossley:2015evo}
M.~Crossley, P.~Glorioso and H.~Liu, \emph{{Effective field theory of
  dissipative fluids}},
  \href{http://dx.doi.org/10.1007/JHEP09(2017)095}{\emph{JHEP} {\bfseries 09}
  (2017) 095}, [\href{https://arxiv.org/abs/1511.03646}{{\ttfamily
  1511.03646}}].

\bibitem{Haehl:2015uoc}
F.~M. Haehl, R.~Loganayagam and M.~Rangamani, \emph{{Topological sigma models
  \& dissipative hydrodynamics}},
  \href{http://dx.doi.org/10.1007/JHEP04(2016)039}{\emph{JHEP} {\bfseries 04}
  (2016) 039}, [\href{https://arxiv.org/abs/1511.07809}{{\ttfamily
  1511.07809}}].

\bibitem{Haehl:2016pec}
F.~M. Haehl, R.~Loganayagam and M.~Rangamani, \emph{{Schwinger-Keldysh
  formalism. Part I: BRST symmetries and superspace}},
  \href{http://dx.doi.org/10.1007/JHEP06(2017)069}{\emph{JHEP} {\bfseries 06}
  (2017) 069}, [\href{https://arxiv.org/abs/1610.01940}{{\ttfamily
  1610.01940}}].

\bibitem{Haehl:2016uah}
F.~M. Haehl, R.~Loganayagam and M.~Rangamani, \emph{{Schwinger-Keldysh
  formalism. Part II: thermal equivariant cohomology}},
  \href{http://dx.doi.org/10.1007/JHEP06(2017)070}{\emph{JHEP} {\bfseries 06}
  (2017) 070}, [\href{https://arxiv.org/abs/1610.01941}{{\ttfamily
  1610.01941}}].

\bibitem{Glorioso:2016gsa}
P.~Glorioso and H.~Liu, \emph{{The second law of thermodynamics from symmetry
  and unitarity}},  \href{https://arxiv.org/abs/1612.07705}{{\ttfamily
  1612.07705}}.

\bibitem{Jensen:2017kzi}
K.~Jensen, N.~Pinzani-Fokeeva and A.~Yarom, \emph{{Dissipative hydrodynamics in
  superspace}},  \href{https://arxiv.org/abs/1701.07436}{{\ttfamily
  1701.07436}}.

\bibitem{Gao:2017bqf}
P.~Gao and H.~Liu, \emph{{Emergent Supersymmetry in Local Equilibrium
  Systems}},  \href{https://arxiv.org/abs/1701.07445}{{\ttfamily 1701.07445}}.

\bibitem{Glorioso:2017fpd}
P.~Glorioso, M.~Crossley and H.~Liu, \emph{{Effective field theory of
  dissipative fluids (II): classical limit, dynamical KMS symmetry and entropy
  current}}, \href{http://dx.doi.org/10.1007/JHEP09(2017)096}{\emph{JHEP}
  {\bfseries 09} (2017) 096},
  [\href{https://arxiv.org/abs/1701.07817}{{\ttfamily 1701.07817}}].

\bibitem{Glorioso:2017lcn}
P.~Glorioso, H.~Liu and S.~Rajagopal, \emph{{Global Anomalies, Discrete
  Symmetries, and Hydrodynamic Effective Actions}},
  \href{https://arxiv.org/abs/1710.03768}{{\ttfamily 1710.03768}}.

\bibitem{Geracie:2017uku}
M.~Geracie, F.~M. Haehl, R.~Loganayagam, P.~Narayan, D.~M. Ramirez and
  M.~Rangamani, \emph{{Schwinger-Keldysh superspace in quantum mechanics}},
  \href{https://arxiv.org/abs/1712.04459}{{\ttfamily 1712.04459}}.

\bibitem{Heller:2015dha}
M.~P. Heller and M.~Spalinski, \emph{{Hydrodynamics Beyond the Gradient
  Expansion: Resurgence and Resummation}},
  \href{http://dx.doi.org/10.1103/PhysRevLett.115.072501}{\emph{Phys. Rev.
  Lett.} {\bfseries 115} (2015) 072501},
  [\href{https://arxiv.org/abs/1503.07514}{{\ttfamily 1503.07514}}].

\bibitem{Basar:2015ava}
G.~Basar and G.~V. Dunne, \emph{{Hydrodynamics, resurgence, and
  transasymptotics}},
  \href{http://dx.doi.org/10.1103/PhysRevD.92.125011}{\emph{Phys. Rev.}
  {\bfseries D92} (2015) 125011},
  [\href{https://arxiv.org/abs/1509.05046}{{\ttfamily 1509.05046}}].

\bibitem{Aniceto:2015mto}
I.~Aniceto and M.~Spaliński, \emph{{Resurgence in Extended Hydrodynamics}},
  \href{http://dx.doi.org/10.1103/PhysRevD.93.085008}{\emph{Phys. Rev.}
  {\bfseries D93} (2016) 085008},
  [\href{https://arxiv.org/abs/1511.06358}{{\ttfamily 1511.06358}}].

\bibitem{Buchel:2016cbj}
A.~Buchel, M.~P. Heller and J.~Noronha, \emph{{Entropy Production,
  Hydrodynamics, and Resurgence in the Primordial Quark-Gluon Plasma from
  Holography}}, \href{http://dx.doi.org/10.1103/PhysRevD.94.106011}{\emph{Phys.
  Rev.} {\bfseries D94} (2016) 106011},
  [\href{https://arxiv.org/abs/1603.05344}{{\ttfamily 1603.05344}}].

\bibitem{Spalinski:2017mel}
M.~Spaliński, \emph{{On the hydrodynamic attractor of Yang–Mills plasma}},
  \href{http://dx.doi.org/10.1016/j.physletb.2017.11.059}{\emph{Phys. Lett.}
  {\bfseries B776} (2018) 468--472},
  [\href{https://arxiv.org/abs/1708.01921}{{\ttfamily 1708.01921}}].

\bibitem{Emparan:2009cs}
R.~Emparan, T.~Harmark, V.~Niarchos and N.~A. Obers, \emph{{World-Volume
  Effective Theory for Higher-Dimensional Black Holes}},
  \href{http://dx.doi.org/10.1103/PhysRevLett.102.191301}{\emph{Phys. Rev.
  Lett.} {\bfseries 102} (2009) 191301},
  [\href{https://arxiv.org/abs/0902.0427}{{\ttfamily 0902.0427}}].

\bibitem{Emparan:2009at}
R.~Emparan, T.~Harmark, V.~Niarchos and N.~A. Obers, \emph{{Essentials of
  Blackfold Dynamics}},
  \href{http://dx.doi.org/10.1007/JHEP03(2010)063}{\emph{JHEP} {\bfseries 03}
  (2010) 063}, [\href{https://arxiv.org/abs/0910.1601}{{\ttfamily 0910.1601}}].

\bibitem{Emparan:2011br}
R.~Emparan, \emph{{Blackfolds}},
  \href{https://arxiv.org/abs/1106.2021}{{\ttfamily 1106.2021}}.

\bibitem{Camps:2012hw}
J.~Camps and R.~Emparan, \emph{{Derivation of the blackfold effective theory}},
  \href{http://dx.doi.org/10.1007/JHEP03(2012)038,
  10.1007/JHEP06(2012)155}{\emph{JHEP} {\bfseries 03} (2012) 038},
  [\href{https://arxiv.org/abs/1201.3506}{{\ttfamily 1201.3506}}].

\bibitem{Lehner:2010pn}
L.~Lehner and F.~Pretorius, \emph{{Black Strings, Low Viscosity Fluids, and
  Violation of Cosmic Censorship}},
  \href{http://dx.doi.org/10.1103/PhysRevLett.105.101102}{\emph{Phys. Rev.
  Lett.} {\bfseries 105} (2010) 101102},
  [\href{https://arxiv.org/abs/1006.5960}{{\ttfamily 1006.5960}}].

\bibitem{poisson_2004}
E.~Poisson, \emph{A Relativist's Toolkit: The Mathematics of Black-Hole
  Mechanics}.
\newblock Cambridge University Press, 2004,
  \href{http://dx.doi.org/10.1017/CBO9780511606601}{10.1017/CBO9780511606601}.

\bibitem{Lahiri:2007ae}
S.~Lahiri and S.~Minwalla, \emph{{Plasmarings as dual black rings}},
  \href{http://dx.doi.org/10.1088/1126-6708/2008/05/001}{\emph{JHEP} {\bfseries
  05} (2008) 001}, [\href{https://arxiv.org/abs/0705.3404}{{\ttfamily
  0705.3404}}].

\bibitem{Bhattacharya:2009gm}
J.~Bhattacharya and S.~Lahiri, \emph{{Lumps of plasma in arbitrary
  dimensions}}, \href{http://dx.doi.org/10.1007/JHEP08(2010)073}{\emph{JHEP}
  {\bfseries 08} (2010) 073},
  [\href{https://arxiv.org/abs/0903.4734}{{\ttfamily 0903.4734}}].

\bibitem{Gibbons:1976ue}
G.~W. Gibbons and S.~W. Hawking, \emph{{Action Integrals and Partition
  Functions in Quantum Gravity}},
  \href{http://dx.doi.org/10.1103/PhysRevD.15.2752}{\emph{Phys. Rev.}
  {\bfseries D15} (1977) 2752--2756}.

\end{thebibliography}\endgroup
\bibliographystyle{JHEP}

\end{document}